\documentclass[twocolumn]{aa}  
\usepackage{longtable}
\usepackage{pdflscape} 
\usepackage{xcolor}
\usepackage{listings}
\usepackage{natbib}
\usepackage{subcaption}
\bibpunct{(}{)}{;}{a}{}{,} 
\usepackage{booktabs,caption}
\usepackage[flushleft]{threeparttable}
\usepackage{adjustbox}
\newcommand\gaia{\textit{Gaia}}

\definecolor{codegreen}{rgb}{0,0.6,0}
\definecolor{codegray}{rgb}{0.5,0.5,0.5}
\definecolor{codepurple}{rgb}{0.58,0,0.82}
\definecolor{backcolour}{rgb}{0.95,0.95,0.92}

\lstdefinestyle{mystyle}{
    commentstyle=\color{codegreen},
    keywordstyle=\color{magenta},
    numberstyle=\tiny\color{codegray},
    stringstyle=\color{codepurple},
    basicstyle=\ttfamily,
    breakatwhitespace=false,         
    breaklines=true,                 
    captionpos=b,                    
    keepspaces=true,                 
    numbers=none,                    
    showspaces=false,                
    showstringspaces=false,
    showtabs=false,                  
    tabsize=2
}
\title{Cluster Ages to Reconstruct the Milky Way Assembly (CARMA)\\I. The final word on the origin of NGC~6388 and NGC~6441}
\titlerunning{CARMA I.}
\authorrunning{ Massari et al.}
\author{Davide Massari \inst{1}, Fernando Aguado-Agelet \inst{2,3}, Matteo Monelli \inst{3,4,5}, Santi Cassisi \inst{6,7}, Elena Pancino \inst{8}, Sara Saracino \inst{9}, Carme Gallart \inst{3,4}, Tom\'as Ruiz-Lara \inst{10}, Emma Fern\'andez-Alvar \inst{3,4}, Francisco Surot \inst{3,4},  Amalie Stokholm \inst{11, 1, 12}, Maurizio Salaris \inst{9,6}, Andrea Miglio\inst{11}, Edoardo Ceccarelli \inst{1,11}
}
\institute{INAF - Osservatorio di Astrofisica e Scienza dello Spazio di Bologna, Via Gobetti 93/3, I-40129 Bologna, Italy\\ \email{davide.massari@inaf.it}
             \and
              atlanTTic, Universidade de Vigo, Escola de Enxeñar\'ia de Telecomunicaci\'on, 36310, Vigo, Spain 
             \and
             Universidad de La Laguna, Avda. Astrof\'isico Fco. S\'anchez, E-38205 La Laguna, Tenerife, Spain
             \and
             Instituto de Astrof\'isica de Canarias, Calle V\'ia L\'actea s/n, E-38206 La Laguna, Tenerife, Spain
             \and 
             INAF – Osservatorio Astronomico di Roma, Via Frascati 33, 00078 Monte Porzio Catone, Roma, Italy
             \and
             INAF – Osservatorio Astronomico di Abruzzo, Via M. Maggini, 64100 Teramo, Italy
             \and
             INFN - Sezione di Pisa, Universit\'a di Pisa, Largo Pontecorvo 3, 56127 Pisa, Italy
             \and
             INAF - Osservatorio Astrofisico di Arcetri, Largo E. Fermi 5, I-50125 Firenze, Italy
             \and
             Astrophysics Research Institute, Liverpool John Moores University, 146 Brownlow Hill, Liverpool L3 5RF, UK
             \and
             Universidad de Granada, Departamento de Física Teórica y del Cosmos, Campus Fuente Nueva, Edificio Mecenas, 18071 Granada, Spain
             \and
             Dipartimento di Fisica e Astronomia, Universit\`a degli Studi di Bologna, Via Piero Gobetti 93/2, 40129 Bologna, Italy
             \and 
             Stellar Astrophysics Centre, Department of Physics and Astronomy, Aarhus University, Ny Munkegade 120, DK-8000 Aarhus C, Denmark
}

\abstract{
We present CARMA, the Cluster Ages to Reconstruct the Milky Way Assembly project, that aims at determining precise and accurate age measurements for the entire system of known Galactic globular clusters and at using them to trace the most significant merger events experienced by the Milky Way. The strength of CARMA relies on the use of homogeneous photometry, theoretical isochrones, and statistical methods, that will enable to define a systematic-free chronological scale for the complete sample of Milky Way globulars. In this paper we describe the CARMA framework in detail, and present a first application on a sample of six metal-rich globular clusters with the aim of putting the final word on the debated origin of NGC~6388 and NGC~6441. Our results demonstrate that this pair of clusters is coeval with other four systems having a clear in-situ origin. Moreover, their location in the age-metallicity plane matches the one occupied by in-situ field stars. Such an accurate age comparison enabled by the CARMA methodology rules out the possibility that NGC~6388 and NGC~6441 have been accreted as part of a past merger event.
}


\begin{document}
\flushbottom
\maketitle
\thispagestyle{empty}

\section{Introduction}

The precise six-dimensional phase space information provided by the ESA-{\it Gaia} space mission \citep[][]{edr3, gaiadr3} in terms of position, parallax, proper motion and line-of-sight velocity measurements has revolutionised our view of the early history of the Milky Way \citep[see][for a review]{helmi20}. In combination with chemical abundances provided by large spectroscopic surveys such as APOGEE \citep{apogee}, GALAH \citep{galah}, the Gaia ESO Survey \citep{gaiaeso}, H3 \citep{h3},  RAVE \citep{rave}, SEGUE \citep{segue} and LAMOST \citep{lamost}, this groundbreaking wealth of information has enabled to disentangle the contribution of several past merger events in building-up the halo of our Galaxy.

 The picture stemming out from the analysis conducted on this unprecedented data set identified as the main component of the local halo the debris of the Gaia-Enceladus-Sausage (GES) dwarf galaxy \citep{helmi18, belokurov18}. GES is unanimously recognised as the latest significant merger event experienced by the Milky Way, also responsible for dynamically heating the proto-thick disk that was already in place at the time of GES accretion \citep[e.g., ][]{dimatteo19, gallart19, belokurov20}. Other than GES, and in addition to the obvious contribution by the Sagittarius dwarf Galaxy \citep{ibata94}, other prominent substructures have been discovered as over-densities showing some degree of coherency in the chemo-dynamical space populated by halo stars and globular clusters, such as the Helmi streams \citep{helmi99, koppelmanh99}, Sequoia \citep{myeong19}, Thamnos \citep{koppelman19} and Kraken \citep{kruijssen19, massari19}. A first merger tree of the Milky Way based on the properties of these progenitors has been depicted by \cite{kruijssen20}.

With the constant improvement of the quality and completeness of the available data, this early picture has become more and more complex. Many new substructures in the dynamical space have been discovered \citep[see e.g.][]{yuan20, naidu20, necib20, aguado21, horta21, refiorentin21, mardini22, malhan22, myeong22, oria22, tenachi22, ruizlara22, dodd23, mikkola23}. Yet, their interpretation as independent merger events, rather than spurts from already known progenitors or even in-situ structures originated from the Milky Way disk, is made very challenging by the fact that $i$) these structures often dynamically overlap, and $ii$) the  distribution in energy of the debris of progenitors that sink rapidly shows multiple bumps and wrinckles, originated at each stripping event \citep{amarante22, khoperskov22}. In these cases, the addition of the chemical information is crucial, and yet sometimes not sufficient to unambiguously solve the most controversial cases \citep[e.g.][]{monty20, feuillet21, malhan22, horta23, monty23}. Additional information is therefore required, and one key ingredient could be provided by stellar ages.

Despite their importance, estimating accurate stellar ages remains a challenging task in astrophysics \citep[][]{soderblom2010}.
Many of the current limitations stem from our poor understanding of some of the physical processes involved in stellar evolution and the difficulty in controlling systematic effects on the observational measurements \citep[][]{lebreton14, lebreton14b}. 
Most age-dating methods rely on measurements of stellar properties such as luminosity, chemical composition, and surface temperature (which can be affected by various factors such as stellar activity or atmospheric dynamics), and on their comparison with the predictions of stellar evolution models. Uncertainties in our knowledge of the underlying physics and assumptions made in the models (both stellar evolution ones and synthetic spectra) also translate in uncertainties in the age estimates.
Age measurements for individual stars in the old regime are becoming more and more precise, though. On the one hand, this is thanks to the availability of homogeneous photometric and spectroscopic data for large samples of stars, that help in overcoming some of the systematic effects \citep[see e.g.,][]{gallart19, xiang22}. On the other hand, the development of asteroseismology has led to age measurement precisions of $\sim10$\% \citep[see e.g.,][]{montalban21}. The asteroseismic age-dating technique, that does not require large sample of stars to achieve high precision, and that depends more weakly on the estimate of stellar photospheric properties, is particularly promising in this respect \citep[see e.g.,][]{verma2022,tailo22}, but further refinement and testing are still required. 

Among the tracers of the Milky Way assembly history, those for which age can be measured in the most precise and accurate way are globular clusters (GCs). When considering relative ages, the current best measurements can achieve a precision of the order of $\sim500$ Myr \citep{vandenberg13}. With these precise measurements, the age-metallicity relation (AMR) of Milky Way GCs has proved to be a powerful tool to assess the origin of GCs as in-situ or accreted stellar systems \citep[see e.g.,][]{forbes10, leaman13, kruijssen19, massari19, callingham22}. Even more precise measurements might lead to distinguish the different progenitors of the accreted ones, and in turn to precisely characterise their accretion time \citep{kruijssen20}.
Unfortunately, precise GC age measurements are limited to relatively small samples. Different age indicators, photometric catalogues, assumptions on distance and reddening, and theoretical models are examples of the many sources of systematic uncertainties that affect different compilations of GC ages. By comparing the results from \citet{forbes10} and \citet{vandenberg13}, \citet{massari19} demonstrated that these systematic errors can add up to differences of 2 or more Gyrs, and that they are not trivial to properly take into account and correct for, as they might depend on the GC age itself as well as on the GC metallicity.

The objective of CARMA (Cluster Ages to Reconstruct the Milky-way Assembly) is to overcome this limitation by building up the first complete and homogeneous catalog of Milky Way GC ages. In combination with the GCs dynamical and chemical properties, this will enable the complete characterisation of the accretted and in-situ population of Milky Way GCs, by solving the cases with still ambiguous origin \citep[see e.g.][]{minelli21, carretta22}, and at the same time to characterise the progenitors of the past merger events in terms of accretion time thanks to the analysis of their AMRs, and from the analysis of the colour-magnitude diagrams (CMDs) of the progenitor systems through CMD-fitting \citep[e.g.,][]{gallart19, ruizlara22b}, whenever possible.

In this first paper of the series we present the method that CARMA will employ to determine the GCs ages, and we describe its first application to put the final word on the determination of the origin of two GCs, namely NGC~6388 and NGC~6441, whose in-situ or accreted nature cannot be unambiguously determined by using neither the dynamical information nor their chemical properties. To this end, we derive homogeneous ages for these two GCs and for four additional GCs of similar metallicity ([Fe/H]$\simeq-0.5$), and for which there is agreement in the literature about their in-situ origin, based on dynamical \citep[see][]{massari19, forbes20, bajkova20, callingham22} and chemical \citep[][]{minelli21} evidence. As additional evidence, we also compare the ages and metallicities of these clusters with the two age-metallicity sequences of the Milky Way kinematic halo (one associated to GES, the other to heated early disk) derived using {\it Gaia} DR3 data and updated methodology with respect to \cite{gallart19}.

The paper is organised as follows. Section \ref{method} presents the method adopted by CARMA to determine GC ages, including a description of the theoretical models and of the isochrone-fitting algorithm employed for this goal. In Section \ref{application} the proof-of-concept scientific case, concerning the origin of the pair of GCs NGC~6388 and NGC~6441, is discussed together with the data used in the analysis. The results of the investigation are presented in Section \ref{results}, and the final conclusions on the origin of the two GCs are drawn in Section \ref{conclusion}.

\section{The method}\label{method}

In this Section we present the method employed to determine the age of each GC, which is based on the isochrone fitting code presented in \citet{saracino19}, here refined and updated. We also present a brief summary of the CMD-fitting technique applied here to derive the age-metallicity relation of Milky Way halo field stars, highlighting the differences with respect to the original work by \cite{gallart19}.

\subsection{The stellar evolution framework}\label{models}
CARMA relies on the adoption of the stellar theoretical models provided within the latest release of the Bag of Stellar Tracks and Isochrones (BaSTI) database \citep{hidalgo18, pietrinferni21}. More specifically, for the present analysis we adopt the model sets accounting for the occurrence of diffusive processes \citep[we refer to][for a detailed discussion on the adopted input physics and physical assumptions]{hidalgo18}. For the sake of the project, which is ensuring the highest possible degree of accuracy in terms of age determination, we decided to only use solar-scaled models. This prescription and the use of the global metallicity [M/H] rather than the iron content [Fe/H] as input for the code allow us to avoid making any assumption on the $\alpha$-element abundance of each GC, which would otherwise be based on heterogeneous measurements coming from many different sources. \cite{salaris93} has in fact demonstrated that the impact of different [$\alpha$/Fe] mixtures on theoretical models for population II stars can simply be treated as an additional term on the global metallicity [M/H], according to the relation\footnote{The coefficients of this relation
are slightly different from the ones provided by \cite{salaris93} in order to take into account the use of a different reference solar mixture between the BaSTI models, that are based on the heavy element distribution provided by \cite{caffau11}, whereas the models by \cite{salaris93} were based on the \cite{rossaller} mixture.}

\begin{equation}
    [M/H] = [Fe/H]+\log(0.694\times10^{[\alpha/Fe]}+0.301).
\end{equation}

This is especially true when working in optical-infrared colour combinations adopted in this work, as in these bands the effect of the specific $\alpha$-elements distribution on the bolometric corrections becomes negligible \citep{cassisi04}
Estimates of [$\alpha$/Fe] for the target GCs will only be used a-posteriori to perform a direct comparison of our best-fit metallicity with existing (possibly high-resolution) spectroscopic measurements.

The whole sets of isochrones adopted in the present work have been transferred from the theoretical H-R diagram to the various relevant photometric planes by adopting self-consistent color-temperature ($T_{eff}$) relations and bolometric corrections \citep[for more details we refer to][]{hidalgo18}.

The ratio between the extinction in a given photometric band $A_\lambda$ and $A_V$ depends on the flux distribution of the stellar source, and is in principle dependent on parameters such as $T_{eff}$, surface gravity
($\log~g$), and the chemical composition \citep[see, e.g.,][for a discussion on this issue]{bedin05a}.
When $A_V$ is small, this effect is also small and a single value of $A_\lambda/A_V$ can be safely applied along the whole isochrone. But for those clusters affected by large extinction (tipically for values larger than $E(B-V)=0.10$) it is necessary to consider the variation of $A_\lambda/A_V$ along the isochrones in the fit to the observed CMD. Such an effect is more and more significant in bluer photometric passbands. 

To properly take into account this effect in the procedure of isochrone-fitting to the CMD of those clusters affected by large extinction, we have applied $T_{eff}$-dependent reddening corrections to the magnitudes of the theoretical isochrones. The $T_{eff}$-dependent corrections have been evaluated by adopting the web interface\footnote{http://stev.oapd.inaf.it/cgi-bin/cmd} that implements the prescriptions by \cite{girardi08}, to determine the extinctions in the various photometric passbands, covering a wide range of $T_{eff}$ and values of the interstellar extinction. Fig.~1 shows how the total extinction in the two filters used in this work, namely A$_{F606W}$ and A$_{F814W}$ vary as a function of $T_{eff}$ and $E(B-V)$.

\begin{figure*}[!htb]
\center{
\includegraphics[width=\textwidth]{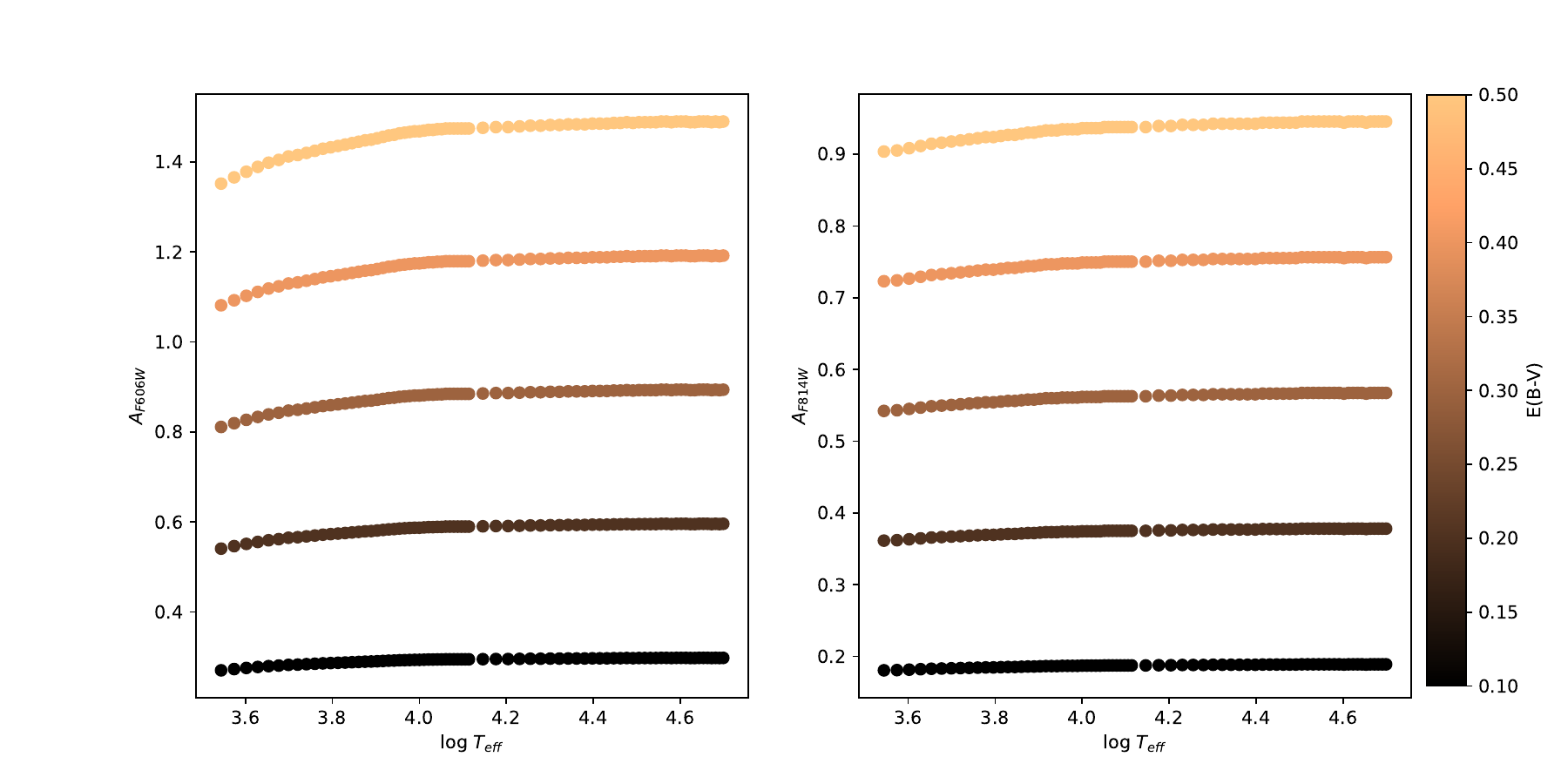}
\caption{Behaviour of the total extinction in the two filters used in this work, A$_{F606W}$ (left-hand panel) and A$_{F814W}$ (right-hand panel), as a function of T$_{eff}$ and colour-coded based on E(B-V).}}\label{fig:teff_red}
\end{figure*} 

\subsection{The isochrone-fitting code}\label{code}

The isochrone fitting algorithm developed within CARMA is a series of procedures that provide age, metallicity, distance and reddening best-fit values within a Markov Chain Monte Carlo (MCMC) statistical framework, thus further associating robust uncertainty estimates to all of the output parameters.
The steps followed by the algorithm are described in detail as follows:

\begin{enumerate}
    \item	The first step of the algorithm consists in the construction of the best CMD to be fit. This includes the application of photometric quality cuts, the selection of GC member stars via kinematic information (whenever available) or via a selection based on the distance from the cluster centre, the correction for differential reddening effects, and the exclusion of obvious photometric binaries along the CMD main sequence. The details of this step applied to the case under study are described in Sect.~\ref{subsec:data_set}. 
    \item	A grid of BaSTI \citep{hidalgo18} theoretical isochrones in the appropriate photometric bands is created with a very fine sampling of the age-metallicity space. The adopted step in [M/H] is of 0.01 dex, whereas the step in age is of 100 Myr, but  a linear interpolation algorithm\footnote{ The interpolation is performed by means of the Python library scipy.interpolate.1d \citep{scipy}. Different choices of interpolation algorithms might introduce further small systematic errors on the age estimates, but CARMA avoids this source of inhomogeneity as well.} enables sampling the parameter space even more finely. As explained in Sect.~\ref{models}, the adopted models have a solar-scaled [$\alpha$/Fe] mixture, they include diffusion effects as well as mass loss, and they are corrected for $T_{eff}$-dependent reddening effects. Finally, they are converted to the observational plane by applying the cluster distance modulus and extinction, by assuming the reddening law by \cite{cardelli89}.
    \item	At this point a fitting function, which measures the goodness-of-fit between the isochrones and the observed CMD, is defined. The portion of the isochrones that is fit to the CMD goes from the bottom of the main sequence to the tip of the red giant branch. Building upon the method presented in \cite{saracino19} our proposed function is the sum of two terms. The first term ($\mathcal{L}_{priors}$) is the likelihood evaluating the consistency of the inferred parameters with the initial priors on [M/H], distance and E(B-V). It has a Gaussian form, and when expressed in natural logarithm can be written as:
    \begin{equation}
        \mathcal{L}_{priors}=\mathcal{L}_{E(B-V)}+\mathcal{L}_{DM}+\mathcal{L}_{[M/H]},
    \end{equation}
    where each individual term is in turn expressed as the natural logarithm of a Gaussian function:
    \begin{equation}
        \mathcal{L}_{x}=-0.5\times (x-x_{prior})^2/x_{std}^2,
    \end{equation}

    with $x=$[E(B-V), DM, [M/H]] and $x_{prior}$ and $x_{std}$ being the adopted values of the priors and their associated uncertainties, respectively. 
    The second term ($\mathcal{L}_{fit}$) is the likelihood associated to the fit of the $i$ individual points of the CMD, and is computed as: 
    \begin{equation}
        \mathcal{L}_{fit} = \sum_{i=1}^{N}[\min(dist_{i})]^{2}/\sigma_{i}^{2},
    \end{equation}
    where $N$ is the total number of stars in the CMD, $\min(dist)$ is the minimum distance between data and model and $\sigma$ is the photometric error.
    After some testing, we realised that in a few cases the fitting algorithm prefers local solutions that are significantly off (by more than 0.5 dex) of the spectroscopic measurements of [M/H] used as prior. Given that the spectroscopic metallicity is an independent and very robust prior, we decided to help the algorithm in avoiding these local solutions, clearly offset even by visual inspection, by assigning greater importance to the first term, with a weight of 1. Conversely, the second term measuring the disparity between the CMD and the isochrone is assigned a lower weight, but not so low as to force the solution on the priors. Rather, we want the MCMC chains to still sample a wide range of parameters, that can deviate from the values of the priors, still providing visually good fits. After some tests, we find that the best weight to assign to the second term in this sense is 0.3, so that:
    \begin{equation}
       \mathcal{L}_{tot} = \mathcal{L}_{priors} + 0.3\times\mathcal{L}_{fit} 
    \end{equation}
    \item	The algorithm then looks for the best-fitting isochrone, meaning the model that minimizes $\mathcal{L}_{tot}$, by exploring the parameter space (age, [M/H], distance and reddening) by means of the emcee \citep{emcee} python package that provides an efficient implementation of the affine-invariant MCMC ensemble planner. 
    \item	Finally, the algorithm provides the best solution in terms of age, [M/H], distance and E(B-V), together with the associated uncertainties (corresponging to the 16th and the 84th percentiles of the posterior distributions) and their correlation parameter. 
\end{enumerate}
The results coming from the application of this algorithm to the GCs under study are shown in the Appendix.

\subsection{The CMD-fitting methodology}\label{sec:CMDftGaia}

The star formation history (SFH), and the distribution of ages and metallicities of the stars in a complex stellar system can be quantitatively retrieved from the comparison of its observed CMD reaching the oldest main sequence turnoff, with theoretical CMDs derived from stellar evolution models, after observational effects are properly taken into account \citep[e.g., ][]{gallart99, dolphin02, cignoni10, monelli10}. The application of this technique to Gaia DR2 data by members of our team is discussed in \cite{gallart19} and \cite{ruizlara20}, while updated procedures, that we call CMDft.Gaia are introduced in \cite{ruizlara22b} and in more detail in Gallart et al. (in prep). We refer to these works for a detailed description of the methodology, while a brief summary is presented here.

{CMDft.Gaia is a suite of procedures that includes i) the computation of synthetic CMDs in the Gaia bands, adopting a given set of stellar evolution models, IMF and a parameterization of the binary star population; ii) the simulation in the synthetic CMDs of the observational errors and completeness affecting the observed CMD after quality and reddening cuts; and iii) the derivation of the SFH with $Dir$SFH, which finds the combination of simple stellar populations (SSPs) that best fits the observed CMD. A distinctive feature of $Dir$SFH is that it defines the SSPs with a $Dir$ichlet tesselation \citep[][]{green78} of the synthetic CMD from a grid of seed points within the available range of ages and metallicities. The final SFH is derived as the weighted average of a large number (of the order of 100) individual solutions obtained by slightly modifying the grid of age and metallicity seed points.

\section{Application: the origin of NGC~6388 and NGC~6441}\label{application}

By using a combination of dynamical properties and AMR data, \cite{massari19} associated the 151 GCs having {\it Gaia} DR2 proper motions to their most likely galaxy progenitor, being this the Milky Way or one of the past accretion events described in the Introduction. Some of these associations are naturally uncertain, as for example the location of a GC in the (E, L$_{z}$) integrals of motion space is sometimes at the boundary between regions associated with different progenitors. Among these uncertain associations, those concerning the pair NGC~6388-NGC~6441 have been of particular interest in the literature. According to \cite{massari19}, NGC~6388 is an in-situ GC, associated with the bulge of our Galaxy, while NGC~6441 has been accreted during the merger event involving the Kraken dwarf galaxy\footnote{Kraken is referred to as Low-Energy group in the nomenclature by \cite{massari19}}. These associations did not take into account any age estimate as the authors demonstrated that for [Fe/H]$\gtrsim-0.5$ (as is the case for these two GCs) the systematic uncertainties among different compilations of age measurements can reach up to more than 2 Gyr. They are thus based purely on orbital properties computed using the distances provided by the Harris catalog \citep{harris96}. When adopting the distances provided by \cite{baumgardt21}, instead, NGC~6388 orbital apocentre increases from 3.4 kpc to 4.2 kpc, and the cluster enters the region occupied by Kraken's GCs. Viceversa, the vertical angular momentum of NGC~6441 increases from L$_{z}=268$ km/s kpc to L$_{z}=519$ km/s kpc, so that the cluster moves from the Kraken to an in-situ association with the Milky Way disk. According to \citet{forbes20} and \citet{callingham22}, both clusters are in-situ.

Such a dynamical ambiguity did not find solution when including the information on the clusters chemistry. By performing a direct relative comparison among the abundances of [Zn/Fe], [Sc/Fe] and [V/Fe] of these two GCs and a sample of undoubtedly in-situ GCs, \cite{minelli21} found that NGC~6388 and NGC~6441 show systematically lower abundances, and hence recognised both clusters as accreted. By investigating the same chemical elements for NGC~6388, \cite{carretta22} instead reached an opposite conclusion, associating the cluster to the Milky Way. Finally, by analysing APOGEE $\alpha$-element abundance, \cite{horta20} suggested a possible accreted origin for NGC~6388.

This open case thus sets a perfect stage for CARMA to demonstrate the importance of accurate and precise age measurements to help unravelling the origin of the entire system of Milky Way GCs, and in turn to contribute in reconstructing our Galaxy assembly history. In particular, we propose to determine accurate relative ages for a sample including these two GCs with ambiguous origin, as well as four clusters in the same metallicity range ([Fe/H]$\sim-0.5$) whose origin is unambiguously in-situ, according to all the indicators (dynamics and chemistry) and to the different studies in the literature \citep{massari19, forbes20, bajkova20, callingham22}. These four clusters are NGC~5927, NGC~6304, NGC~6352 and NGC~6496. If NGC~6388 and NGC~6441 were of accreted origin, according to the behavior of the AMR observed in dwarf galaxies \citep{kruijssen19} they should be about 2 Gyr younger than the sample of in-situ clusters.

\subsection{The globular cluster data set} \label{subsec:data_set}

In order to ensure the highest accuracy on this differential age comparison, for all of the six GCs we used {\it Hubble Space Telescope} photometry taken with the Wide Field Channel of the Advanced Camera for Survey (ACS/WFC) in the F606W and F814W filters. The catalogs produced with the KS2 code \citep{bellini17} have been made public by the HUGS survey \citep{piotto15, nardiello18}, and the photometry comes from observations of the ACS Survey of Galactic Globular Clusters \citep[GO-10775, PI: A. Sarajedini, see][]{sarajedini07}. By following the prescriptions in \cite{bellini17, nardiello18}, we adopted the photometry originated by KS2 method-2, as it is best suited for faint stars and crowded environments. 

Before running the isochrone-fitting code on the photometric catalogs, we took some other actions to perform our age estimates on the best possible photometry.
First of all, the HUGS survey provide proper-motions based probability memberships for all the sources in the catalogs. We thus restricted our analysis only to stars having a probability membership $>90$\%, which is particularly crucial in the regions highly contaminated by field stars such as those populated by NGC~6388 and NGC~6441.

\begin{figure}[thb]
\center{
\includegraphics[width=\columnwidth]{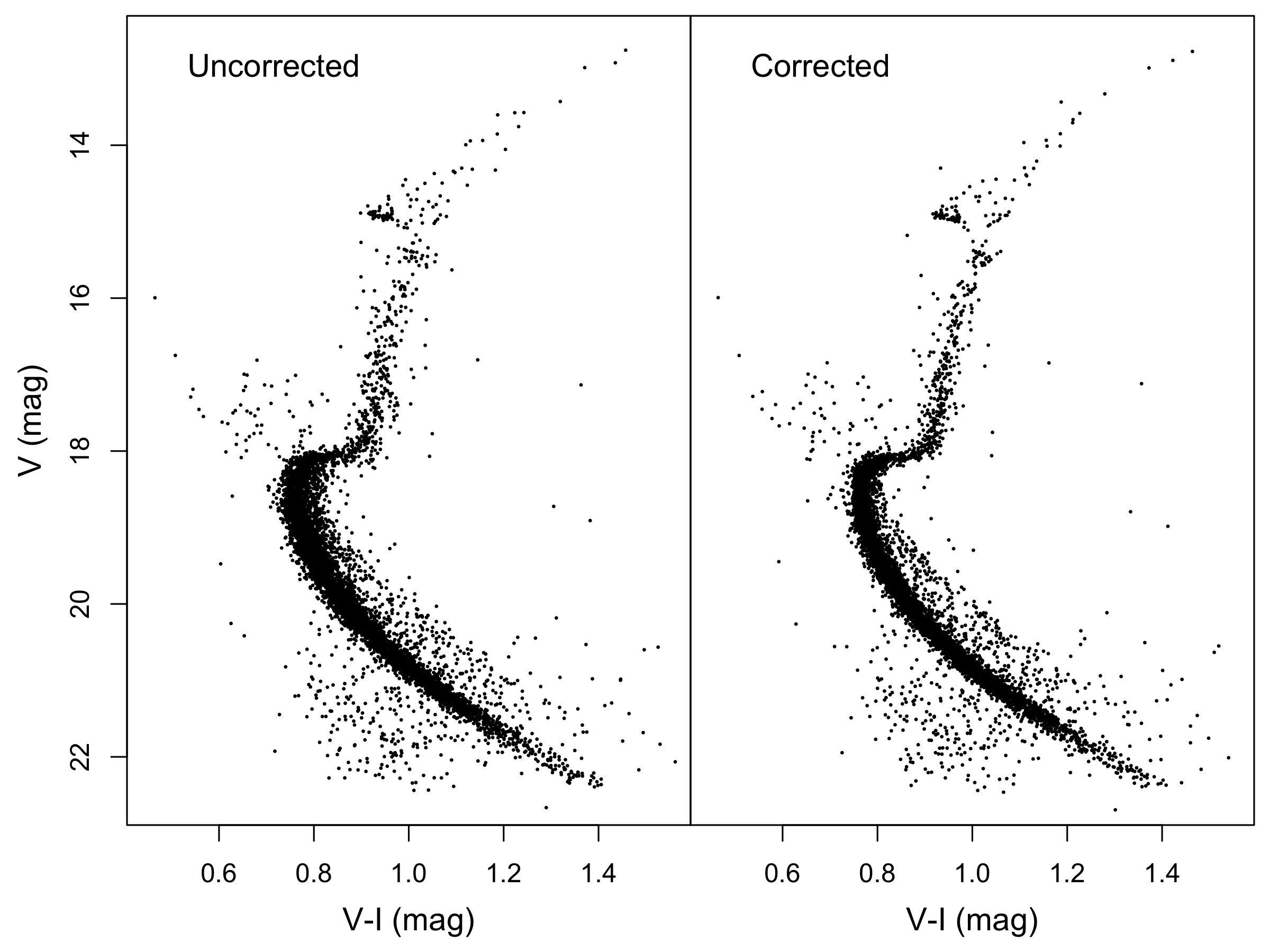}
}
\caption{Example of differential reddening correction in the case of NGC~6352. The original CMD is shown on the left and the corrected one on the right.}
\label{fig:drcmd}
\end{figure} 

After the membership selection, for each cluster we performed a differential reddening correction using the method described by \cite{milone12}. Briefly, we used main sequence stars down to four magnitudes below the turnoff as the reference sample, and paid particular attention in removing the equal-mass binary sequence from it. Main sequence stars are preferred over red giants as they are more numerous and thus ensure higher resolution in the determination of the reddening spatial variation. Then, for each individual cluster member, we estimated the differential reddening value d[E(B--V)] as the median offset from the mean ridge line along the reddening vector, computed among the 60 closest reference neighbors. We adopted an R$_V$=A$_V$/E(B--V) of 3.1 and the reddening law by \cite{cardelli89}.  This process has been repeated iteratively, typically 2-3 times depending on the GC, until the residual d[E(B-V)] values matched the typical photometric error. 
An example of the resulting correction on the CMD can be found in Fig.~\ref{fig:drcmd}. The associated reddening map is instead shown in Fig.~\ref{fig:drmap}.

\begin{figure}[thb]
\center{
\includegraphics[width=\columnwidth]{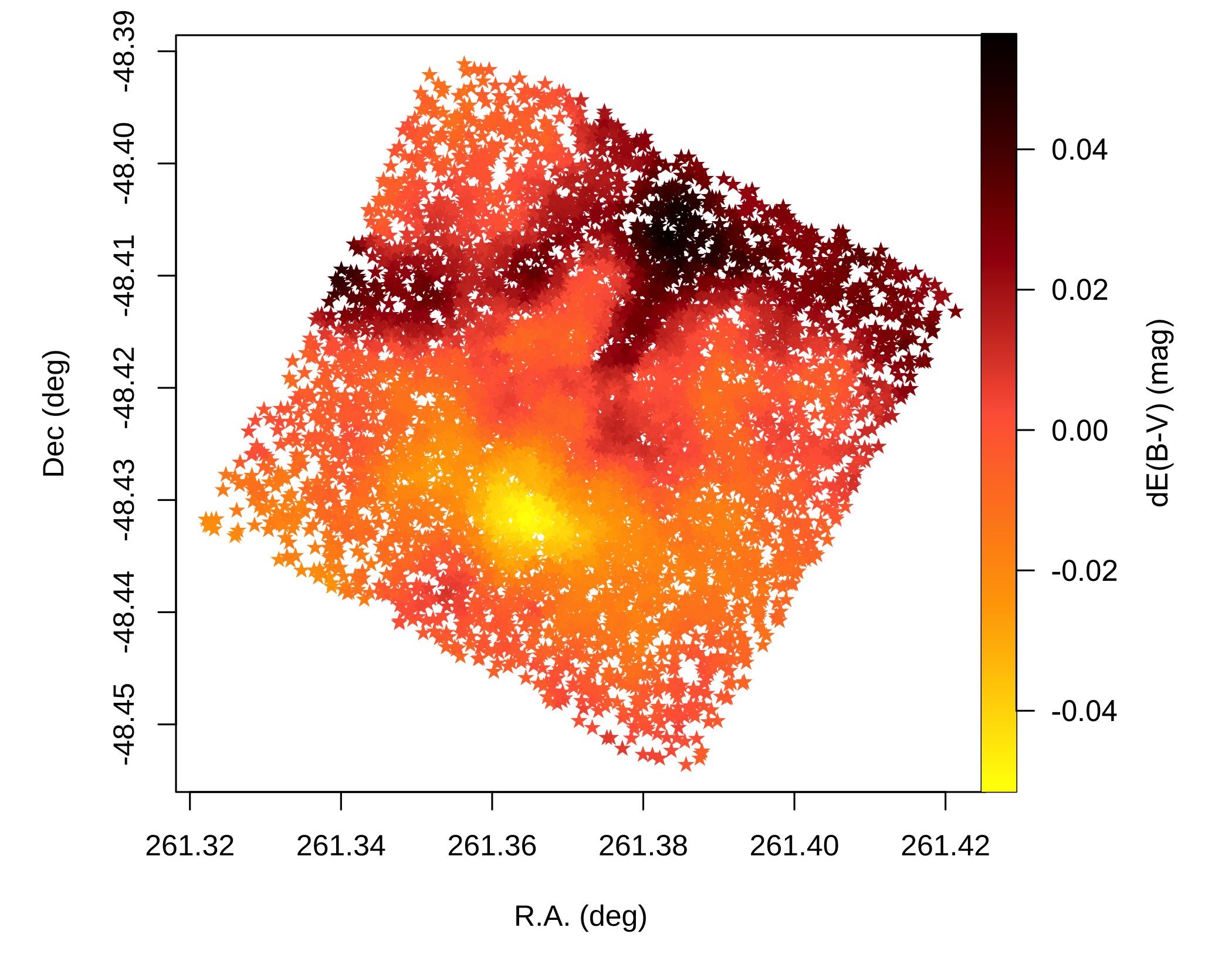}
}
\caption{Example of a differential reddening map for NGC~6352. Each of the stars used to compute the differential reddening correction is plotted in sky coordinates and colored with its value of dE(B--V).}
\label{fig:drmap}
\end{figure} 

The differential reddening-corrected, membership-selected catalogs have then been further cut by excluding the innermost regions of each cluster, with a distance cut that ranged between a projected radius R=20 arcsec and R=60 arcsec depending on each GC density and core radius size. Such a cut introduces a double advantage. On the one hand, it excludes sources in the most crowded regions, where the photometry is be affected by bad measurements. On the other hand, given that we are solely interested in the age measurement, we do not want the complexity of GC stellar populations \citep[the so-called multiple-population phenomenon, see][for a review]{gratton19} to affect our estimates. With only a few exceptions \citep[see e.g.,][]{leitinger23}, all the observational results agree in finding that regardless of the kind of complexity involved (from the iron spread, to the helium variation, and the (anti-)correlation affecting lighter elements, see e.g. \cite{pancino03, piotto07, massari14, milone17}, the first primordial population of GC is less centrally concentrated than any other peculiar one. The choice of colour-magnitude diagrams (CMDs) in the F606W and F814W filters already ensures that only possible iron- and helium-spreads should affect the CMD by broadening the evolutionary sequences and thus altering the age estimates. The radial distance cut further reduces this possibility by preferentially excluding peculiar populations \citep[see e.g.,][]{sbordone11, cassisi13, cassisi20}.

Nonetheless, NGC~6388 and NGC~6441 are known to be extremely peculiar GCs. In fact, \cite{bellini13} demonstrated that the two clusters share a similar helium spread, that reaches values of $\Delta$Y$\simeq0.07$ \citep[see also][]{caloi07, busso07}, but despite their similar metallicity and helium content, their CMDs in the optical and ultra-violet bands display differences that can likely be ascribed to peculiar C,N,O abundances. For what concerns this study, these complex stellar populations manifest as split sub-giant branches (SGB) and broad red giant branches in the clusters optical CMD, and most importantly, these photometric features are not entirely erased by the radial distance cut. This is why we followed the prescriptions by \cite{milone17} \citep[see also][]{milone15} and the HUGS photometry in the required bands to create a chromosome map for each of the two clusters and to select only stars belonging to the primordial populations. Fig.~\ref{fig:chromosome} shows an example of how such a selection works in the region of the (m$_{F814W}$, m$_{F606W}$-m$_{F814W}$) CMD around the SGB, that is the most peculiar sequence of these two clusters in the optical CMD. Stars excluded from the analysis are those shown in cyan, that in fact tend to occupy a region that is redder and fainter than the primordial population (black symbols), and would thus affect the isochrone fitting.

\begin{figure*}[ht!]
\center{
\includegraphics[width=\columnwidth]{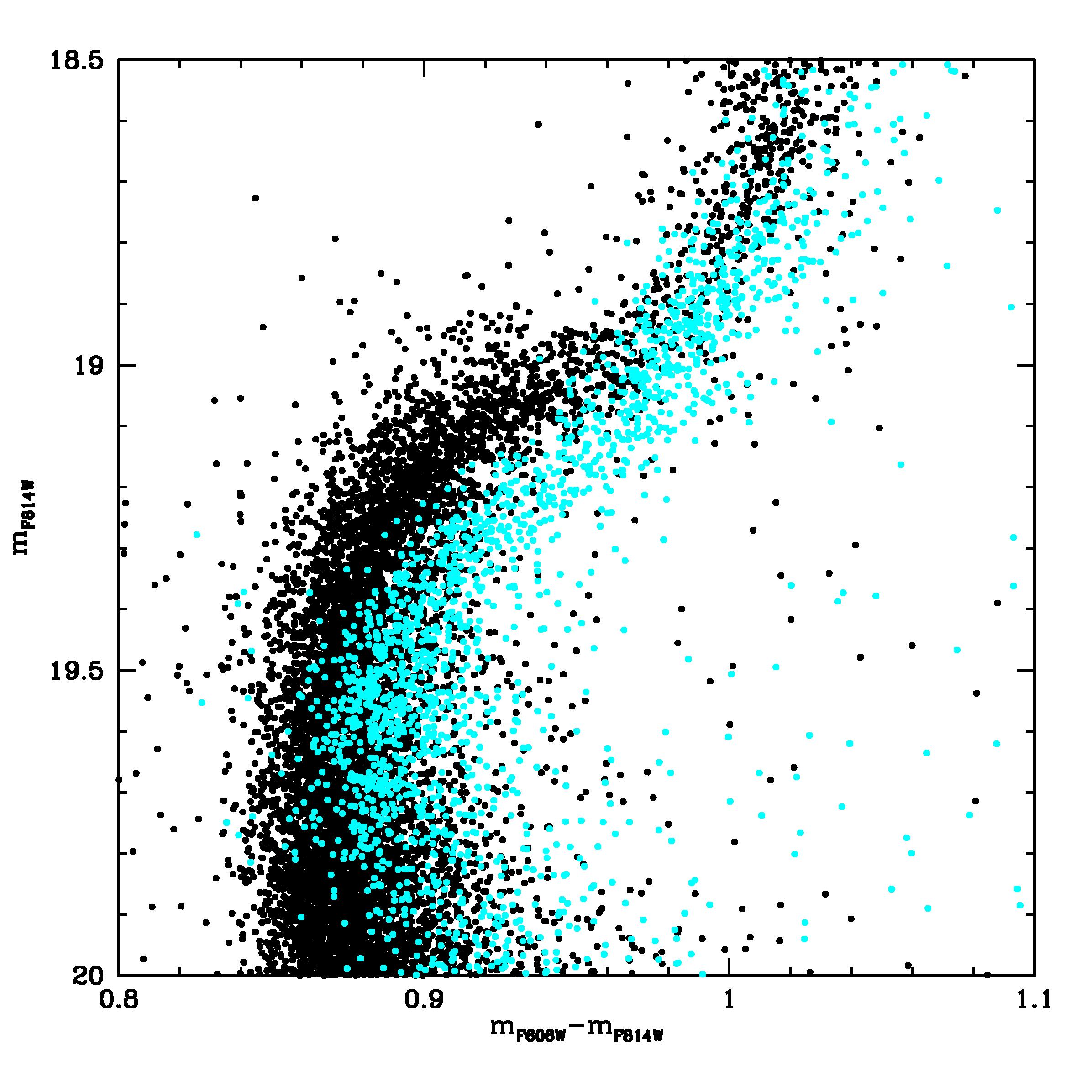}
\includegraphics[width=\columnwidth]{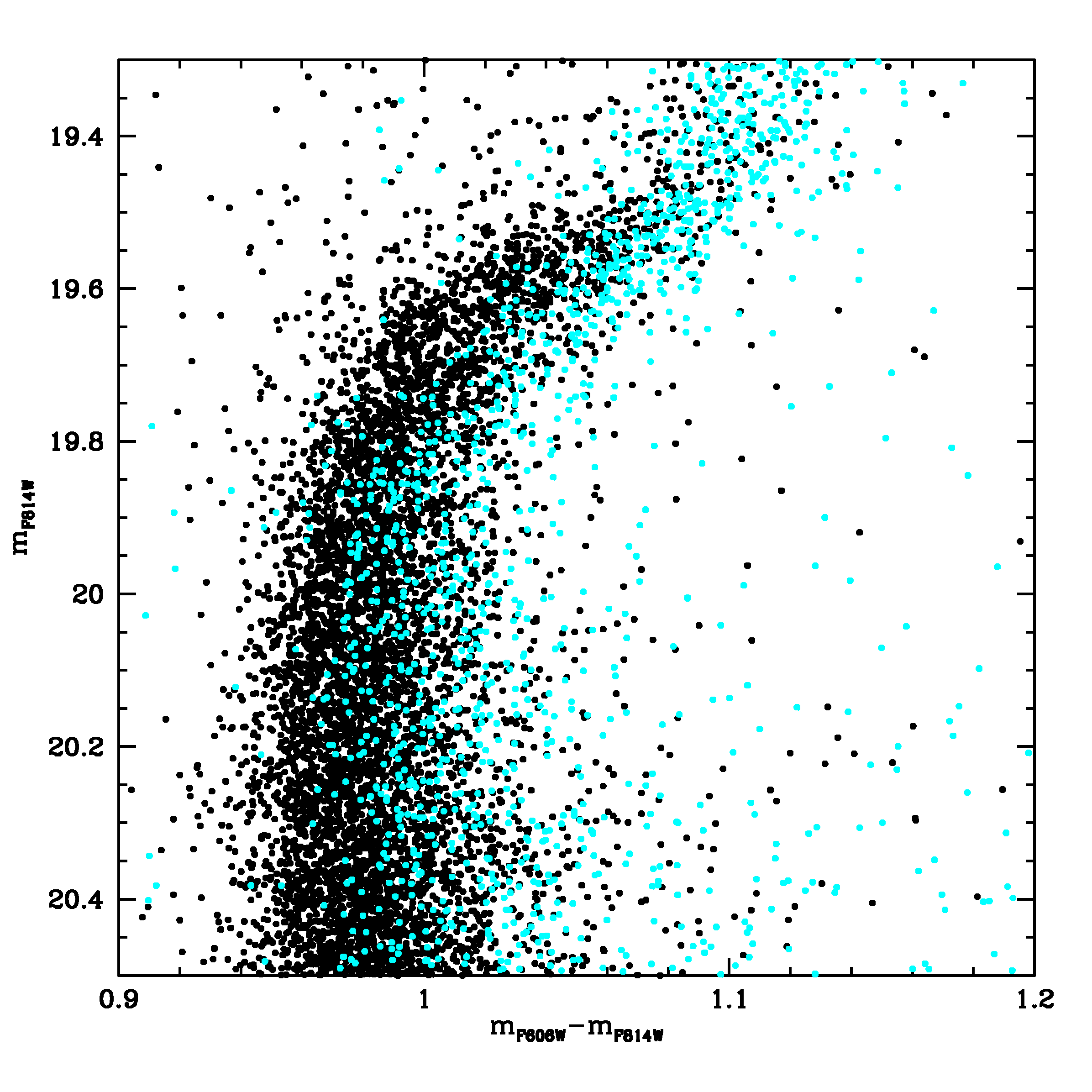}
}
\caption{Zoom-in around the region of the SGB for the (m$_{F814W}$, m$_{F606W}$-m$_{F814W}$) CMDs of NGC~6388 (left-hand panel) and NGC~6441 (right-hand panel). Stars highlighted in cyan are those selected via chromosome maps as belonging to chemically peculiar populations, and are therefore excluded from the analysis. Black symbols are instead stars belonging to the primordial population.}
\label{fig:chromosome}
\end{figure*} 

\subsection{The halo field dataset and its SFH derivation} \label{subsec:halo_data_set}

From {\it Gaia} DR3 \citep[][]{gaiadr3}, we have selected stars belonging to the Milky Way halo within a cylinder of heliocentric radius 1 kpc and distance from the plane |Z|$<3.5$ kpc, by adopting a cut in tangential velocities  V$_{\rm T}>200$ km/s \citep[as defined from proper motions in][]{gaiababu18}. This CMD presents the typical double sequence of stars indicating the presence of an in-situ and an accreted population \citep[][]{helmi18, gallart19}. We have considered only stars with small relative parallax error (\verb|parallax_over_error| $>$ 5 in order to derive their distances directly by inversion of the parallax \citep[after applying the global zero-point indicated in][]{lindegren21}, and we have corrected their colors and magnitudes for extinction using the 3D extinction maps by \cite{green19} with the recipes presented in \cite{gaiababu18}.

The synthetic CMD used to derive the SFH and associated age-metallicity relations has been calculated with the BaSTI solar-scaled models (see Sec.\ref{models}), assuming a Kroupa initial mass fraction \citep[][]{kroupa01}, a fraction of unresolved binaries ($\beta$) of 30\%, and a minimum mass ratio for binaries (q) of 0.1.

As a consistency check, we have also derived alternative SFHs by i) correcting the reddening of the observed stars with the 3D extinction map by \cite{lallement18}; ii) adopting a fraction of unresolved binaries of 50\%. The resulting age-metallicity relations are similar, with small differences that do not affect the conclusions of this work.

\section{Results}\label{results}

The result of the isochrone fitting for each of the six GCs under analysis is shown in the Appendix in Figs.A1-A6.
The isochrone fitting code run using Gaussian priors on the input paramaters, centered on the metallicity (assuming solar-scaled [$\alpha$/Fe] mixture), colour-excess and distance modulus values provided in \cite{harris96}, and with dispersion $\sigma_{[M/H]}=0.1$, $\sigma_{E(B-V)}=0.05$ and $\sigma_{[DM]}=0.1$, respectively.
Finally, we run the code twice per each GC, once on the (m$_{F814W}$, m$_{F606W}$-m$_{F814W}$) CMD and once on the (m$_{F606W}$, m$_{F606W}$-m$_{F814W}$) CMD. The results we present here are the average value of the two, while the overall uncertainties are computed such to encompass the upper and lower limits of both runs combined (all the uncertainties $<0.01$ have been conservatively rounded-up to that value). Table~\ref{tab:results} summarises these results.

\begin{table*}[!htbp]
\centering
\caption{\label{tab:results} Results of the isochrone fitting. }
{
    \begin{tabular}{lcccc}
Name & [M/H]  &   E(B-V)  &  DM  &  age  \\  
     &        &    [mag]    &  [mag] &  [Gyr] 
\\
\hline
\\
\vspace{0.2 cm}
NGC5927 & -0.42$^{+0.03}_{-0.03}$ & 0.42$^{+0.01}_{-0.01}$ & 14.59$^{+0.01}_{-0.01}$  & 12.33$^{+0.14}_{-0.14}$\\
\vspace{0.2 cm}
NGC6304 & -0.40$^{+0.04}_{-0.04}$ &  0.49$^{+0.01}_{-0.02}$ & 14.01$^{+0.02}_{-0.01}$ & 13.07$^{+0.29}_{-0.49}$ \\
\vspace{0.2 cm}
NGC6352 & -0.48$^{+0.04}_{-0.03}$ &  0.27$^{+0.01}_{-0.02}$ & 13.71$^{+0.01}_{-0.02}$ & 11.91$^{+0.14}_{-0.14}$\\
\vspace{0.2 cm}
NGC6388 & -0.46$^{+0.02}_{-0.02}$ &  0.36$^{+0.01}_{-0.01}$ & 15.36$^{+0.03}_{-0.03}$ & 11.88$^{+0.49}_{-0.52}$\\
\vspace{0.2 cm}
NGC6441 & -0.53$^{+0.02}_{-0.02}$ &  0.46$^{+0.01}_{-0.01}$ & 15.64$^{+0.02}_{-0.03}$ & 13.11$^{+0.20}_{-0.29}$\\
\vspace{0.2 cm}
NGC6496 & -0.47$^{+0.03}_{-0.02}$ &  0.24$^{+0.01}_{-0.02}$ & 14.93$^{+0.01}_{-0.02}$ & 13.12$^{+0.20}_{-0.15}$\\
\hline
    \end{tabular}
}
\tablefoot{ The CMD fits and corner plots are shown in the Appendix.}
\end{table*}

As a first sanity check, we compare the output metallicity, E(B-V) and DM with existing estimates from the literature. For the metallicity, we adopt the catalog by \cite{harris96} as a reference. As explained in Sect.~\ref{models}, the output of the isochrone fitting is [M/H], and corresponds to [Fe/H] only in case of solar-scaled $\alpha$-element abundances (see Equation 1). We thus determined the [$\alpha$/Fe] value that minimizes the average difference between our results and the literature ones, finding that [$\alpha$/Fe]$=0.08$ makes the difference null, as shown in Fig.\ref{fig:feh_diff}. Given that these GCs are rather metal-rich, and the typical behavior of the [$\alpha$/Fe] abundance ratio shows a decrease towards solar values in the metal-rich regime \citep[see e.g.,][]{horta20}, [$\alpha$/Fe]$\simeq0.1$ seems a rather reasonable value. The scatter of the distribution ($\sigma$=0.07) further guarantees that the adopted priors are not too stringent. Overall, this advocates for a successful sanity check on the clusters metallicity. 

\begin{figure}[ht!]
\center{
\includegraphics[width=\columnwidth]{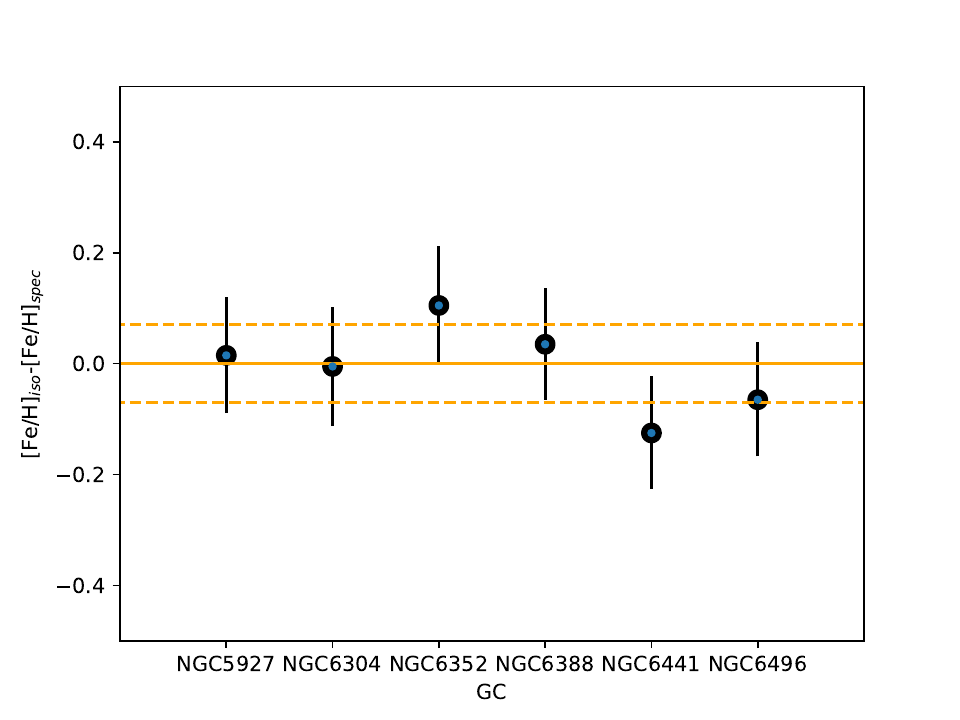}
}
\caption{Difference between the output metallicities [Fe/H] computed by assuming [$\alpha$/Fe]$=0.08$ and literature spectroscopic estimates for each GC, as provided by \cite{harris96}. The scatter around the mean value is 0.07 dex. Errorbars are given by the sum in quadrature between our uncertainties and those by \cite{harris96}, that amount to $\sim0.1$ dex.}
\label{fig:feh_diff}
\end{figure} 

The same sanity check performed on colour excess and distance modulus values from the literature provides good results as well. Fig.~\ref{fig:ebvdist} shows the difference between our findings and other E(B-V) and DM estimates coming from isochrone fitting methods (contrarily to the adopted priors, that come from multiple different techniques). In particular, the values for the four in-situ GCs have been taken from \cite{vandenberg13}, while those for NGC~6388 have been provided in \cite{moretti09} and those for NGC~6441 in \cite{baumgardt21}. Concerning the colour excess (see the left-hand panel) the mean difference is only $-0.006$ mag, and the dispersion around the mean is $0.01$ mag. In the plot, the error-bars have been computed as the sum in quadrature between the uncertainties estimated by our code and a 5\% uncertainty, which is the typical value associated to literature estimates. As for DM values (see the right-hand panel of Fig.~\ref{fig:ebvdist}), the mean difference is zero, with a small dispersion of $0.02$. In this case, the uncertainties shown in the plot are the sum in quadrature between ours and a typical literature value of 0.05 mag.

\begin{figure*}[ht!]
\center{
\includegraphics[width=\columnwidth]{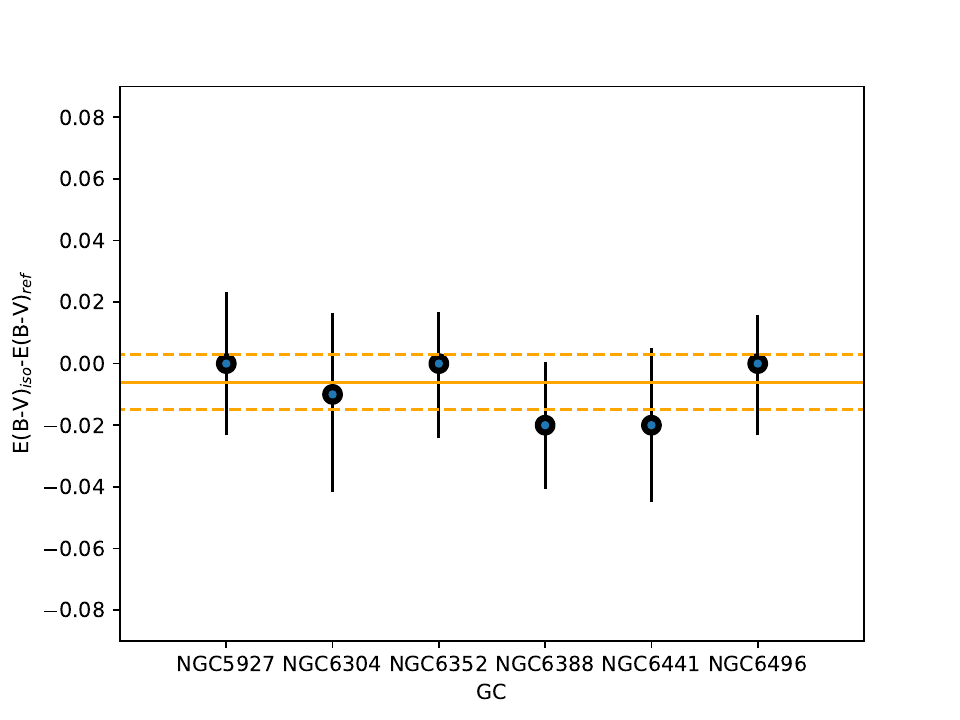}
\includegraphics[width=\columnwidth]{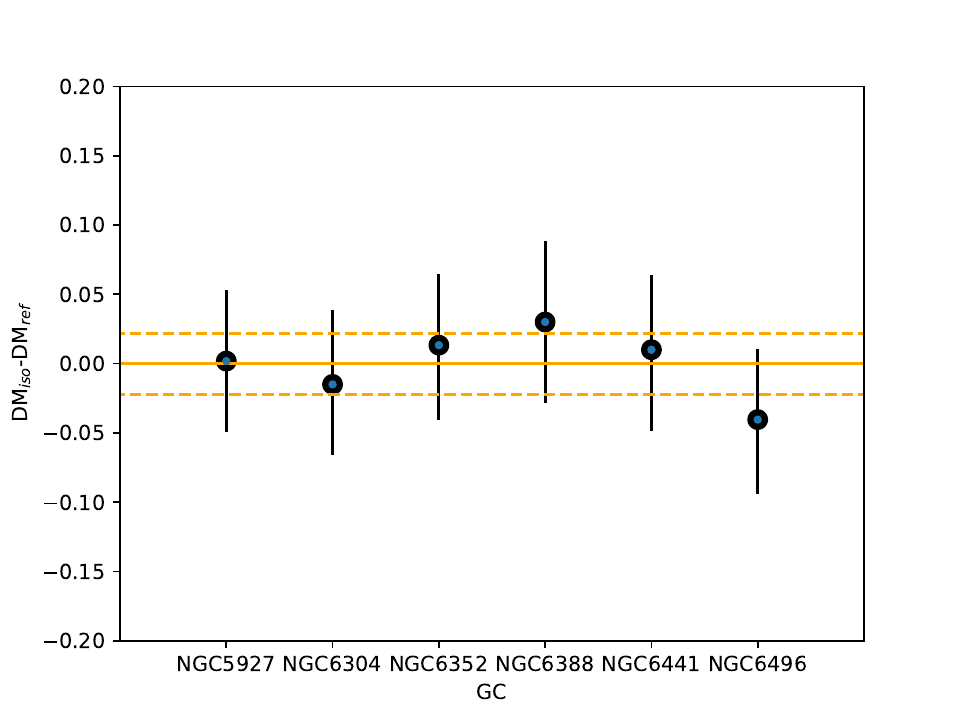}
}
\caption{Difference between the absolute reddening value (left-hand panel) and distance modulus (right-hand panel) coming from our best fit solution and literature ones.}
\label{fig:ebvdist}
\end{figure*} 

All these successful consistency checks demonstrate that the isochrone fitting has worked properly, by converging to reasonable solutions. In the following we focus on the age estimates, with the important remark that, without an appropriate independent absolute zero-point (that we currently miss), these isochrone-fitting absolute ages have to be interpreted {\it in a relative} sense, by focusing on age differences, rather than on their absolute values. 


\begin{figure}[ht!]
\center{
\includegraphics[width=\columnwidth]{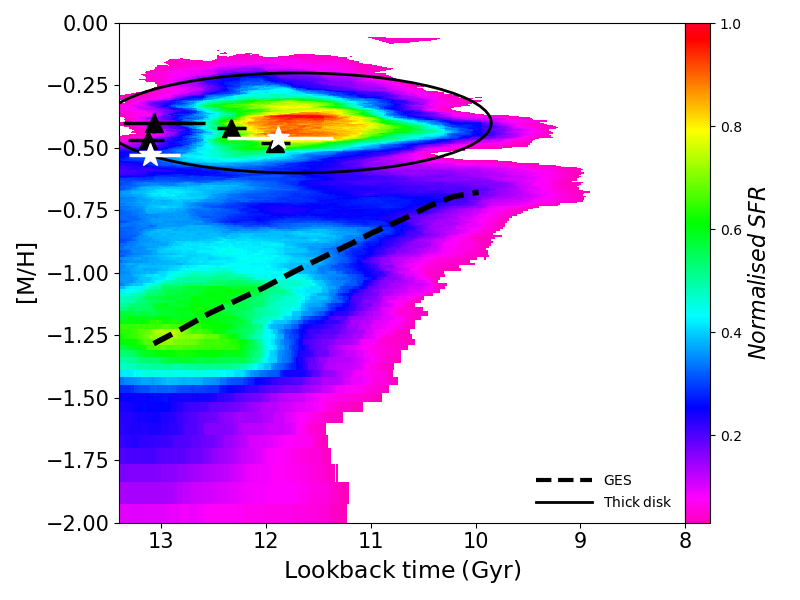}
}
\caption{Location of the six GCs under study in the age-metallicity plane, overplotted on the age-metallicity relation derived from the halo CMD. GCs with undoubted in-situ origin are shown as black symbols, while NGC~6388 and NGC~6441 are marked in white. The location of GES (black dashed line) and of the Milky Way thick disk (black ellipse) are shown for sake of comparison.}
\label{fig:age}
\end{figure} 

Fig.~\ref{fig:age} shows the location of the six GCs analysed here in the age-metallicity plane, overplotted on the age-metallicity relation derived from the halo field population. 
The first immediate feature that stands out is that all of the six GCs have rather similar ages. The youngest is NGC~6388, but its age is perfectly consistent within a 1$\sigma$ uncertainty with the age of at least two of the in-situ GCs, namely NGC~5927 and NGC~6352. The other cluster with debated origin is NGC~6441, that is even older, and with an age consistent with that of NGC~6304 and NGC~6496. By looking at the nominal uncertainties, that take into account the interplay of all of the parameters involved in the fit, it might be tempting to distinguish the six GCs in two groups with strikingly identical age, one group (NGC~6441, NGC~6304 and NGC~6496) with a mean $<t>=13.10$ Gyr (and a tiny dispersion of $\sigma_{t}=0.03$ Gyr), and a second group (NGC~6388, NGC~5927 and NGC~6352) with $<t>=12.05$ Gyr and $\sigma_{t}=0.25$ Gyr. If this second group consists of the youngest among the in-situ GCs, as CARMA will determine when the sample of GC ages becomes more and more complete, they might precisely pin-point an important event in the Milky Way merger history that has halted the in-situ GC formation. In fact, the quoted numbers are not dissimilar from the estimated accretion time of the Kraken \citep{kruijssen20} or of the GES \citep{helmi18, gallart19, montalban21, xiang22, ciuca23} merger events. For the moment though, given that we focus here on {\it age differences}, we prefer not to make strong claims, and to leave this possible evidence as an interesting feature to be more thoroughly investigated in the future.

In order to put a final word on the origin of the two debated GCs NGC~6388 and NGC~6441, in Fig.~\ref{fig:age} we compare their position on the age-metallicity plane with the age-metallicity relation for the halo field population, which clearly shows the signature of GES at low metallicity ([M/H]< -0.75) and that of the heated early disk in-situ population at the highest metallicity end. GES stars show a clearly distinct age-metallicity relation with [M/H] $\simeq$ -1.25 at 13 Gyr ago and [M/H] $\simeq$ -0.8 at 12 Gyr ago, while the in-situ disk remains $\sim 13$ Gyr old up to [M/H]$\simeq-0.7$ and then becomes younger at higher metallicity. Very clearly, the six GCs studied here are all entirely consistent with the pattern expected by an in-situ formation. In particular, to be located on the AMR of an accreted system like GES, NGC~6388 and NGC~6441 should be about 2 Gyr younger (or more) than the in-situ GCs, which is excluded at a 2.5$\sigma$ and at a 7$\sigma$ level, respectively. We can therefore conclude that at high significance both the clusters were born in the Milky Way.

\section{Summary and Conclusions}\label{conclusion}

In this first paper of a series, we have presented the CARMA project by highlighting its methods and objectives, and we have shown an immediate application to solve the case of the debated origin of the pair of GCs NGC~6388 and NGC~6441.

CARMA aims at providing the first complete and homogeneous compilation of GCs ages to date, targeting precision and accuracy $<0.5$ Gyr. To do so, we have developed an isochrone fitting code that builds upon the work of \cite{saracino19} and provides age estimates and uncertainties in a statistically robust MCMC framework.
By using differential reddening corrected, and kinematically decontaminated photometry, which is then properly treated in order to exclude chemically peculiar stars and bad measurements, CARMA code finds the best solution among a fine grid of BaSTI theoretical isochrones corrected for temperature-dependent extinction effects, and provides homogeneous estimates of [M/H], E(B-V), DM and age.

As a proof-of-concept study we have applied our method to six Milky Way GCs. Four out of these six have a well established in-situ origin, while the other two, namely NGC~6388 and NGC~6441 have widely debated origins in the literature. Our results show that NGC~6388 and NGC~6441 are coeval with other in-situ GCs, and a comparison between their location in the age-metallicity plane 
and that of the sequences derived for GES and the heated, in-situ early disk indicate at high significance that they have formed in-situ as well.
The youngest among these in-situ GCs might pin-point the moment when GC formation in the Milky Way disk was suppressed (or when a second peak of GC formation was triggered) by the merger event with a large dwarf galaxy such as Kraken or Gaia-Enceladus-Sausage, but further investigation, as well as independent absolute age calibrators \citep[like those that will be provided by the asteroseismic mission HAYDN,][]{haydn}, are required to make more robust claims in this sense.

These findings demonstrate the power of precise and accurate GC age measurements in the context of reconstructing the Milky Way assembly history. Future investigations by CARMA will target GCs associated to different external progenitors, including the Milky Way itself, to disentangle uncertain associations and put precise constraints on the accretion time and mass of the related merger events.

\section*{Acknowledgements}

We thank the anonymous referee for comments and suggestions that improved the quality of our paper.
DM, SC, EP and AM acknowledge financial support from PRIN-MIUR-22: CHRONOS: adjusting the clock(s) to unveil the CHRONO-chemo-dynamical Structure of the Galaxy” (PI: S. Cassisi). AS acknowledges support from the European Research Council Consolidator Grant funding scheme (project ASTEROCHRONOMETRY, G.A. n. 772293, \url{http://www.asterochronometry.eu}).
Funding for the Stellar Astrophysics Centre is provided by The Danish National Research Foundation (Grant agreement No.~DNRF106). SS acknowledges funding from STFC under the grant no. R276234.
CG, EFA, TRL and SC acknowledge support from the Agencia Estatal de Investigación del Ministerio de Ciencia e Innovación (AEI-MCINN) under grant "At the forefront of Galactic Archaeology: evolution of the luminous and dark matter components of the Milky Way and Local Group dwarf galaxies in the Gaia era" with reference PID2020-118778GB-I00/10.13039/501100011033. CG also acknowledge support from the Severo Ochoa program through CEX2019-000920-S.
TRL acknowledges support from Juan de la Cierva fellowship (IJC2020-043742-I), financed by MCIN/AEI/10.13039/501100011033

This work has made use of data from the European Space Agency (ESA) mission
\gaia\ (\url{https://www.cosmos.esa.int/gaia}), processed by the \gaia\
Data Processing and Analysis Consortium (DPAC,
\url{https://www.cosmos.esa.int/web/gaia/dpac/consortium}). Funding for the DPAC
has been provided by national institutions, in particular the institutions
participating in the \gaia\ Multilateral Agreement.
This project has received funding from the European Research Council (ERC) under the European Union’s Horizon 2020 research and innovation programme (grant agreement No. 804240) for S.S. and Á.S. M.M. acknowledges support from the Agencia Estatal de Investigaci\'on del Ministerio de Ciencia e Innovaci\'on (MCIN/AEI) under the grant "RR Lyrae stars, a lighthouse to distant galaxies and early galaxy evolution" and the European Regional Development Fun (ERDF) with reference PID2021-127042OB-I00, and from the Spanish Ministry of Science and Innovation (MICINN) through the Spanish State Research Agency, under Severo Ochoa Programe 2020-2023 (CEX2019-000920-S).
. P.J. acknowledges support from the Swiss National Science Foundation. 

\bibliographystyle{aa}
\bibliography{refs}

\begin{thebibliography}{106}
\expandafter\ifx\csname natexlab\endcsname\relax\def\natexlab#1{#1}\fi

\bibitem[{{Aguado} {et~al.}(2021){Aguado}, {Myeong}, {Belokurov}, {Evans},
  {Koposov}, {Allende Prieto}, {Lanfranchi}, {Matteucci}, {Shetrone},
  {Sbordone}, {Navarrete}, {Gonz{\'a}lez Hern{\'a}ndez}, {Chanam{\'e}},
  {Peralta de Arriba}, \& {Yuan}}]{aguado21}
{Aguado}, D.~S., {Myeong}, G.~C., {Belokurov}, V., {et~al.} 2021, \mnras, 500,
  889

\bibitem[{{Amarante} {et~al.}(2022){Amarante}, {Debattista}, {Beraldo e Silva},
  {Laporte}, \& {Deg}}]{amarante22}
{Amarante}, J. A.~S., {Debattista}, V.~P., {Beraldo e Silva}, L., {Laporte}, C.
  F.~P., \& {Deg}, N. 2022, \apj, 937, 12

\bibitem[{{Bajkova} {et~al.}(2020){Bajkova}, {Carraro}, {Korchagin},
  {Budanova}, \& {Bobylev}}]{bajkova20}
{Bajkova}, A.~T., {Carraro}, G., {Korchagin}, V.~I., {Budanova}, N.~O., \&
  {Bobylev}, V.~V. 2020, \apj, 895, 69

\bibitem[{{Baumgardt} \& {Vasiliev}(2021)}]{baumgardt21}
{Baumgardt}, H. \& {Vasiliev}, E. 2021, \mnras, 505, 5957

\bibitem[{{Bedin} {et~al.}(2005){Bedin}, {Cassisi}, {Castelli}, {Piotto},
  {Anderson}, {Salaris}, {Momany}, \& {Pietrinferni}}]{bedin05a}
{Bedin}, L.~R., {Cassisi}, S., {Castelli}, F., {et~al.} 2005, \mnras, 357, 1038

\bibitem[{{Bellini} {et~al.}(2017){Bellini}, {Anderson}, {Bedin}, {King}, {van
  der Marel}, {Piotto}, \& {Cool}}]{bellini17}
{Bellini}, A., {Anderson}, J., {Bedin}, L.~R., {et~al.} 2017, \apj, 842, 6

\bibitem[{{Bellini} {et~al.}(2013){Bellini}, {Piotto}, {Milone}, {King},
  {Renzini}, {Cassisi}, {Anderson}, {Bedin}, {Nardiello}, {Pietrinferni}, \&
  {Sarajedini}}]{bellini13}
{Bellini}, A., {Piotto}, G., {Milone}, A.~P., {et~al.} 2013, \apj, 765, 32

\bibitem[{{Belokurov} {et~al.}(2018){Belokurov}, {Erkal}, {Evans}, {Koposov},
  \& {Deason}}]{belokurov18}
{Belokurov}, V., {Erkal}, D., {Evans}, N.~W., {Koposov}, S.~E., \& {Deason},
  A.~J. 2018, \mnras, 478, 611

\bibitem[{{Belokurov} {et~al.}(2020){Belokurov}, {Sanders}, {Fattahi}, {Smith},
  {Deason}, {Evans}, \& {Grand}}]{belokurov20}
{Belokurov}, V., {Sanders}, J.~L., {Fattahi}, A., {et~al.} 2020, \mnras, 494,
  3880

\bibitem[{{Busso} {et~al.}(2007){Busso}, {Cassisi}, {Piotto}, {Castellani},
  {Romaniello}, {Catelan}, {Djorgovski}, {Recio Blanco}, {Renzini}, {Rich},
  {Sweigart}, \& {Zoccali}}]{busso07}
{Busso}, G., {Cassisi}, S., {Piotto}, G., {et~al.} 2007, \aap, 474, 105

\bibitem[{{Caffau} {et~al.}(2011){Caffau}, {Ludwig}, {Steffen}, {Freytag}, \&
  {Bonifacio}}]{caffau11}
{Caffau}, E., {Ludwig}, H.~G., {Steffen}, M., {Freytag}, B., \& {Bonifacio}, P.
  2011, \solphys, 268, 255

\bibitem[{{Callingham} {et~al.}(2022){Callingham}, {Cautun}, {Deason}, {Frenk},
  {Grand}, \& {Marinacci}}]{callingham22}
{Callingham}, T.~M., {Cautun}, M., {Deason}, A.~J., {et~al.} 2022, \mnras, 513,
  4107

\bibitem[{{Caloi} \& {D'Antona}(2007)}]{caloi07}
{Caloi}, V. \& {D'Antona}, F. 2007, \aap, 463, 949

\bibitem[{{Cardelli} {et~al.}(1989){Cardelli}, {Clayton}, \&
  {Mathis}}]{cardelli89}
{Cardelli}, J.~A., {Clayton}, G.~C., \& {Mathis}, J.~S. 1989, \apj, 345, 245

\bibitem[{{Carretta} \& {Bragaglia}(2022)}]{carretta22}
{Carretta}, E. \& {Bragaglia}, A. 2022, \aap, 660, L1

\bibitem[{{Cassisi} {et~al.}(2013){Cassisi}, {Mucciarelli}, {Pietrinferni},
  {Salaris}, \& {Ferguson}}]{cassisi13}
{Cassisi}, S., {Mucciarelli}, A., {Pietrinferni}, A., {Salaris}, M., \&
  {Ferguson}, J. 2013, \aap, 554, A19

\bibitem[{{Cassisi} \& {Salaris}(2020)}]{cassisi20}
{Cassisi}, S. \& {Salaris}, M. 2020, \aapr, 28, 5

\bibitem[{{Cassisi} {et~al.}(2004){Cassisi}, {Salaris}, {Castelli}, \&
  {Pietrinferni}}]{cassisi04}
{Cassisi}, S., {Salaris}, M., {Castelli}, F., \& {Pietrinferni}, A. 2004, \apj,
  616, 498

\bibitem[{{Cignoni} \& {Tosi}(2010)}]{cignoni10}
{Cignoni}, M. \& {Tosi}, M. 2010, Advances in Astronomy, 2010, 158568

\bibitem[{{Ciuc{\u{a}}} {et~al.}(2023){Ciuc{\u{a}}}, {Kawata}, {Ting}, {Grand},
  {Miglio}, {Hayden}, {Baba}, {Fragkoudi}, {Monty}, {Buder}, \&
  {Freeman}}]{ciuca23}
{Ciuc{\u{a}}}, I., {Kawata}, D., {Ting}, Y.-S., {et~al.} 2023, \mnras
  [\eprint[arXiv]{2211.01006}]

\bibitem[{{Conroy} {et~al.}(2019){Conroy}, {Bonaca}, {Cargile}, {Johnson},
  {Caldwell}, {Naidu}, {Zaritsky}, {Fabricant}, {Moran}, {Rhee},
  {Szentgyorgyi}, {Berlind}, {Calkins}, {Kattner}, \& {Ly}}]{h3}
{Conroy}, C., {Bonaca}, A., {Cargile}, P., {et~al.} 2019, \apj, 883, 107

\bibitem[{{Cui} {et~al.}(2012){Cui}, {Zhao}, {Chu}, {Li}, {Li}, {Zhang}, {Su},
  {Yao}, {Wang}, {Xing}, {Li}, {Zhu}, {Wang}, {Gu}, {Luo}, {Xu}, {Zhang},
  {Liu}, {Zhang}, {Yang}, {Cao}, {Chen}, {Chen}, {Chen}, {Chen}, {Chu}, {Feng},
  {Gong}, {Hou}, {Hu}, {Hu}, {Hu}, {Jia}, {Jiang}, {Jiang}, {Jiang}, {Jin},
  {Li}, {Li}, {Li}, {Liu}, {Liu}, {Lu}, {Mao}, {Men}, {Qi}, {Qi}, {Shi},
  {Tang}, {Tao}, {Wang}, {Wang}, {Wang}, {Wang}, {Wang}, {Wang}, {Wang},
  {Wang}, {Wang}, {Wang}, {Wang}, {Wang}, {Xu}, {Xu}, {Yang}, {Yu}, {Yuan},
  {Yuan}, {Zhai}, {Zhang}, {Zhang}, {Zhang}, {Zhao}, {Zhou}, {Zhou}, {Zhu}, \&
  {Zou}}]{lamost}
{Cui}, X.-Q., {Zhao}, Y.-H., {Chu}, Y.-Q., {et~al.} 2012, Research in Astronomy
  and Astrophysics, 12, 1197

\bibitem[{{De Silva} {et~al.}(2015){De Silva}, {Freeman}, {Bland-Hawthorn},
  {Martell}, {de Boer}, {Asplund}, {Keller}, {Sharma}, {Zucker}, {Zwitter},
  {Anguiano}, {Bacigalupo}, {Bayliss}, {Beavis}, {Bergemann}, {Campbell},
  {Cannon}, {Carollo}, {Casagrande}, {Casey}, {Da Costa}, {D'Orazi}, {Dotter},
  {Duong}, {Heger}, {Ireland}, {Kafle}, {Kos}, {Lattanzio}, {Lewis}, {Lin},
  {Lind}, {Munari}, {Nataf}, {O'Toole}, {Parker}, {Reid}, {Schlesinger},
  {Sheinis}, {Simpson}, {Stello}, {Ting}, {Traven}, {Watson}, {Wittenmyer},
  {Yong}, \& {{\v{Z}}erjal}}]{galah}
{De Silva}, G.~M., {Freeman}, K.~C., {Bland-Hawthorn}, J., {et~al.} 2015,
  \mnras, 449, 2604

\bibitem[{{Di Matteo} {et~al.}(2019){Di Matteo}, {Haywood}, {Lehnert}, {Katz},
  {Khoperskov}, {Snaith}, {G{\'o}mez}, \& {Robichon}}]{dimatteo19}
{Di Matteo}, P., {Haywood}, M., {Lehnert}, M.~D., {et~al.} 2019, \aap, 632, A4

\bibitem[{{Dodd} {et~al.}(2023){Dodd}, {Callingham}, {Helmi}, {Matsuno},
  {Ruiz-Lara}, {Balbinot}, \& {L{\"o}vdal}}]{dodd23}
{Dodd}, E., {Callingham}, T.~M., {Helmi}, A., {et~al.} 2023, \aap, 670, L2

\bibitem[{{Dolphin}(2002)}]{dolphin02}
{Dolphin}, A.~E. 2002, \mnras, 332, 91

\bibitem[{{Feuillet} {et~al.}(2021){Feuillet}, {Sahlholdt}, {Feltzing}, \&
  {Casagrande}}]{feuillet21}
{Feuillet}, D.~K., {Sahlholdt}, C.~L., {Feltzing}, S., \& {Casagrande}, L.
  2021, \mnras, 508, 1489

\bibitem[{{Forbes}(2020)}]{forbes20}
{Forbes}, D.~A. 2020, \mnras, 493, 847

\bibitem[{{Forbes} \& {Bridges}(2010)}]{forbes10}
{Forbes}, D.~A. \& {Bridges}, T. 2010, \mnras, 404, 1203

\bibitem[{{Foreman-Mackey} {et~al.}(2019){Foreman-Mackey}, {Farr}, {Sinha},
  {Archibald}, {Hogg}, {Sanders}, {Zuntz}, {Williams}, {Nelson}, {de
  Val-Borro}, {Erhardt}, {Pashchenko}, \& {Pla}}]{emcee}
{Foreman-Mackey}, D., {Farr}, W., {Sinha}, M., {et~al.} 2019, The Journal of
  Open Source Software, 4, 1864

\bibitem[{{Gaia Collaboration} {et~al.}(2018){Gaia Collaboration}, {Babusiaux},
  {van Leeuwen}, {Barstow}, {Jordi}, {Vallenari}, {Bossini}, {Bressan},
  {Cantat-Gaudin}, {van Leeuwen}, {Brown}, {Prusti}, {de Bruijne},
  {Bailer-Jones}, {Biermann}, {Evans}, {Eyer}, {Jansen}, {Klioner}, {Lammers},
  {Lindegren}, {Luri}, {Mignard}, {Panem}, {Pourbaix}, {Randich}, {Sartoretti},
  {Siddiqui}, {Soubiran}, {Walton}, {Arenou}, {Bastian}, {Cropper}, {Drimmel},
  {Katz}, {Lattanzi}, {Bakker}, {Cacciari}, {Casta{\~n}eda}, {Chaoul}, {Cheek},
  {De Angeli}, {Fabricius}, {Guerra}, {Holl}, {Masana}, {Messineo}, {Mowlavi},
  {Nienartowicz}, {Panuzzo}, {Portell}, {Riello}, {Seabroke}, {Tanga},
  {Th{\'e}venin}, {Gracia-Abril}, {Comoretto}, {Garcia-Reinaldos}, {Teyssier},
  {Altmann}, {Andrae}, {Audard}, {Bellas-Velidis}, {Benson}, {Berthier},
  {Blomme}, {Burgess}, {Busso}, {Carry}, {Cellino}, {Clementini}, {Clotet},
  {Creevey}, {Davidson}, {De Ridder}, {Delchambre}, {Dell'Oro}, {Ducourant},
  {Fern{\'a}ndez-Hern{\'a}ndez}, {Fouesneau}, {Fr{\'e}mat}, {Galluccio},
  {Garc{\'\i}a-Torres}, {Gonz{\'a}lez-N{\'u}{\~n}ez}, {Gonz{\'a}lez-Vidal},
  {Gosset}, {Guy}, {Halbwachs}, {Hambly}, {Harrison}, {Hern{\'a}ndez},
  {Hestroffer}, {Hodgkin}, {Hutton}, {Jasniewicz}, {Jean-Antoine-Piccolo},
  {Jordan}, {Korn}, {Krone-Martins}, {Lanzafame}, {Lebzelter}, {L{\"o}ffler},
  {Manteiga}, {Marrese}, {Mart{\'\i}n-Fleitas}, {Moitinho}, {Mora}, {Muinonen},
  {Osinde}, {Pancino}, {Pauwels}, {Petit}, {Recio-Blanco}, {Richards},
  {Rimoldini}, {Robin}, {Sarro}, {Siopis}, {Smith}, {Sozzetti}, {S{\"u}veges},
  {Torra}, {van Reeven}, {Abbas}, {Abreu Aramburu}, {Accart}, {Aerts},
  {Altavilla}, {{\'A}lvarez}, {Alvarez}, {Alves}, {Anderson}, {Andrei},
  {Anglada Varela}, {Antiche}, {Antoja}, {Arcay}, {Astraatmadja}, {Bach},
  {Baker}, {Balaguer-N{\'u}{\~n}ez}, {Balm}, {Barache}, {Barata}, {Barbato},
  {Barblan}, {Barklem}, {Barrado}, {Barros}, {Bartholom{\'e} Mu{\~n}oz},
  {Bassilana}, {Becciani}, {Bellazzini}, {Berihuete}, {Bertone}, {Bianchi},
  {Bienaym{\'e}}, {Blanco-Cuaresma}, {Boch}, {Boeche}, {Bombrun}, {Borrachero},
  {Bouquillon}, {Bourda}, {Bragaglia}, {Bramante}, {Breddels}, {Brouillet},
  {Br{\"u}semeister}, {Brugaletta}, {Bucciarelli}, {Burlacu}, {Busonero},
  {Butkevich}, {Buzzi}, {Caffau}, {Cancelliere}, {Cannizzaro}, {Carballo},
  {Carlucci}, {Carrasco}, {Casamiquela}, {Castellani}, {Castro-Ginard},
  {Charlot}, {Chemin}, {Chiavassa}, {Cocozza}, {Costigan}, {Cowell}, {Crifo},
  {Crosta}, {Crowley}, {Cuypers}, {Dafonte}, {Damerdji}, {Dapergolas}, {David},
  {David}, {de Laverny}, {De Luise}, {De March}, {de Martino}, {de Souza}, {de
  Torres}, {Debosscher}, {del Pozo}, {Delbo}, {Delgado}, {Delgado}, {Diakite},
  {Diener}, {Distefano}, {Dolding}, {Drazinos}, {Dur{\'a}n}, {Edvardsson},
  {Enke}, {Eriksson}, {Esquej}, {Eynard Bontemps}, {Fabre}, {Fabrizio},
  {Faigler}, {Falc{\~a}o}, {Farr{\`a}s Casas}, {Federici}, {Fedorets},
  {Fernique}, {Figueras}, {Filippi}, {Findeisen}, {Fonti}, {Fraile}, {Fraser},
  {Fr{\'e}zouls}, {Gai}, {Galleti}, {Garabato}, {Garc{\'\i}a-Sedano},
  {Garofalo}, {Garralda}, {Gavel}, {Gavras}, {Gerssen}, {Geyer}, {Giacobbe},
  {Gilmore}, {Girona}, {Giuffrida}, {Glass}, {Gomes}, {Granvik}, {Gueguen},
  {Guerrier}, {Guiraud}, {Guti{\'e}}, {Haigron}, {Hatzidimitriou}, {Hauser},
  {Haywood}, {Heiter}, {Helmi}, {Heu}, {Hilger}, {Hobbs}, {Hofmann}, {Holland},
  {Huckle}, {Hypki}, {Icardi}, {Jan{\ss}en}, {Jevardat de Fombelle}, {Jonker},
  {Juh{\'a}sz}, {Julbe}, {Karampelas}, {Kewley}, {Klar}, {Kochoska}, {Kohley},
  {Kolenberg}, {Kontizas}, {Kontizas}, {Koposov}, {Kordopatis},
  {Kostrzewa-Rutkowska}, {Koubsky}, {Lambert}, {Lanza}, {Lasne}, {Lavigne}, {Le
  Fustec}, {Le Poncin-Lafitte}, {Lebreton}, {Leccia}, {Leclerc},
  {Lecoeur-Taibi}, {Lenhardt}, {Leroux}, {Liao}, {Licata}, {Lindstr{\o}m},
  {Lister}, {Livanou}, {Lobel}, {L{\'o}pez}, {Managau}, {Mann}, {Mantelet},
  {Marchal}, {Marchant}, {Marconi}, {Marinoni}, {Marschalk{\'o}}, {Marshall},
  {Martino}, {Marton}, {Mary}, {Massari}, {Matijevi{\v{c}}}, {Mazeh},
  {McMillan}, {Messina}, {Michalik}, {Millar}, {Molina}, {Molinaro},
  {Moln{\'a}r}, {Montegriffo}, {Mor}, {Morbidelli}, {Morel}, {Morris},
  {Mulone}, {Muraveva}, {Musella}, {Nelemans}, {Nicastro}, {Noval},
  {O'Mullane}, {Ord{\'e}novic}, {Ord{\'o}{\~n}ez-Blanco}, {Osborne}, {Pagani},
  {Pagano}, {Pailler}, {Palacin}, {Palaversa}, {Panahi}, {Pawlak},
  {Piersimoni}, {Pineau}, {Plachy}, {Plum}, {Poggio}, {Poujoulet},
  {Pr{\v{s}}a}, {Pulone}, {Racero}, {Ragaini}, {Rambaux}, {Ramos-Lerate},
  {Regibo}, {Reyl{\'e}}, {Riclet}, {Ripepi}, {Riva}, {Rivard}, {Rixon},
  {Roegiers}, {Roelens}, {Romero-G{\'o}mez}, {Rowell}, {Royer}, {Ruiz-Dern},
  {Sadowski}, {Sagrist{\`a} Sell{\'e}s}, {Sahlmann}, {Salgado}, {Salguero},
  {Sanna}, {Santana-Ros}, {Sarasso}, {Savietto}, {Schultheis}, {Sciacca},
  {Segol}, {Segovia}, {S{\'e}gransan}, {Shih}, {Siltala}, {Silva}, {Smart},
  {Smith}, {Solano}, {Solitro}, {Sordo}, {Soria Nieto}, {Souchay}, {Spagna},
  {Spoto}, {Stampa}, {Steele}, {Steidelm{\"u}ller}, {Stephenson}, {Stoev},
  {Suess}, {Surdej}, {Szabados}, {Szegedi-Elek}, {Tapiador}, {Taris}, {Tauran},
  {Taylor}, {Teixeira}, {Terrett}, {Teyssandier}, {Thuillot}, {Titarenko},
  {Torra Clotet}, {Turon}, {Ulla}, {Utrilla}, {Uzzi}, {Vaillant}, {Valentini},
  {Valette}, {van Elteren}, {Van Hemelryck}, {Vaschetto}, {Vecchiato},
  {Veljanoski}, {Viala}, {Vicente}, {Vogt}, {von Essen}, {Voss}, {Votruba},
  {Voutsinas}, {Walmsley}, {Weiler}, {Wertz}, {Wevers}, {Wyrzykowski},
  {Yoldas}, {{\v{Z}}erjal}, {Ziaeepour}, {Zorec}, {Zschocke}, {Zucker},
  {Zurbach}, \& {Zwitter}}]{gaiababu18}
{Gaia Collaboration}, {Babusiaux}, C., {van Leeuwen}, F., {et~al.} 2018, \aap,
  616, A10

\bibitem[{{Gaia Collaboration} {et~al.}(2021){Gaia Collaboration}, {Brown},
  {Vallenari}, {Prusti}, {de Bruijne}, {Babusiaux}, {Biermann}, {Creevey},
  {Evans}, {Eyer}, {Hutton}, {Jansen}, {Jordi}, {Klioner}, {Lammers},
  {Lindegren}, {Luri}, {Mignard}, {Panem}, {Pourbaix}, {Randich}, {Sartoretti},
  {Soubiran}, {Walton}, {Arenou}, {Bailer-Jones}, {Bastian}, {Cropper},
  {Drimmel}, {Katz}, {Lattanzi}, {van Leeuwen}, {Bakker}, {Cacciari},
  {Casta{\~n}eda}, {De Angeli}, {Ducourant}, {Fabricius}, {Fouesneau},
  {Fr{\'e}mat}, {Guerra}, {Guerrier}, {Guiraud}, {Jean-Antoine Piccolo},
  {Masana}, {Messineo}, {Mowlavi}, {Nicolas}, {Nienartowicz}, {Pailler},
  {Panuzzo}, {Riclet}, {Roux}, {Seabroke}, {Sordo}, {Tanga}, {Th{\'e}venin},
  {Gracia-Abril}, {Portell}, {Teyssier}, {Altmann}, {Andrae}, {Bellas-Velidis},
  {Benson}, {Berthier}, {Blomme}, {Brugaletta}, {Burgess}, {Busso}, {Carry},
  {Cellino}, {Cheek}, {Clementini}, {Damerdji}, {Davidson}, {Delchambre},
  {Dell'Oro}, {Fern{\'a}ndez-Hern{\'a}ndez}, {Galluccio}, {Garc{\'\i}a-Lario},
  {Garcia-Reinaldos}, {Gonz{\'a}lez-N{\'u}{\~n}ez}, {Gosset}, {Haigron},
  {Halbwachs}, {Hambly}, {Harrison}, {Hatzidimitriou}, {Heiter},
  {Hern{\'a}ndez}, {Hestroffer}, {Hodgkin}, {Holl}, {Jan{\ss}en}, {Jevardat de
  Fombelle}, {Jordan}, {Krone-Martins}, {Lanzafame}, {L{\"o}ffler}, {Lorca},
  {Manteiga}, {Marchal}, {Marrese}, {Moitinho}, {Mora}, {Muinonen}, {Osborne},
  {Pancino}, {Pauwels}, {Petit}, {Recio-Blanco}, {Richards}, {Riello},
  {Rimoldini}, {Robin}, {Roegiers}, {Rybizki}, {Sarro}, {Siopis}, {Smith},
  {Sozzetti}, {Ulla}, {Utrilla}, {van Leeuwen}, {van Reeven}, {Abbas}, {Abreu
  Aramburu}, {Accart}, {Aerts}, {Aguado}, {Ajaj}, {Altavilla}, {{\'A}lvarez},
  {{\'A}lvarez Cid-Fuentes}, {Alves}, {Anderson}, {Anglada Varela}, {Antoja},
  {Audard}, {Baines}, {Baker}, {Balaguer-N{\'u}{\~n}ez}, {Balbinot}, {Balog},
  {Barache}, {Barbato}, {Barros}, {Barstow}, {Bartolom{\'e}}, {Bassilana},
  {Bauchet}, {Baudesson-Stella}, {Becciani}, {Bellazzini}, {Bernet}, {Bertone},
  {Bianchi}, {Blanco-Cuaresma}, {Boch}, {Bombrun}, {Bossini}, {Bouquillon},
  {Bragaglia}, {Bramante}, {Breedt}, {Bressan}, {Brouillet}, {Bucciarelli},
  {Burlacu}, {Busonero}, {Butkevich}, {Buzzi}, {Caffau}, {Cancelliere},
  {C{\'a}novas}, {Cantat-Gaudin}, {Carballo}, {Carlucci}, {Carnerero},
  {Carrasco}, {Casamiquela}, {Castellani}, {Castro-Ginard}, {Castro Sampol},
  {Chaoul}, {Charlot}, {Chemin}, {Chiavassa}, {Cioni}, {Comoretto}, {Cooper},
  {Cornez}, {Cowell}, {Crifo}, {Crosta}, {Crowley}, {Dafonte}, {Dapergolas},
  {David}, {David}, {de Laverny}, {De Luise}, {De March}, {De Ridder}, {de
  Souza}, {de Teodoro}, {de Torres}, {del Peloso}, {del Pozo}, {Delbo},
  {Delgado}, {Delgado}, {Delisle}, {Di Matteo}, {Diakite}, {Diener},
  {Distefano}, {Dolding}, {Eappachen}, {Edvardsson}, {Enke}, {Esquej}, {Fabre},
  {Fabrizio}, {Faigler}, {Fedorets}, {Fernique}, {Fienga}, {Figueras},
  {Fouron}, {Fragkoudi}, {Fraile}, {Franke}, {Gai}, {Garabato},
  {Garcia-Gutierrez}, {Garc{\'\i}a-Torres}, {Garofalo}, {Gavras}, {Gerlach},
  {Geyer}, {Giacobbe}, {Gilmore}, {Girona}, {Giuffrida}, {Gomel}, {Gomez},
  {Gonzalez-Santamaria}, {Gonz{\'a}lez-Vidal}, {Granvik},
  {Guti{\'e}rrez-S{\'a}nchez}, {Guy}, {Hauser}, {Haywood}, {Helmi}, {Hidalgo},
  {Hilger}, {H{\l}adczuk}, {Hobbs}, {Holland}, {Huckle}, {Jasniewicz},
  {Jonker}, {Juaristi Campillo}, {Julbe}, {Karbevska}, {Kervella}, {Khanna},
  {Kochoska}, {Kontizas}, {Kordopatis}, {Korn}, {Kostrzewa-Rutkowska},
  {Kruszy{\'n}ska}, {Lambert}, {Lanza}, {Lasne}, {Le Campion}, {Le Fustec},
  {Lebreton}, {Lebzelter}, {Leccia}, {Leclerc}, {Lecoeur-Taibi}, {Liao},
  {Licata}, {Lindstr{\o}m}, {Lister}, {Livanou}, {Lobel}, {Madrero Pardo},
  {Managau}, {Mann}, {Marchant}, {Marconi}, {Marcos Santos}, {Marinoni},
  {Marocco}, {Marshall}, {Martin Polo}, {Mart{\'\i}n-Fleitas}, {Masip},
  {Massari}, {Mastrobuono-Battisti}, {Mazeh}, {McMillan}, {Messina},
  {Michalik}, {Millar}, {Mints}, {Molina}, {Molinaro}, {Moln{\'a}r},
  {Montegriffo}, {Mor}, {Morbidelli}, {Morel}, {Morris}, {Mulone}, {Munoz},
  {Muraveva}, {Murphy}, {Musella}, {Noval}, {Ord{\'e}novic}, {Orr{\`u}},
  {Osinde}, {Pagani}, {Pagano}, {Palaversa}, {Palicio}, {Panahi}, {Pawlak},
  {Pe{\~n}alosa Esteller}, {Penttil{\"a}}, {Piersimoni}, {Pineau}, {Plachy},
  {Plum}, {Poggio}, {Poretti}, {Poujoulet}, {Pr{\v{s}}a}, {Pulone}, {Racero},
  {Ragaini}, {Rainer}, {Raiteri}, {Rambaux}, {Ramos}, {Ramos-Lerate}, {Re
  Fiorentin}, {Regibo}, {Reyl{\'e}}, {Ripepi}, {Riva}, {Rixon}, {Robichon},
  {Robin}, {Roelens}, {Rohrbasser}, {Romero-G{\'o}mez}, {Rowell}, {Royer},
  {Rybicki}, {Sadowski}, {Sagrist{\`a} Sell{\'e}s}, {Sahlmann}, {Salgado},
  {Salguero}, {Samaras}, {Sanchez Gimenez}, {Sanna}, {Santove{\~n}a},
  {Sarasso}, {Schultheis}, {Sciacca}, {Segol}, {Segovia}, {S{\'e}gransan},
  {Semeux}, {Shahaf}, {Siddiqui}, {Siebert}, {Siltala}, {Slezak}, {Smart},
  {Solano}, {Solitro}, {Souami}, {Souchay}, {Spagna}, {Spoto}, {Steele},
  {Steidelm{\"u}ller}, {Stephenson}, {S{\"u}veges}, {Szabados}, {Szegedi-Elek},
  {Taris}, {Tauran}, {Taylor}, {Teixeira}, {Thuillot}, {Tonello}, {Torra},
  {Torra}, {Turon}, {Unger}, {Vaillant}, {van Dillen}, {Vanel}, {Vecchiato},
  {Viala}, {Vicente}, {Voutsinas}, {Weiler}, {Wevers}, {Wyrzykowski}, {Yoldas},
  {Yvard}, {Zhao}, {Zorec}, {Zucker}, {Zurbach}, \& {Zwitter}}]{edr3}
{Gaia Collaboration}, {Brown}, A.~G.~A., {Vallenari}, A., {et~al.} 2021, \aap,
  649, A1

\bibitem[{{Gaia Collaboration} {et~al.}(2023){Gaia Collaboration}, {Vallenari},
  {Brown}, {Prusti}, {de Bruijne}, {Arenou}, {Babusiaux}, {Biermann},
  {Creevey}, {Ducourant}, {Evans}, {Eyer}, {Guerra}, {Hutton}, {Jordi},
  {Klioner}, {Lammers}, {Lindegren}, {Luri}, {Mignard}, {Panem}, {Pourbaix},
  {Randich}, {Sartoretti}, {Soubiran}, {Tanga}, {Walton}, {Bailer-Jones},
  {Bastian}, {Drimmel}, {Jansen}, {Katz}, {Lattanzi}, {van Leeuwen}, {Bakker},
  {Cacciari}, {Casta{\~n}eda}, {De Angeli}, {Fabricius}, {Fouesneau},
  {Fr{\'e}mat}, {Galluccio}, {Guerrier}, {Heiter}, {Masana}, {Messineo},
  {Mowlavi}, {Nicolas}, {Nienartowicz}, {Pailler}, {Panuzzo}, {Riclet}, {Roux},
  {Seabroke}, {Sordo}, {Th{\'e}venin}, {Gracia-Abril}, {Portell}, {Teyssier},
  {Altmann}, {Andrae}, {Audard}, {Bellas-Velidis}, {Benson}, {Berthier},
  {Blomme}, {Burgess}, {Busonero}, {Busso}, {C{\'a}novas}, {Carry}, {Cellino},
  {Cheek}, {Clementini}, {Damerdji}, {Davidson}, {de Teodoro}, {Nu{\~n}ez
  Campos}, {Delchambre}, {Dell'Oro}, {Esquej}, {Fern{\'a}ndez-Hern{\'a}ndez},
  {Fraile}, {Garabato}, {Garc{\'\i}a-Lario}, {Gosset}, {Haigron}, {Halbwachs},
  {Hambly}, {Harrison}, {Hern{\'a}ndez}, {Hestroffer}, {Hodgkin}, {Holl},
  {Jan{\ss}en}, {Jevardat de Fombelle}, {Jordan}, {Krone-Martins}, {Lanzafame},
  {L{\"o}ffler}, {Marchal}, {Marrese}, {Moitinho}, {Muinonen}, {Osborne},
  {Pancino}, {Pauwels}, {Recio-Blanco}, {Reyl{\'e}}, {Riello}, {Rimoldini},
  {Roegiers}, {Rybizki}, {Sarro}, {Siopis}, {Smith}, {Sozzetti}, {Utrilla},
  {van Leeuwen}, {Abbas}, {{\'A}brah{\'a}m}, {Abreu Aramburu}, {Aerts},
  {Aguado}, {Ajaj}, {Aldea-Montero}, {Altavilla}, {{\'A}lvarez}, {Alves},
  {Anders}, {Anderson}, {Anglada Varela}, {Antoja}, {Baines}, {Baker},
  {Balaguer-N{\'u}{\~n}ez}, {Balbinot}, {Balog}, {Barache}, {Barbato},
  {Barros}, {Barstow}, {Bartolom{\'e}}, {Bassilana}, {Bauchet}, {Becciani},
  {Bellazzini}, {Berihuete}, {Bernet}, {Bertone}, {Bianchi}, {Binnenfeld},
  {Blanco-Cuaresma}, {Blazere}, {Boch}, {Bombrun}, {Bossini}, {Bouquillon},
  {Bragaglia}, {Bramante}, {Breedt}, {Bressan}, {Brouillet}, {Brugaletta},
  {Bucciarelli}, {Burlacu}, {Butkevich}, {Buzzi}, {Caffau}, {Cancelliere},
  {Cantat-Gaudin}, {Carballo}, {Carlucci}, {Carnerero}, {Carrasco},
  {Casamiquela}, {Castellani}, {Castro-Ginard}, {Chaoul}, {Charlot}, {Chemin},
  {Chiaramida}, {Chiavassa}, {Chornay}, {Comoretto}, {Contursi}, {Cooper},
  {Cornez}, {Cowell}, {Crifo}, {Cropper}, {Crosta}, {Crowley}, {Dafonte},
  {Dapergolas}, {David}, {David}, {de Laverny}, {De Luise}, {De March}, {De
  Ridder}, {de Souza}, {de Torres}, {del Peloso}, {del Pozo}, {Delbo},
  {Delgado}, {Delisle}, {Demouchy}, {Dharmawardena}, {Di Matteo}, {Diakite},
  {Diener}, {Distefano}, {Dolding}, {Edvardsson}, {Enke}, {Fabre}, {Fabrizio},
  {Faigler}, {Fedorets}, {Fernique}, {Fienga}, {Figueras}, {Fournier},
  {Fouron}, {Fragkoudi}, {Gai}, {Garcia-Gutierrez}, {Garcia-Reinaldos},
  {Garc{\'\i}a-Torres}, {Garofalo}, {Gavel}, {Gavras}, {Gerlach}, {Geyer},
  {Giacobbe}, {Gilmore}, {Girona}, {Giuffrida}, {Gomel}, {Gomez},
  {Gonz{\'a}lez-N{\'u}{\~n}ez}, {Gonz{\'a}lez-Santamar{\'\i}a},
  {Gonz{\'a}lez-Vidal}, {Granvik}, {Guillout}, {Guiraud},
  {Guti{\'e}rrez-S{\'a}nchez}, {Guy}, {Hatzidimitriou}, {Hauser}, {Haywood},
  {Helmer}, {Helmi}, {Sarmiento}, {Hidalgo}, {Hilger}, {H{\l}adczuk}, {Hobbs},
  {Holland}, {Huckle}, {Jardine}, {Jasniewicz}, {Jean-Antoine Piccolo},
  {Jim{\'e}nez-Arranz}, {Jorissen}, {Juaristi Campillo}, {Julbe}, {Karbevska},
  {Kervella}, {Khanna}, {Kontizas}, {Kordopatis}, {Korn}, {K{\'o}sp{\'a}l},
  {Kostrzewa-Rutkowska}, {Kruszy{\'n}ska}, {Kun}, {Laizeau}, {Lambert},
  {Lanza}, {Lasne}, {Le Campion}, {Lebreton}, {Lebzelter}, {Leccia}, {Leclerc},
  {Lecoeur-Taibi}, {Liao}, {Licata}, {Lindstr{\o}m}, {Lister}, {Livanou},
  {Lobel}, {Lorca}, {Loup}, {Madrero Pardo}, {Magdaleno Romeo}, {Managau},
  {Mann}, {Manteiga}, {Marchant}, {Marconi}, {Marcos}, {Marcos Santos},
  {Mar{\'\i}n Pina}, {Marinoni}, {Marocco}, {Marshall}, {Martin Polo},
  {Mart{\'\i}n-Fleitas}, {Marton}, {Mary}, {Masip}, {Massari},
  {Mastrobuono-Battisti}, {Mazeh}, {McMillan}, {Messina}, {Michalik}, {Millar},
  {Mints}, {Molina}, {Molinaro}, {Moln{\'a}r}, {Monari}, {Mongui{\'o}},
  {Montegriffo}, {Montero}, {Mor}, {Mora}, {Morbidelli}, {Morel}, {Morris},
  {Muraveva}, {Murphy}, {Musella}, {Nagy}, {Noval}, {Oca{\~n}a}, {Ogden},
  {Ordenovic}, {Osinde}, {Pagani}, {Pagano}, {Palaversa}, {Palicio},
  {Pallas-Quintela}, {Panahi}, {Payne-Wardenaar}, {Pe{\~n}alosa Esteller},
  {Penttil{\"a}}, {Pichon}, {Piersimoni}, {Pineau}, {Plachy}, {Plum}, {Poggio},
  {Pr{\v{s}}a}, {Pulone}, {Racero}, {Ragaini}, {Rainer}, {Raiteri}, {Rambaux},
  {Ramos}, {Ramos-Lerate}, {Re Fiorentin}, {Regibo}, {Richards}, {Rios Diaz},
  {Ripepi}, {Riva}, {Rix}, {Rixon}, {Robichon}, {Robin}, {Robin}, {Roelens},
  {Rogues}, {Rohrbasser}, {Romero-G{\'o}mez}, {Rowell}, {Royer}, {Ruz Mieres},
  {Rybicki}, {Sadowski}, {S{\'a}ez N{\'u}{\~n}ez}, {Sagrist{\`a} Sell{\'e}s},
  {Sahlmann}, {Salguero}, {Samaras}, {Sanchez Gimenez}, {Sanna},
  {Santove{\~n}a}, {Sarasso}, {Schultheis}, {Sciacca}, {Segol}, {Segovia},
  {S{\'e}gransan}, {Semeux}, {Shahaf}, {Siddiqui}, {Siebert}, {Siltala},
  {Silvelo}, {Slezak}, {Slezak}, {Smart}, {Snaith}, {Solano}, {Solitro},
  {Souami}, {Souchay}, {Spagna}, {Spina}, {Spoto}, {Steele},
  {Steidelm{\"u}ller}, {Stephenson}, {S{\"u}veges}, {Surdej}, {Szabados},
  {Szegedi-Elek}, {Taris}, {Taylor}, {Teixeira}, {Tolomei}, {Tonello}, {Torra},
  {Torra}, {Torralba Elipe}, {Trabucchi}, {Tsounis}, {Turon}, {Ulla}, {Unger},
  {Vaillant}, {van Dillen}, {van Reeven}, {Vanel}, {Vecchiato}, {Viala},
  {Vicente}, {Voutsinas}, {Weiler}, {Wevers}, {Wyrzykowski}, {Yoldas}, {Yvard},
  {Zhao}, {Zorec}, {Zucker}, \& {Zwitter}}]{gaiadr3}
{Gaia Collaboration}, {Vallenari}, A., {Brown}, A.~G.~A., {et~al.} 2023, \aap,
  674, A1

\bibitem[{{Gallart} {et~al.}(2019){Gallart}, {Bernard}, {Brook}, {Ruiz-Lara},
  {Cassisi}, {Hill}, \& {Monelli}}]{gallart19}
{Gallart}, C., {Bernard}, E.~J., {Brook}, C.~B., {et~al.} 2019, Nature
  Astronomy, 3, 932

\bibitem[{{Gallart} {et~al.}(1999){Gallart}, {Freedman}, {Aparicio},
  {Bertelli}, \& {Chiosi}}]{gallart99}
{Gallart}, C., {Freedman}, W.~L., {Aparicio}, A., {Bertelli}, G., \& {Chiosi},
  C. 1999, \aj, 118, 2245

\bibitem[{{Gilmore} {et~al.}(2012){Gilmore}, {Randich}, {Asplund}, {Binney},
  {Bonifacio}, {Drew}, {Feltzing}, {Ferguson}, {Jeffries}, {Micela},
  {Negueruela}, {Prusti}, {Rix}, {Vallenari}, {Alfaro}, {Allende-Prieto},
  {Babusiaux}, {Bensby}, {Blomme}, {Bragaglia}, {Flaccomio}, {Fran{\c{c}}ois},
  {Irwin}, {Koposov}, {Korn}, {Lanzafame}, {Pancino}, {Paunzen},
  {Recio-Blanco}, {Sacco}, {Smiljanic}, {Van Eck}, {Walton}, {Aden}, {Aerts},
  {Affer}, {Alcala}, {Altavilla}, {Alves}, {Antoja}, {Arenou}, {Argiroffi},
  {Asensio Ramos}, {Bailer-Jones}, {Balaguer-Nunez}, {Bayo}, {Barbuy},
  {Barisevicius}, {Barrado y Navascues}, {Battistini}, {Bellas Velidis},
  {Bellazzini}, {Belokurov}, {Bergemann}, {Bertelli}, {Biazzo}, {Bienayme},
  {Bland-Hawthorn}, {Boeche}, {Bonito}, {Boudreault}, {Bouvier}, {Brandao},
  {Brown}, {de Bruijne}, {Burleigh}, {Caballero}, {Caffau}, {Calura},
  {Capuzzo-Dolcetta}, {Caramazza}, {Carraro}, {Casagrande}, {Casewell},
  {Chapman}, {Chiappini}, {Chorniy}, {Christlieb}, {Cignoni}, {Cocozza},
  {Colless}, {Collet}, {Collins}, {Correnti}, {Covino}, {Crnojevic}, {Cropper},
  {Cunha}, {Damiani}, {David}, {Delgado}, {Duffau}, {Edvardsson}, {Eldridge},
  {Enke}, {Eriksson}, {Evans}, {Eyer}, {Famaey}, {Fellhauer}, {Ferreras},
  {Figueras}, {Fiorentino}, {Flynn}, {Folha}, {Franciosini}, {Frasca},
  {Freeman}, {Fremat}, {Friel}, {Gaensicke}, {Gameiro}, {Garzon}, {Geier},
  {Geisler}, {Gerhard}, {Gibson}, {Gomboc}, {Gomez}, {Gonzalez-Fernandez},
  {Gonzalez Hernandez}, {Gosset}, {Grebel}, {Greimel}, {Groenewegen},
  {Grundahl}, {Guarcello}, {Gustafsson}, {Hadrava}, {Hatzidimitriou}, {Hambly},
  {Hammersley}, {Hansen}, {Haywood}, {Heber}, {Heiter}, {Held}, {Helmi},
  {Hensler}, {Herrero}, {Hill}, {Hodgkin}, {Huelamo}, {Huxor}, {Ibata},
  {Jackson}, {de Jong}, {Jonker}, {Jordan}, {Jordi}, {Jorissen}, {Katz},
  {Kawata}, {Keller}, {Kharchenko}, {Klement}, {Klutsch}, {Knude}, {Koch},
  {Kochukhov}, {Kontizas}, {Koubsky}, {Lallement}, {de Laverny}, {van Leeuwen},
  {Lemasle}, {Lewis}, {Lind}, {Lindstrom}, {Lobel}, {Lopez Santiago}, {Lucas},
  {Ludwig}, {Lueftinger}, {Magrini}, {Maiz Apellaniz}, {Maldonado}, {Marconi},
  {Marino}, {Martayan}, {Martinez-Valpuesta}, {Matijevic}, {McMahon},
  {Messina}, {Meyer}, {Miglio}, {Mikolaitis}, {Minchev}, {Minniti}, {Moitinho},
  {Momany}, {Monaco}, {Montalto}, {Monteiro}, {Monier}, {Montes}, {Mora},
  {Moraux}, {Morel}, {Mowlavi}, {Mucciarelli}, {Munari}, {Napiwotzki},
  {Nardetto}, {Naylor}, {Naze}, {Nelemans}, {Okamoto}, {Ortolani}, {Pace},
  {Palla}, {Palous}, {Parker}, {Penarrubia}, {Pillitteri}, {Piotto}, {Posbic},
  {Prisinzano}, {Puzeras}, {Quirrenbach}, {Ragaini}, {Read}, {Read}, {Reyle},
  {De Ridder}, {Robichon}, {Robin}, {Roeser}, {Romano}, {Royer}, {Ruchti},
  {Ruzicka}, {Ryan}, {Ryde}, {Santos}, {Sanz Forcada}, {Sarro Baro},
  {Sbordone}, {Schilbach}, {Schmeja}, {Schnurr}, {Schoenrich}, {Scholz},
  {Seabroke}, {Sharma}, {De Silva}, {Smith}, {Solano}, {Sordo}, {Soubiran},
  {Sousa}, {Spagna}, {Steffen}, {Steinmetz}, {Stelzer}, {Stempels},
  {Tabernero}, {Tautvaisiene}, {Thevenin}, {Torra}, {Tosi}, {Tolstoy}, {Turon},
  {Walker}, {Wambsganss}, {Worley}, {Venn}, {Vink}, {Wyse}, {Zaggia},
  {Zeilinger}, {Zoccali}, {Zorec}, {Zucker}, {Zwitter}, \& {Gaia-ESO Survey
  Team}}]{gaiaeso}
{Gilmore}, G., {Randich}, S., {Asplund}, M., {et~al.} 2012, The Messenger, 147,
  25

\bibitem[{{Girardi} {et~al.}(2008){Girardi}, {Dalcanton}, {Williams}, {de
  Jong}, {Gallart}, {Monelli}, {Groenewegen}, {Holtzman}, {Olsen}, {Seth},
  {Weisz}, \& {ANGST/ANGRRR Collaboration}}]{girardi08}
{Girardi}, L., {Dalcanton}, J., {Williams}, B., {et~al.} 2008, \pasp, 120, 583

\bibitem[{{Gratton} {et~al.}(2019){Gratton}, {Bragaglia}, {Carretta},
  {D'Orazi}, {Lucatello}, \& {Sollima}}]{gratton19}
{Gratton}, R., {Bragaglia}, A., {Carretta}, E., {et~al.} 2019, \aapr, 27, 8

\bibitem[{{Green} {et~al.}(2019){Green}, {Schlafly}, {Zucker}, {Speagle}, \&
  {Finkbeiner}}]{green19}
{Green}, G.~M., {Schlafly}, E., {Zucker}, C., {Speagle}, J.~S., \&
  {Finkbeiner}, D. 2019, \apj, 887, 93

\bibitem[{Green \& Sibson(1978)}]{green78}
Green, P.~J. \& Sibson, R. 1978, The Computer Journal, 21, 168

\bibitem[{{Harris}(1996)}]{harris96}
{Harris}, W.~E. 1996, \aj, 112, 1487

\bibitem[{{Helmi}(2020)}]{helmi20}
{Helmi}, A. 2020, \araa, 58, 205

\bibitem[{{Helmi} {et~al.}(2018){Helmi}, {Babusiaux}, {Koppelman}, {Massari},
  {Veljanoski}, \& {Brown}}]{helmi18}
{Helmi}, A., {Babusiaux}, C., {Koppelman}, H.~H., {et~al.} 2018, \nat, 563, 85

\bibitem[{{Helmi} {et~al.}(1999){Helmi}, {White}, {de Zeeuw}, \&
  {Zhao}}]{helmi99}
{Helmi}, A., {White}, S. D.~M., {de Zeeuw}, P.~T., \& {Zhao}, H. 1999, \nat,
  402, 53

\bibitem[{{Hidalgo} {et~al.}(2018){Hidalgo}, {Pietrinferni}, {Cassisi},
  {Salaris}, {Mucciarelli}, {Savino}, {Aparicio}, {Silva Aguirre}, \&
  {Verma}}]{hidalgo18}
{Hidalgo}, S.~L., {Pietrinferni}, A., {Cassisi}, S., {et~al.} 2018, \apj, 856,
  125

\bibitem[{{Horta} {et~al.}(2020){Horta}, {Schiavon}, {Mackereth}, {Beers},
  {Fern{\'a}ndez-Trincado}, {Frinchaboy}, {Garc{\'\i}a-Hern{\'a}ndez},
  {Geisler}, {Hasselquist}, {J{\"o}nsson}, {Lane}, {Majewski},
  {M{\'e}sz{\'a}ros}, {Bidin}, {Nataf}, {Roman-Lopes}, {Nitschelm},
  {Vargas-Gonz{\'a}lez}, \& {Zasowski}}]{horta20}
{Horta}, D., {Schiavon}, R.~P., {Mackereth}, J.~T., {et~al.} 2020, \mnras, 493,
  3363

\bibitem[{{Horta} {et~al.}(2021){Horta}, {Schiavon}, {Mackereth}, {Pfeffer},
  {Mason}, {Kisku}, {Fragkoudi}, {Allende Prieto}, {Cunha}, {Hasselquist},
  {Holtzman}, {Majewski}, {Nataf}, {O'Connell}, {Schultheis}, \&
  {Smith}}]{horta21}
{Horta}, D., {Schiavon}, R.~P., {Mackereth}, J.~T., {et~al.} 2021, \mnras, 500,
  1385

\bibitem[{{Horta} {et~al.}(2023){Horta}, {Schiavon}, {Mackereth}, {Weinberg},
  {Hasselquist}, {Feuillet}, {O'Connell}, {Anguiano}, {Allende-Prieto},
  {Beaton}, {Bizyaev}, {Cunha}, {Geisler}, {Garc{\'\i}a-Hern{\'a}ndez},
  {Holtzman}, {J{\"o}nsson}, {Lane}, {Majewski}, {M{\'e}sz{\'a}ros}, {Minniti},
  {Nitschelm}, {Shetrone}, {Smith}, \& {Zasowski}}]{horta23}
{Horta}, D., {Schiavon}, R.~P., {Mackereth}, J.~T., {et~al.} 2023, \mnras, 520,
  5671

\bibitem[{{Ibata} {et~al.}(1994){Ibata}, {Gilmore}, \& {Irwin}}]{ibata94}
{Ibata}, R.~A., {Gilmore}, G., \& {Irwin}, M.~J. 1994, \nat, 370, 194

\bibitem[{{Khoperskov} \& {Gerhard}(2022)}]{khoperskov22}
{Khoperskov}, S. \& {Gerhard}, O. 2022, \aap, 663, A38

\bibitem[{{Koppelman} {et~al.}(2019{\natexlab{a}}){Koppelman}, {Helmi},
  {Massari}, {Price-Whelan}, \& {Starkenburg}}]{koppelman19}
{Koppelman}, H.~H., {Helmi}, A., {Massari}, D., {Price-Whelan}, A.~M., \&
  {Starkenburg}, T.~K. 2019{\natexlab{a}}, \aap, 631, L9

\bibitem[{{Koppelman} {et~al.}(2019{\natexlab{b}}){Koppelman}, {Helmi},
  {Massari}, {Roelenga}, \& {Bastian}}]{koppelmanh99}
{Koppelman}, H.~H., {Helmi}, A., {Massari}, D., {Roelenga}, S., \& {Bastian},
  U. 2019{\natexlab{b}}, \aap, 625, A5

\bibitem[{{Kroupa}(2001)}]{kroupa01}
{Kroupa}, P. 2001, \mnras, 322, 231

\bibitem[{{Kruijssen} {et~al.}(2020){Kruijssen}, {Pfeffer}, {Chevance},
  {Bonaca}, {Trujillo-Gomez}, {Bastian}, {Reina-Campos}, {Crain}, \&
  {Hughes}}]{kruijssen20}
{Kruijssen}, J.~M.~D., {Pfeffer}, J.~L., {Chevance}, M., {et~al.} 2020, \mnras,
  498, 2472

\bibitem[{{Kruijssen} {et~al.}(2019){Kruijssen}, {Pfeffer}, {Reina-Campos},
  {Crain}, \& {Bastian}}]{kruijssen19}
{Kruijssen}, J.~M.~D., {Pfeffer}, J.~L., {Reina-Campos}, M., {Crain}, R.~A., \&
  {Bastian}, N. 2019, \mnras, 486, 3180

\bibitem[{{Lallement} {et~al.}(2018){Lallement}, {Capitanio}, {Ruiz-Dern},
  {Danielski}, {Babusiaux}, {Vergely}, {Elyajouri}, {Arenou}, \&
  {Leclerc}}]{lallement18}
{Lallement}, R., {Capitanio}, L., {Ruiz-Dern}, L., {et~al.} 2018, \aap, 616,
  A132

\bibitem[{{Leaman} {et~al.}(2013){Leaman}, {VandenBerg}, \&
  {Mendel}}]{leaman13}
{Leaman}, R., {VandenBerg}, D.~A., \& {Mendel}, J.~T. 2013, \mnras, 436, 122

\bibitem[{{Lebreton} {et~al.}(2014{\natexlab{a}}){Lebreton}, {Goupil}, \&
  {Montalb{\'a}n}}]{lebreton14}
{Lebreton}, Y., {Goupil}, M.~J., \& {Montalb{\'a}n}, J. 2014{\natexlab{a}}, in
  EAS Publications Series, Vol.~65, EAS Publications Series, ed. Y.~{Lebreton},
  D.~{Valls-Gabaud}, \& C.~{Charbonnel}, 99--176

\bibitem[{{Lebreton} {et~al.}(2014{\natexlab{b}}){Lebreton}, {Goupil}, \&
  {Montalb{\'a}n}}]{lebreton14b}
{Lebreton}, Y., {Goupil}, M.~J., \& {Montalb{\'a}n}, J. 2014{\natexlab{b}}, in
  EAS Publications Series, Vol.~65, EAS Publications Series, ed. Y.~{Lebreton},
  D.~{Valls-Gabaud}, \& C.~{Charbonnel}, 177--223

\bibitem[{{Leitinger} {et~al.}(2023){Leitinger}, {Baumgardt}, {Cabrera-Ziri},
  {Hilker}, \& {Pancino}}]{leitinger23}
{Leitinger}, E., {Baumgardt}, H., {Cabrera-Ziri}, I., {Hilker}, M., \&
  {Pancino}, E. 2023, \mnras, 520, 1456

\bibitem[{{Lindegren} {et~al.}(2021){Lindegren}, {Bastian}, {Biermann},
  {Bombrun}, {de Torres}, {Gerlach}, {Geyer}, {Hern{\'a}ndez}, {Hilger},
  {Hobbs}, {Klioner}, {Lammers}, {McMillan}, {Ramos-Lerate},
  {Steidelm{\"u}ller}, {Stephenson}, \& {van Leeuwen}}]{lindegren21}
{Lindegren}, L., {Bastian}, U., {Biermann}, M., {et~al.} 2021, \aap, 649, A4

\bibitem[{{Majewski} {et~al.}(2017){Majewski}, {Schiavon}, {Frinchaboy},
  {Allende Prieto}, {Barkhouser}, {Bizyaev}, {Blank}, {Brunner}, {Burton},
  {Carrera}, {Chojnowski}, {Cunha}, {Epstein}, {Fitzgerald}, {Garc{\'\i}a
  P{\'e}rez}, {Hearty}, {Henderson}, {Holtzman}, {Johnson}, {Lam}, {Lawler},
  {Maseman}, {M{\'e}sz{\'a}ros}, {Nelson}, {Nguyen}, {Nidever}, {Pinsonneault},
  {Shetrone}, {Smee}, {Smith}, {Stolberg}, {Skrutskie}, {Walker}, {Wilson},
  {Zasowski}, {Anders}, {Basu}, {Beland}, {Blanton}, {Bovy}, {Brownstein},
  {Carlberg}, {Chaplin}, {Chiappini}, {Eisenstein}, {Elsworth}, {Feuillet},
  {Fleming}, {Galbraith-Frew}, {Garc{\'\i}a}, {Garc{\'\i}a-Hern{\'a}ndez},
  {Gillespie}, {Girardi}, {Gunn}, {Hasselquist}, {Hayden}, {Hekker}, {Ivans},
  {Kinemuchi}, {Klaene}, {Mahadevan}, {Mathur}, {Mosser}, {Muna}, {Munn},
  {Nichol}, {O'Connell}, {Parejko}, {Robin}, {Rocha-Pinto}, {Schultheis},
  {Serenelli}, {Shane}, {Silva Aguirre}, {Sobeck}, {Thompson}, {Troup},
  {Weinberg}, \& {Zamora}}]{apogee}
{Majewski}, S.~R., {Schiavon}, R.~P., {Frinchaboy}, P.~M., {et~al.} 2017, \aj,
  154, 94

\bibitem[{{Malhan} {et~al.}(2022){Malhan}, {Ibata}, {Sharma}, {Famaey},
  {Bellazzini}, {Carlberg}, {D'Souza}, {Yuan}, {Martin}, \&
  {Thomas}}]{malhan22}
{Malhan}, K., {Ibata}, R.~A., {Sharma}, S., {et~al.} 2022, \apj, 926, 107

\bibitem[{{Mardini} {et~al.}(2022){Mardini}, {Frebel}, {Chiti}, {Meiron},
  {Brauer}, \& {Ou}}]{mardini22}
{Mardini}, M.~K., {Frebel}, A., {Chiti}, A., {et~al.} 2022, \apj, 936, 78

\bibitem[{{Massari} {et~al.}(2019){Massari}, {Koppelman}, \&
  {Helmi}}]{massari19}
{Massari}, D., {Koppelman}, H.~H., \& {Helmi}, A. 2019, \aap, 630, L4

\bibitem[{{Massari} {et~al.}(2014){Massari}, {Mucciarelli}, {Ferraro},
  {Origlia}, {Rich}, {Lanzoni}, {Dalessandro}, {Valenti}, {Ibata}, {Lovisi},
  {Bellazzini}, \& {Reitzel}}]{massari14}
{Massari}, D., {Mucciarelli}, A., {Ferraro}, F.~R., {et~al.} 2014, \apj, 795,
  22

\bibitem[{{Miglio} {et~al.}(2021){Miglio}, {Girardi}, {Grundahl}, {Mosser},
  {Bastian}, {Bragaglia}, {Brogaard}, {Buldgen}, {Chantereau}, {Chaplin},
  {Chiappini}, {Dupret}, {Eggenberger}, {Gieles}, {Izzard}, {Kawata}, {Karoff},
  {Lagarde}, {Mackereth}, {Magrin}, {Meynet}, {Michel}, {Montalb{\'a}n},
  {Nascimbeni}, {Noels}, {Piotto}, {Ragazzoni}, {Soszy{\'n}ski}, {Tolstoy},
  {Toonen}, {Triaud}, \& {Vincenzo}}]{haydn}
{Miglio}, A., {Girardi}, L., {Grundahl}, F., {et~al.} 2021, Experimental
  Astronomy, 51, 963

\bibitem[{{Mikkola} {et~al.}(2023){Mikkola}, {McMillan}, \&
  {Hobbs}}]{mikkola23}
{Mikkola}, D., {McMillan}, P.~J., \& {Hobbs}, D. 2023, \mnras, 519, 1989

\bibitem[{{Milone} {et~al.}(2015){Milone}, {Marino}, {Piotto}, {Bedin},
  {Anderson}, {Renzini}, {King}, {Bellini}, {Brown}, {Cassisi}, {D'Antona},
  {Jerjen}, {Nardiello}, {Salaris}, {Marel}, {Vesperini}, {Yong}, {Aparicio},
  {Sarajedini}, \& {Zoccali}}]{milone15}
{Milone}, A.~P., {Marino}, A.~F., {Piotto}, G., {et~al.} 2015, \mnras, 447, 927

\bibitem[{{Milone} {et~al.}(2012){Milone}, {Piotto}, {Bedin}, {Aparicio},
  {Anderson}, {Sarajedini}, {Marino}, {Moretti}, {Davies}, {Chaboyer},
  {Dotter}, {Hempel}, {Mar{\'\i}n-Franch}, {Majewski}, {Paust}, {Reid},
  {Rosenberg}, \& {Siegel}}]{milone12}
{Milone}, A.~P., {Piotto}, G., {Bedin}, L.~R., {et~al.} 2012, \aap, 540, A16

\bibitem[{{Milone} {et~al.}(2017){Milone}, {Piotto}, {Renzini}, {Marino},
  {Bedin}, {Vesperini}, {D'Antona}, {Nardiello}, {Anderson}, {King}, {Yong},
  {Bellini}, {Aparicio}, {Barbuy}, {Brown}, {Cassisi}, {Ortolani}, {Salaris},
  {Sarajedini}, \& {van der Marel}}]{milone17}
{Milone}, A.~P., {Piotto}, G., {Renzini}, A., {et~al.} 2017, \mnras, 464, 3636

\bibitem[{{Minelli} {et~al.}(2021){Minelli}, {Mucciarelli}, {Massari},
  {Bellazzini}, {Romano}, \& {Ferraro}}]{minelli21}
{Minelli}, A., {Mucciarelli}, A., {Massari}, D., {et~al.} 2021, \apjl, 918, L32

\bibitem[{{Monelli} {et~al.}(2010){Monelli}, {Gallart}, {Hidalgo}, {Aparicio},
  {Skillman}, {Cole}, {Weisz}, {Mayer}, {Bernard}, {Cassisi}, {Dolphin},
  {Drozdovsky}, \& {Stetson}}]{monelli10}
{Monelli}, M., {Gallart}, C., {Hidalgo}, S.~L., {et~al.} 2010, \apj, 722, 1864

\bibitem[{{Montalb{\'a}n} {et~al.}(2021){Montalb{\'a}n}, {Mackereth}, {Miglio},
  {Vincenzo}, {Chiappini}, {Buldgen}, {Mosser}, {Noels}, {Scuflaire}, {Vrard},
  {Willett}, {Davies}, {Hall}, {Nielsen}, {Khan}, {Rendle}, {van Rossem},
  {Ferguson}, \& {Chaplin}}]{montalban21}
{Montalb{\'a}n}, J., {Mackereth}, J.~T., {Miglio}, A., {et~al.} 2021, Nature
  Astronomy, 5, 640

\bibitem[{{Monty} {et~al.}(2020){Monty}, {Venn}, {Lane}, {Lokhorst}, \&
  {Yong}}]{monty20}
{Monty}, S., {Venn}, K.~A., {Lane}, J. M.~M., {Lokhorst}, D., \& {Yong}, D.
  2020, \mnras, 497, 1236

\bibitem[{{Monty} {et~al.}(2023){Monty}, {Yong}, {Marino}, {Karakas},
  {McKenzie}, {Grundahl}, \& {Mura-Guzm{\'a}n}}]{monty23}
{Monty}, S., {Yong}, D., {Marino}, A.~F., {et~al.} 2023, \mnras, 518, 965

\bibitem[{{Moretti} {et~al.}(2009){Moretti}, {Piotto}, {Arcidiacono}, {Milone},
  {Ragazzoni}, {Falomo}, {Farinato}, {Bedin}, {Anderson}, {Sarajedini},
  {Baruffolo}, {Diolaiti}, {Lombini}, {Brast}, {Donaldson}, {Kolb},
  {Marchetti}, \& {Tordo}}]{moretti09}
{Moretti}, A., {Piotto}, G., {Arcidiacono}, C., {et~al.} 2009, \aap, 493, 539

\bibitem[{{Myeong} {et~al.}(2022){Myeong}, {Belokurov}, {Aguado}, {Evans},
  {Caldwell}, \& {Bradley}}]{myeong22}
{Myeong}, G.~C., {Belokurov}, V., {Aguado}, D.~S., {et~al.} 2022, \apj, 938, 21

\bibitem[{{Myeong} {et~al.}(2019){Myeong}, {Vasiliev}, {Iorio}, {Evans}, \&
  {Belokurov}}]{myeong19}
{Myeong}, G.~C., {Vasiliev}, E., {Iorio}, G., {Evans}, N.~W., \& {Belokurov},
  V. 2019, \mnras, 488, 1235

\bibitem[{{Naidu} {et~al.}(2020){Naidu}, {Conroy}, {Bonaca}, {Johnson}, {Ting},
  {Caldwell}, {Zaritsky}, \& {Cargile}}]{naidu20}
{Naidu}, R.~P., {Conroy}, C., {Bonaca}, A., {et~al.} 2020, \apj, 901, 48

\bibitem[{{Nardiello} {et~al.}(2018){Nardiello}, {Libralato}, {Piotto},
  {Anderson}, {Bellini}, {Aparicio}, {Bedin}, {Cassisi}, {Granata}, {King},
  {Lucertini}, {Marino}, {Milone}, {Ortolani}, {Platais}, \& {van der
  Marel}}]{nardiello18}
{Nardiello}, D., {Libralato}, M., {Piotto}, G., {et~al.} 2018, \mnras, 481,
  3382

\bibitem[{{Necib} {et~al.}(2020){Necib}, {Ostdiek}, {Lisanti}, {Cohen},
  {Freytsis}, {Garrison-Kimmel}, {Hopkins}, {Wetzel}, \& {Sanderson}}]{necib20}
{Necib}, L., {Ostdiek}, B., {Lisanti}, M., {et~al.} 2020, Nature Astronomy, 4,
  1078

\bibitem[{{Oria} {et~al.}(2022){Oria}, {Tenachi}, {Ibata}, {Famaey}, {Yuan},
  {Arentsen}, {Martin}, \& {Viswanathan}}]{oria22}
{Oria}, P.-A., {Tenachi}, W., {Ibata}, R., {et~al.} 2022, \apjl, 936, L3

\bibitem[{{Pancino} {et~al.}(2003){Pancino}, {Seleznev}, {Ferraro},
  {Bellazzini}, \& {Piotto}}]{pancino03}
{Pancino}, E., {Seleznev}, A., {Ferraro}, F.~R., {Bellazzini}, M., \& {Piotto},
  G. 2003, \mnras, 345, 683

\bibitem[{{Pietrinferni} {et~al.}(2021){Pietrinferni}, {Hidalgo}, {Cassisi},
  {Salaris}, {Savino}, {Mucciarelli}, {Verma}, {Silva Aguirre}, {Aparicio}, \&
  {Ferguson}}]{pietrinferni21}
{Pietrinferni}, A., {Hidalgo}, S., {Cassisi}, S., {et~al.} 2021, \apj, 908, 102

\bibitem[{{Piotto} {et~al.}(2007){Piotto}, {Bedin}, {Anderson}, {King},
  {Cassisi}, {Milone}, {Villanova}, {Pietrinferni}, \& {Renzini}}]{piotto07}
{Piotto}, G., {Bedin}, L.~R., {Anderson}, J., {et~al.} 2007, \apjl, 661, L53

\bibitem[{{Piotto} {et~al.}(2015){Piotto}, {Milone}, {Bedin}, {Anderson},
  {King}, {Marino}, {Nardiello}, {Aparicio}, {Barbuy}, {Bellini}, {Brown},
  {Cassisi}, {Cool}, {Cunial}, {Dalessandro}, {D'Antona}, {Ferraro}, {Hidalgo},
  {Lanzoni}, {Monelli}, {Ortolani}, {Renzini}, {Salaris}, {Sarajedini}, {van
  der Marel}, {Vesperini}, \& {Zoccali}}]{piotto15}
{Piotto}, G., {Milone}, A.~P., {Bedin}, L.~R., {et~al.} 2015, \aj, 149, 91

\bibitem[{{Re Fiorentin} {et~al.}(2021){Re Fiorentin}, {Spagna}, {Lattanzi}, \&
  {Cignoni}}]{refiorentin21}
{Re Fiorentin}, P., {Spagna}, A., {Lattanzi}, M.~G., \& {Cignoni}, M. 2021,
  \apjl, 907, L16

\bibitem[{{Ross} \& {Aller}(1976)}]{rossaller}
{Ross}, J.~E. \& {Aller}, L.~H. 1976, Science, 191, 1223

\bibitem[{{Ruiz-Lara} {et~al.}(2020){Ruiz-Lara}, {Gallart}, {Bernard}, \&
  {Cassisi}}]{ruizlara20}
{Ruiz-Lara}, T., {Gallart}, C., {Bernard}, E.~J., \& {Cassisi}, S. 2020, Nature
  Astronomy, 4, 965

\bibitem[{{Ruiz-Lara} {et~al.}(2022{\natexlab{a}}){Ruiz-Lara}, {Helmi},
  {Gallart}, {Surot}, \& {Cassisi}}]{ruizlara22b}
{Ruiz-Lara}, T., {Helmi}, A., {Gallart}, C., {Surot}, F., \& {Cassisi}, S.
  2022{\natexlab{a}}, \aap, 668, L10

\bibitem[{{Ruiz-Lara} {et~al.}(2022{\natexlab{b}}){Ruiz-Lara}, {Matsuno},
  {L{\"o}vdal}, {Helmi}, {Dodd}, \& {Koppelman}}]{ruizlara22}
{Ruiz-Lara}, T., {Matsuno}, T., {L{\"o}vdal}, S.~S., {et~al.}
  2022{\natexlab{b}}, \aap, 665, A58

\bibitem[{{Salaris} {et~al.}(1993){Salaris}, {Chieffi}, \&
  {Straniero}}]{salaris93}
{Salaris}, M., {Chieffi}, A., \& {Straniero}, O. 1993, \apj, 414, 580

\bibitem[{{Saracino} {et~al.}(2019){Saracino}, {Dalessandro}, {Ferraro},
  {Lanzoni}, {Geisler}, {Cohen}, {Bellini}, {Vesperini}, {Salaris}, {Cassisi},
  {Pietrinferni}, {Origlia}, {Mauro}, {Villanova}, \& {Moni
  Bidin}}]{saracino19}
{Saracino}, S., {Dalessandro}, E., {Ferraro}, F.~R., {et~al.} 2019, \apj, 874,
  86

\bibitem[{{Sarajedini} {et~al.}(2007){Sarajedini}, {Bedin}, {Chaboyer},
  {Dotter}, {Siegel}, {Anderson}, {Aparicio}, {King}, {Majewski},
  {Mar{\'\i}n-Franch}, {Piotto}, {Reid}, \& {Rosenberg}}]{sarajedini07}
{Sarajedini}, A., {Bedin}, L.~R., {Chaboyer}, B., {et~al.} 2007, \aj, 133, 1658

\bibitem[{{Sbordone} {et~al.}(2011){Sbordone}, {Salaris}, {Weiss}, \&
  {Cassisi}}]{sbordone11}
{Sbordone}, L., {Salaris}, M., {Weiss}, A., \& {Cassisi}, S. 2011, \aap, 534,
  A9

\bibitem[{{Soderblom}(2010)}]{soderblom2010}
{Soderblom}, D.~R. 2010, \araa, 48, 581

\bibitem[{{Steinmetz} {et~al.}(2006){Steinmetz}, {Zwitter}, {Siebert},
  {Watson}, {Freeman}, {Munari}, {Campbell}, {Williams}, {Seabroke}, {Wyse},
  {Parker}, {Bienaym{\'e}}, {Roeser}, {Gibson}, {Gilmore}, {Grebel}, {Helmi},
  {Navarro}, {Burton}, {Cass}, {Dawe}, {Fiegert}, {Hartley}, {Russell},
  {Saunders}, {Enke}, {Bailin}, {Binney}, {Bland-Hawthorn}, {Boeche}, {Dehnen},
  {Eisenstein}, {Evans}, {Fiorucci}, {Fulbright}, {Gerhard}, {Jauregi}, {Kelz},
  {Mijovi{\'c}}, {Minchev}, {Parmentier}, {Pe{\~n}arrubia}, {Quillen}, {Read},
  {Ruchti}, {Scholz}, {Siviero}, {Smith}, {Sordo}, {Veltz}, {Vidrih}, {von
  Berlepsch}, {Boyle}, \& {Schilbach}}]{rave}
{Steinmetz}, M., {Zwitter}, T., {Siebert}, A., {et~al.} 2006, \aj, 132, 1645

\bibitem[{{Tailo} {et~al.}(2022){Tailo}, {Corsaro}, {Miglio}, {Montalb{\'a}n},
  {Brogaard}, {Milone}, {Stokholm}, {Casali}, \& {Bragaglia}}]{tailo22}
{Tailo}, M., {Corsaro}, E., {Miglio}, A., {et~al.} 2022, \aap, 662, L7

\bibitem[{{Tenachi} {et~al.}(2022){Tenachi}, {Oria}, {Ibata}, {Famaey}, {Yuan},
  {Arentsen}, {Martin}, \& {Viswanathan}}]{tenachi22}
{Tenachi}, W., {Oria}, P.-A., {Ibata}, R., {et~al.} 2022, \apjl, 935, L22

\bibitem[{{VandenBerg} {et~al.}(2013){VandenBerg}, {Brogaard}, {Leaman}, \&
  {Casagrande}}]{vandenberg13}
{VandenBerg}, D.~A., {Brogaard}, K., {Leaman}, R., \& {Casagrande}, L. 2013,
  \apj, 775, 134

\bibitem[{{Verma} {et~al.}(2022){Verma}, {R{\o}rsted}, {Serenelli}, {Aguirre
  B{\o}rsen-Koch}, {Winther}, \& {Stokholm}}]{verma2022}
{Verma}, K., {R{\o}rsted}, J.~L., {Serenelli}, A.~M., {et~al.} 2022, \mnras,
  515, 1492

\bibitem[{Virtanen {et~al.}(2020)Virtanen, Gommers, Oliphant, Haberland, Reddy,
  Cournapeau, Burovski, Peterson, Weckesser, Bright, {van der Walt}, Brett,
  Wilson, Millman, Mayorov, Nelson, Jones, Kern, Larson, Carey, Polat, Feng,
  Moore, {VanderPlas}, Laxalde, Perktold, Cimrman, Henriksen, Quintero, Harris,
  Archibald, Ribeiro, Pedregosa, {van Mulbregt}, \& {SciPy 1.0
  Contributors}}]{scipy}
Virtanen, P., Gommers, R., Oliphant, T.~E., {et~al.} 2020, Nature Methods, 17,
  261

\bibitem[{{Xiang} \& {Rix}(2022)}]{xiang22}
{Xiang}, M. \& {Rix}, H.-W. 2022, \nat, 603, 599

\bibitem[{{Yanny} {et~al.}(2009){Yanny}, {Rockosi}, {Newberg}, {Knapp},
  {Adelman-McCarthy}, {Alcorn}, {Allam}, {Allende Prieto}, {An}, {Anderson},
  {Anderson}, {Bailer-Jones}, {Bastian}, {Beers}, {Bell}, {Belokurov},
  {Bizyaev}, {Blythe}, {Bochanski}, {Boroski}, {Brinchmann}, {Brinkmann},
  {Brewington}, {Carey}, {Cudworth}, {Evans}, {Evans}, {Gates}, {G{\"a}nsicke},
  {Gillespie}, {Gilmore}, {Nebot Gomez-Moran}, {Grebel}, {Greenwell}, {Gunn},
  {Jordan}, {Jordan}, {Harding}, {Harris}, {Hendry}, {Holder}, {Ivans},
  {Ivezi{\v{c}}}, {Jester}, {Johnson}, {Kent}, {Kleinman}, {Kniazev},
  {Krzesinski}, {Kron}, {Kuropatkin}, {Lebedeva}, {Lee}, {French Leger},
  {L{\'e}pine}, {Levine}, {Lin}, {Long}, {Loomis}, {Lupton}, {Malanushenko},
  {Malanushenko}, {Margon}, {Martinez-Delgado}, {McGehee}, {Monet}, {Morrison},
  {Munn}, {Neilsen}, {Nitta}, {Norris}, {Oravetz}, {Owen}, {Padmanabhan},
  {Pan}, {Peterson}, {Pier}, {Platson}, {Re Fiorentin}, {Richards}, {Rix},
  {Schlegel}, {Schneider}, {Schreiber}, {Schwope}, {Sibley}, {Simmons},
  {Snedden}, {Allyn Smith}, {Stark}, {Stauffer}, {Steinmetz}, {Stoughton},
  {SubbaRao}, {Szalay}, {Szkody}, {Thakar}, {Sivarani}, {Tucker}, {Uomoto},
  {Vanden Berk}, {Vidrih}, {Wadadekar}, {Watters}, {Wilhelm}, {Wyse}, {Yarger},
  \& {Zucker}}]{segue}
{Yanny}, B., {Rockosi}, C., {Newberg}, H.~J., {et~al.} 2009, \aj, 137, 4377

\bibitem[{{Yuan} {et~al.}(2020){Yuan}, {Myeong}, {Beers}, {Evans}, {Lee},
  {Banerjee}, {Gudin}, {Hattori}, {Li}, {Matsuno}, {Placco}, {Smith},
  {Whitten}, \& {Zhao}}]{yuan20}
{Yuan}, Z., {Myeong}, G.~C., {Beers}, T.~C., {et~al.} 2020, \apj, 891, 39

\end{thebibliography}

\begin{appendix}
\section{Isochrone fitting results}
In this Appendix we show the results of our isochrone fitting algorithm applied to the six GCs under study in this work. Each figure is made up of four panels. The lower ones show the posterior distribution of the parameters of the model, including their correlation, resulting from the fit of the (m$_{F814W}$, m$_{F606W}$-m$_{F814W}$) CMD and of the (m$_{F606W}$, m$_{F606W}$-m$_{F814W}$) CMD. The upper panels overplot the isochrone corresponding to the best-fitting solution to the observed CMDs, where the green symbols mark the stars that were actually used for the fit.

\begin{figure*}
        \centering
        \begin{subfigure}{0.45\textwidth}
            \includegraphics[width=\textwidth]{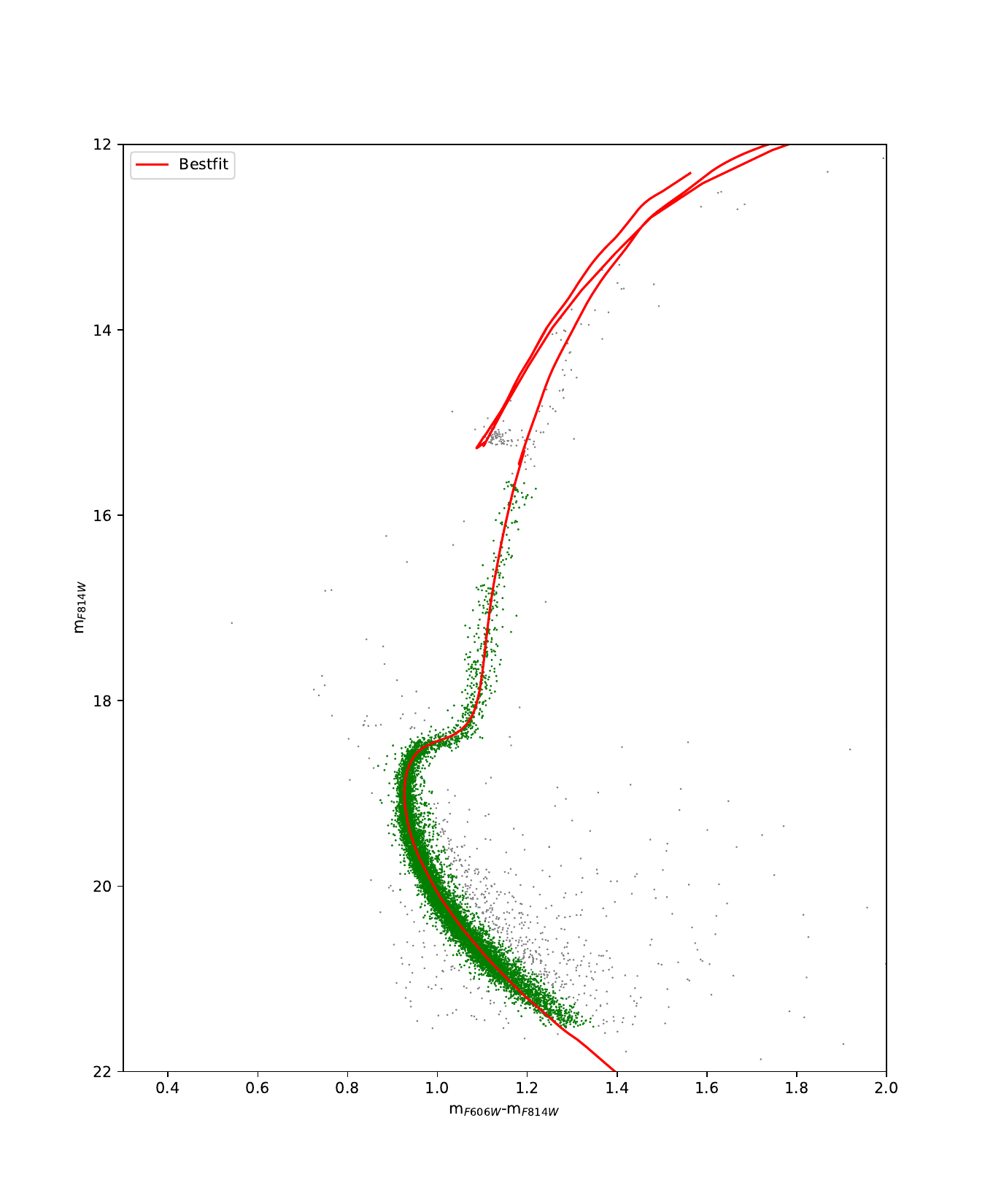}
            \caption[]%
            {{\small }}    
        \end{subfigure}
        \begin{subfigure}{0.45\textwidth}  
            \includegraphics[width=\textwidth]{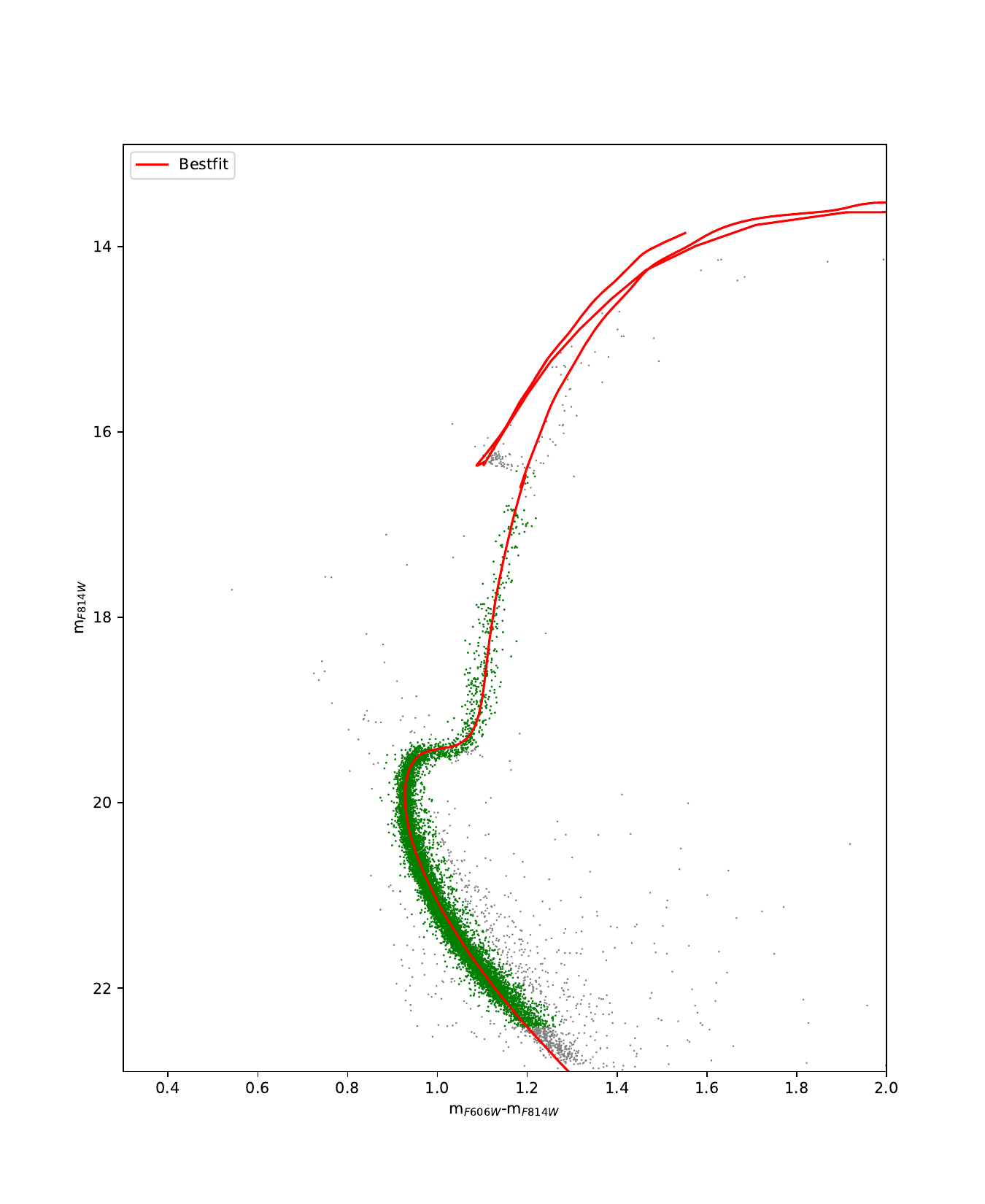}
            \caption[]%
            {{\small }}    
        \end{subfigure}
        \vskip\baselineskip
        \begin{subfigure}{0.45\textwidth}   
            \includegraphics[width=\textwidth]{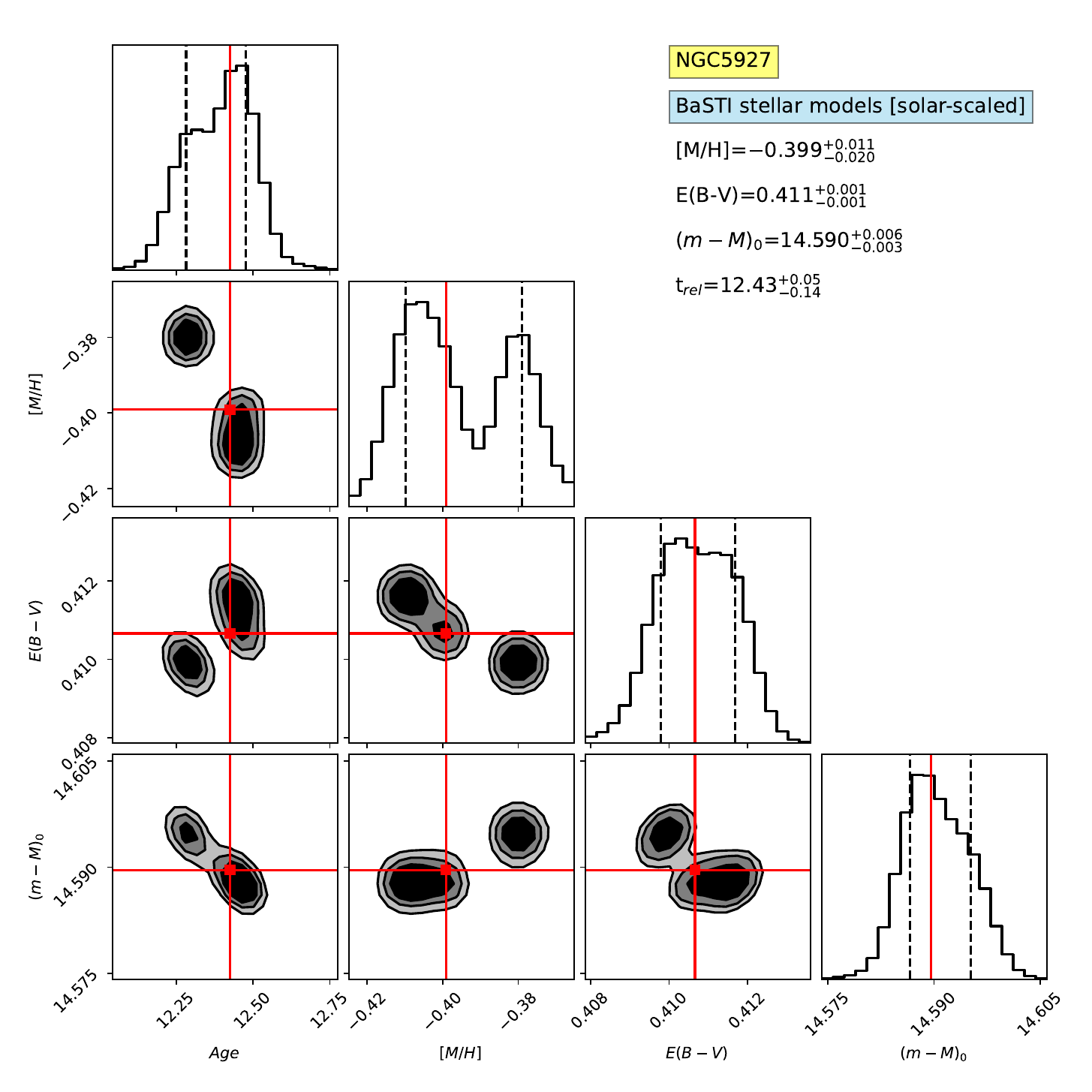}
            \caption[]%
            {{\small }}    
        \end{subfigure}
        \begin{subfigure}{0.45\textwidth}   
            \includegraphics[width=\textwidth]{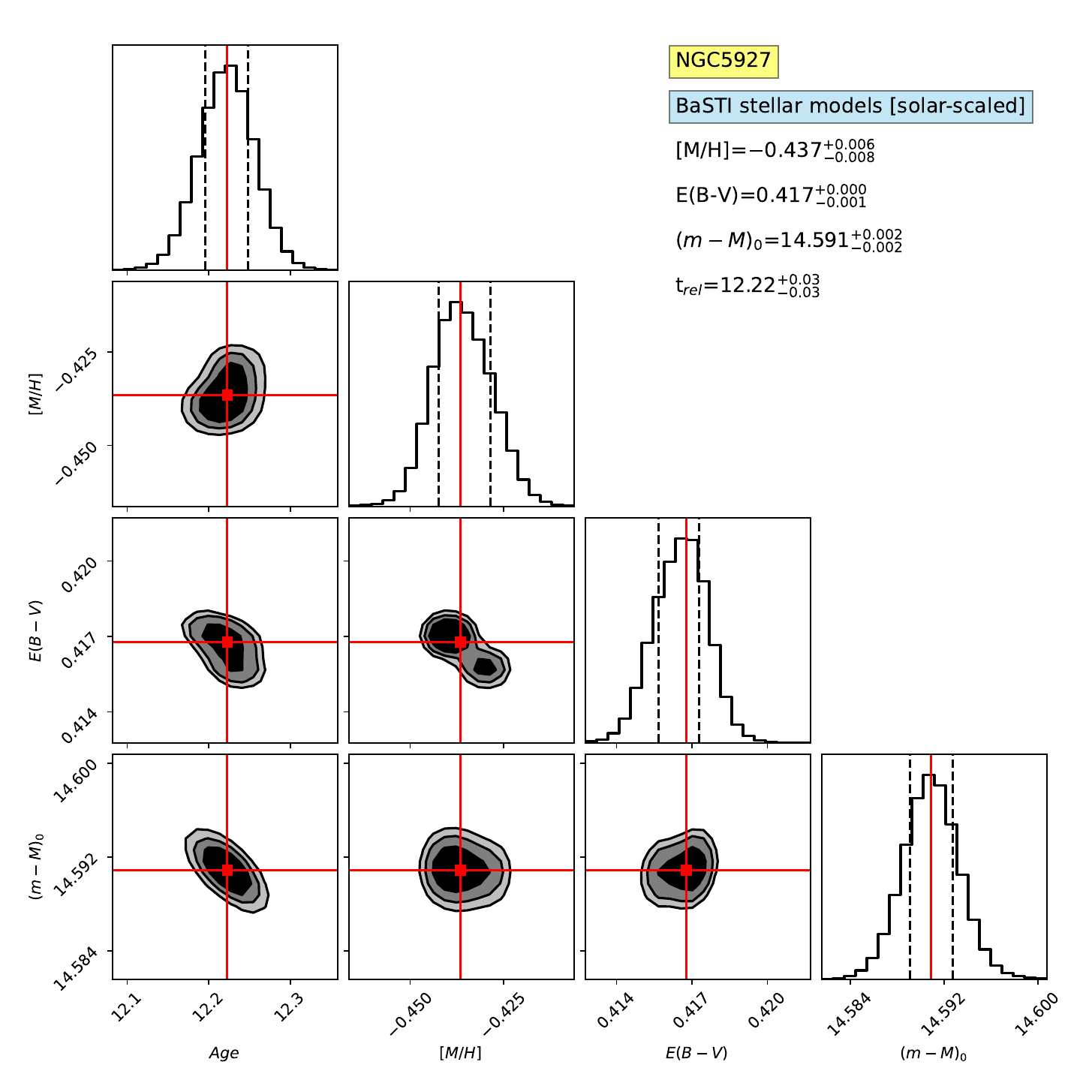}
            \caption[]%
            {{\small }}    
        \end{subfigure}
        \caption[]
        {\small Results for NGC~5927. Panel (a): Best fit model in the (m$_{F814W}$, m$_{F606W}$-m$_{F814W}$) CMD. Panel (b): Best fit model in the (m$_{F606W}$, m$_{F606W}$-m$_{F814W}$) CMD. Panel (c): Posterior distributions for the output parameters and the best-fit solution, quoted in the labels, in the (m$_{F814W}$, m$_{F606W}$-m$_{F814W}$) CMD. Panel (d): Posterior distributions for the output parameters and the best-fit solution, quoted in the labels, in the (m$_{F606W}$, m$_{F606W}$-m$_{F814W}$) CMD.} 
    \end{figure*}

\begin{figure*}
        \centering
        \begin{subfigure}[b]{0.45\textwidth}
            \centering
            \includegraphics[width=\textwidth]{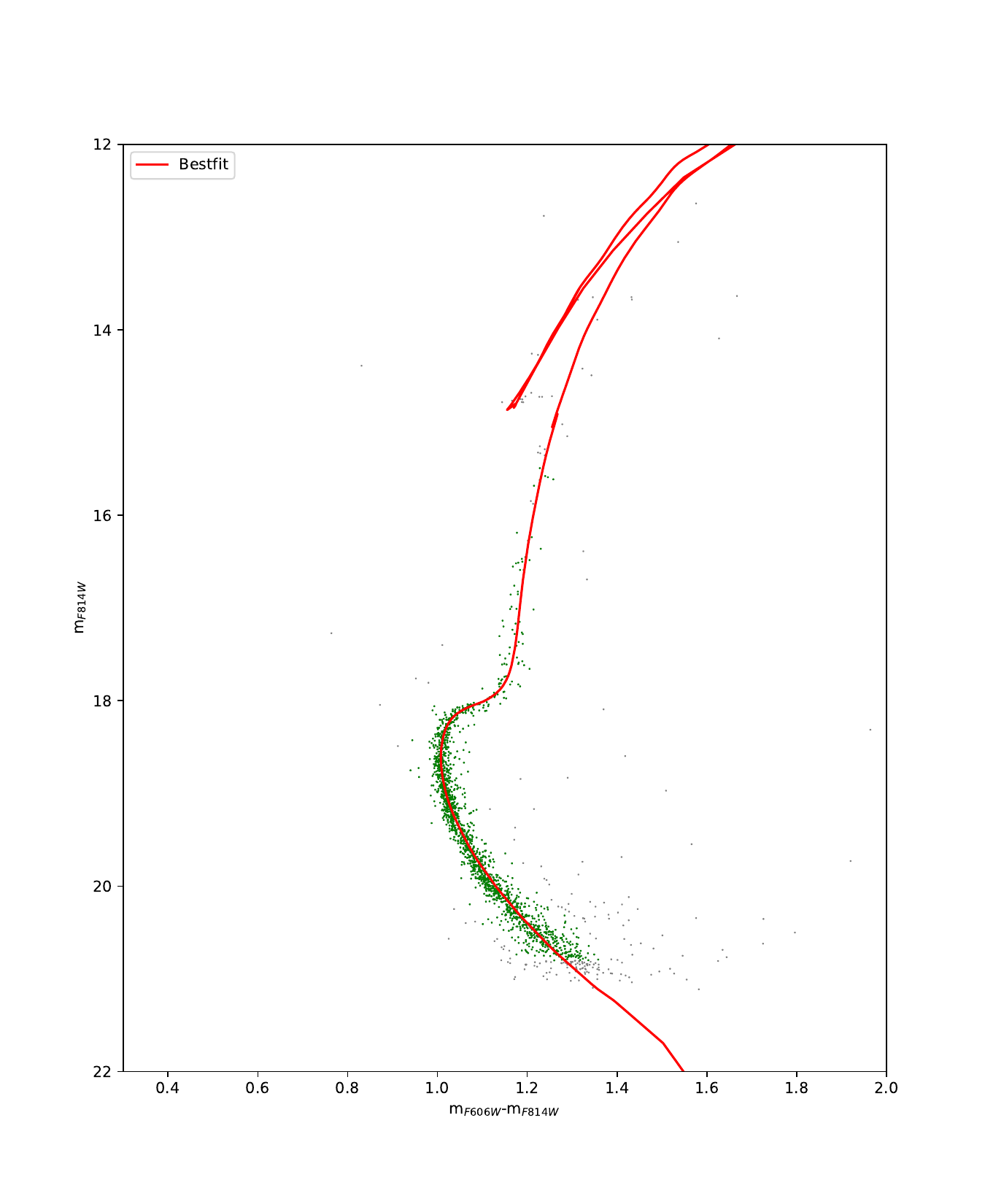}
            \caption[]%
            {{\small }}    
        \end{subfigure}
        \hfill
        \begin{subfigure}[b]{0.45\textwidth}  
            \centering 
            \includegraphics[width=\textwidth]{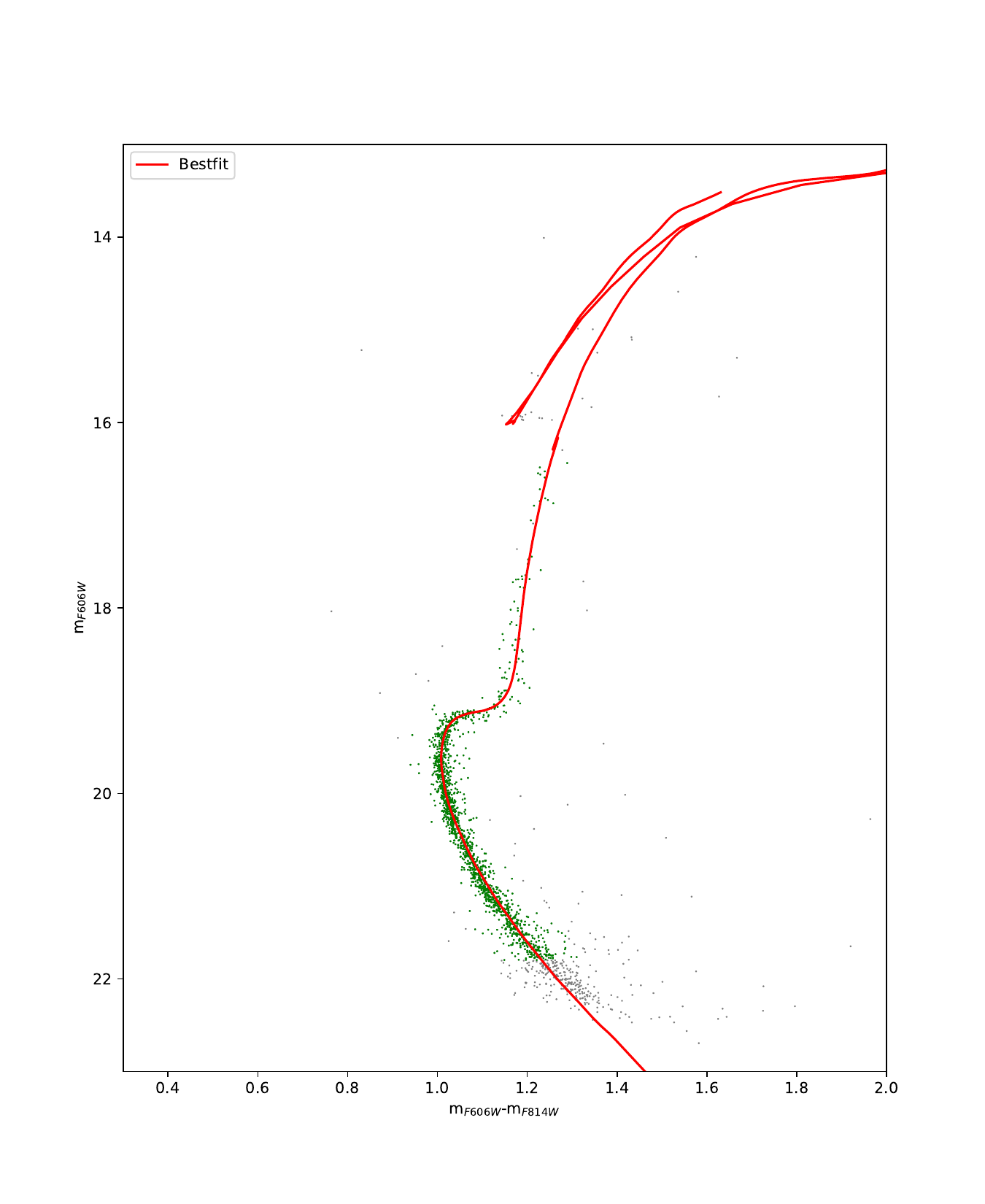}
            \caption[]%
            {{\small }}    
        \end{subfigure}
        \vskip\baselineskip
        \begin{subfigure}[b]{0.45\textwidth}   
            \centering 
            \includegraphics[width=\textwidth]{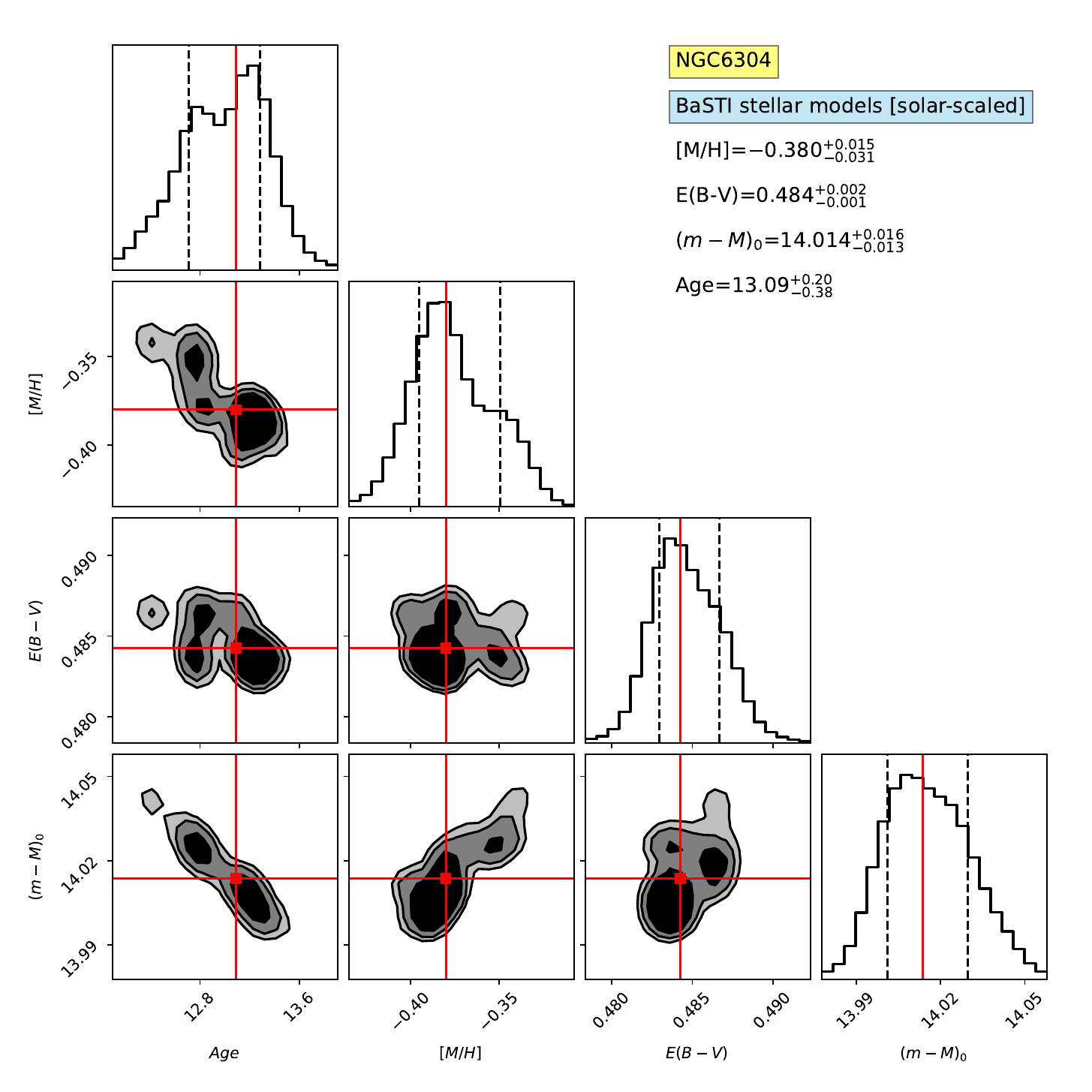}
            \caption[]%
            {{\small }}    
        \end{subfigure}
        \hfill
        \begin{subfigure}[b]{0.45\textwidth}   
            \centering 
            \includegraphics[width=\textwidth]{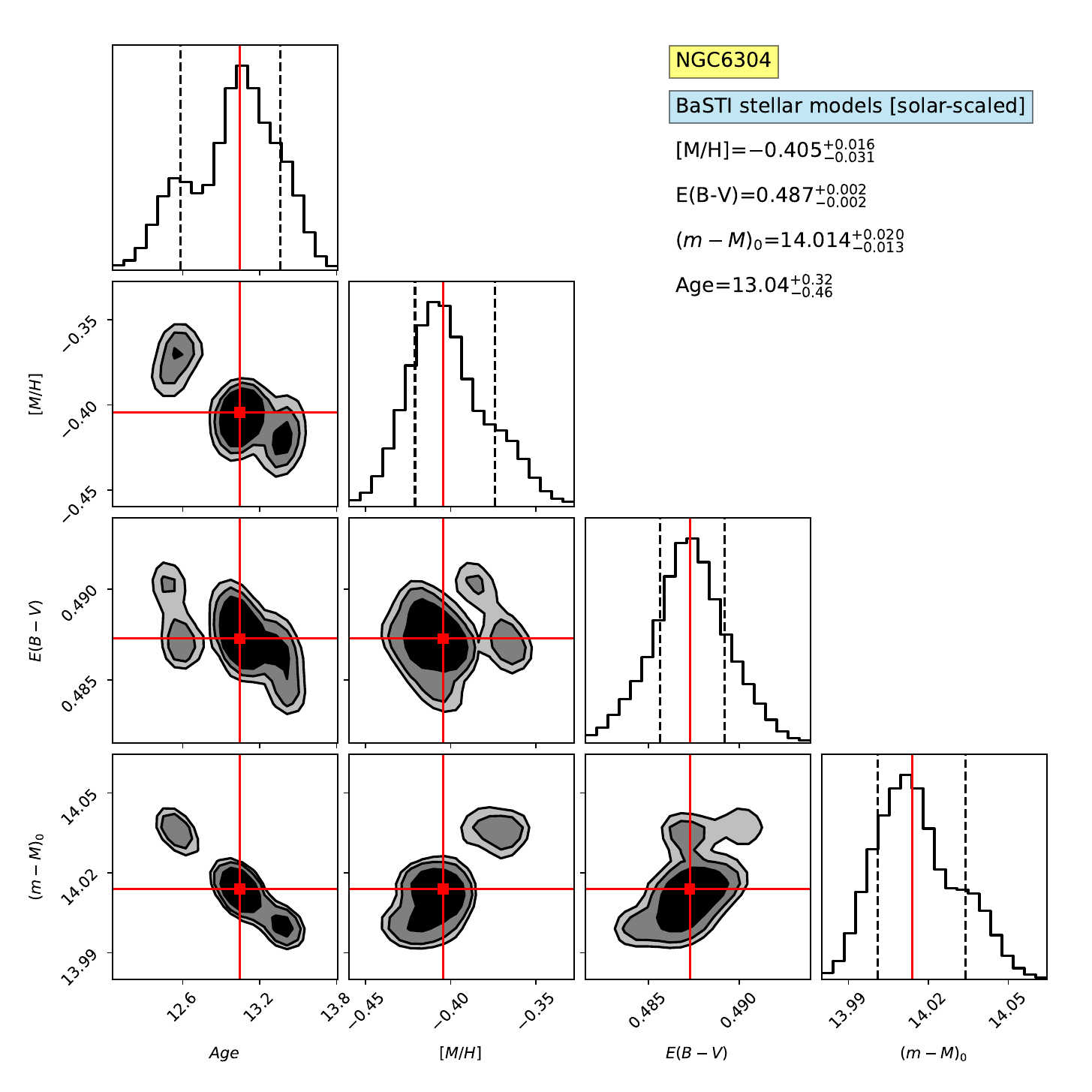}
            \caption[]%
            {{\small }}    
        \end{subfigure}
        \caption[]
        {\small Results for NGC~6304. The meaning of the panels is the same as in Fig.~A.1} 
    \end{figure*}

\begin{figure*}
        \centering
        \begin{subfigure}[b]{0.45\textwidth}
            \centering
            \includegraphics[width=\textwidth]{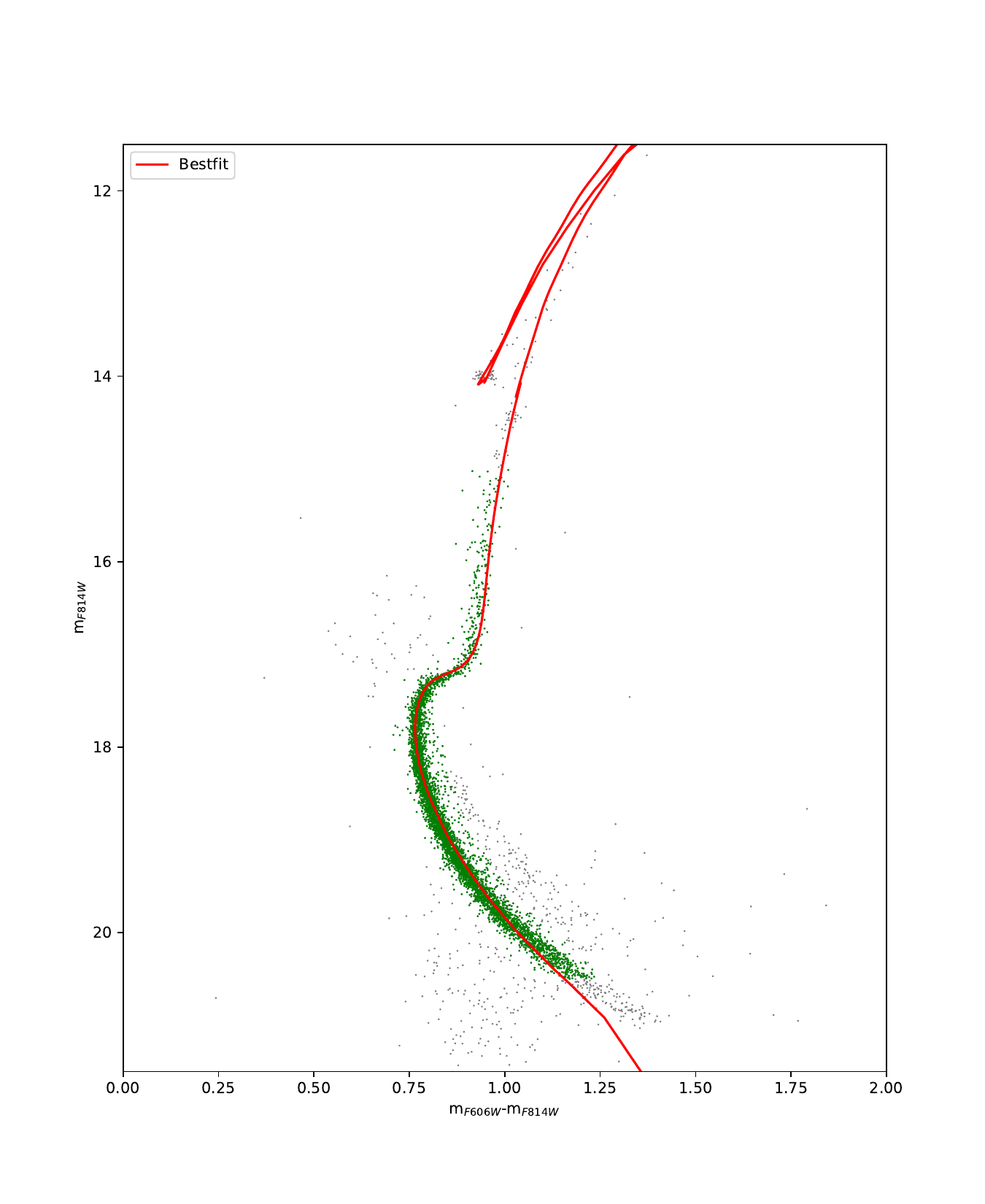}
            \caption[]%
            {{\small }}    
        \end{subfigure}
        \hfill
        \begin{subfigure}[b]{0.45\textwidth}  
            \centering 
            \includegraphics[width=\textwidth]{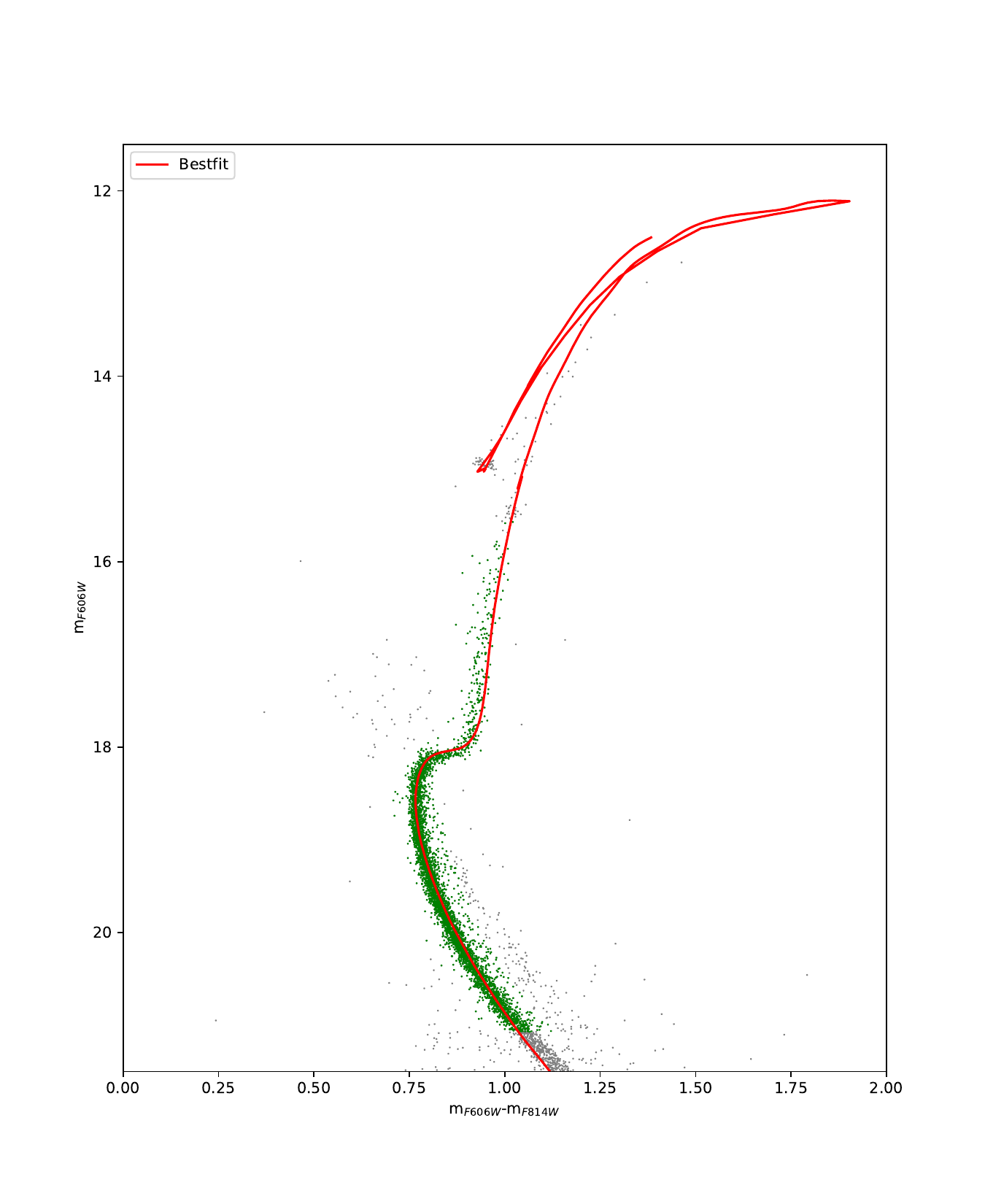}
            \caption[]%
            {{\small }}    
        \end{subfigure}
        \vskip\baselineskip
        \begin{subfigure}[b]{0.45\textwidth}   
            \centering 
            \includegraphics[width=\textwidth]{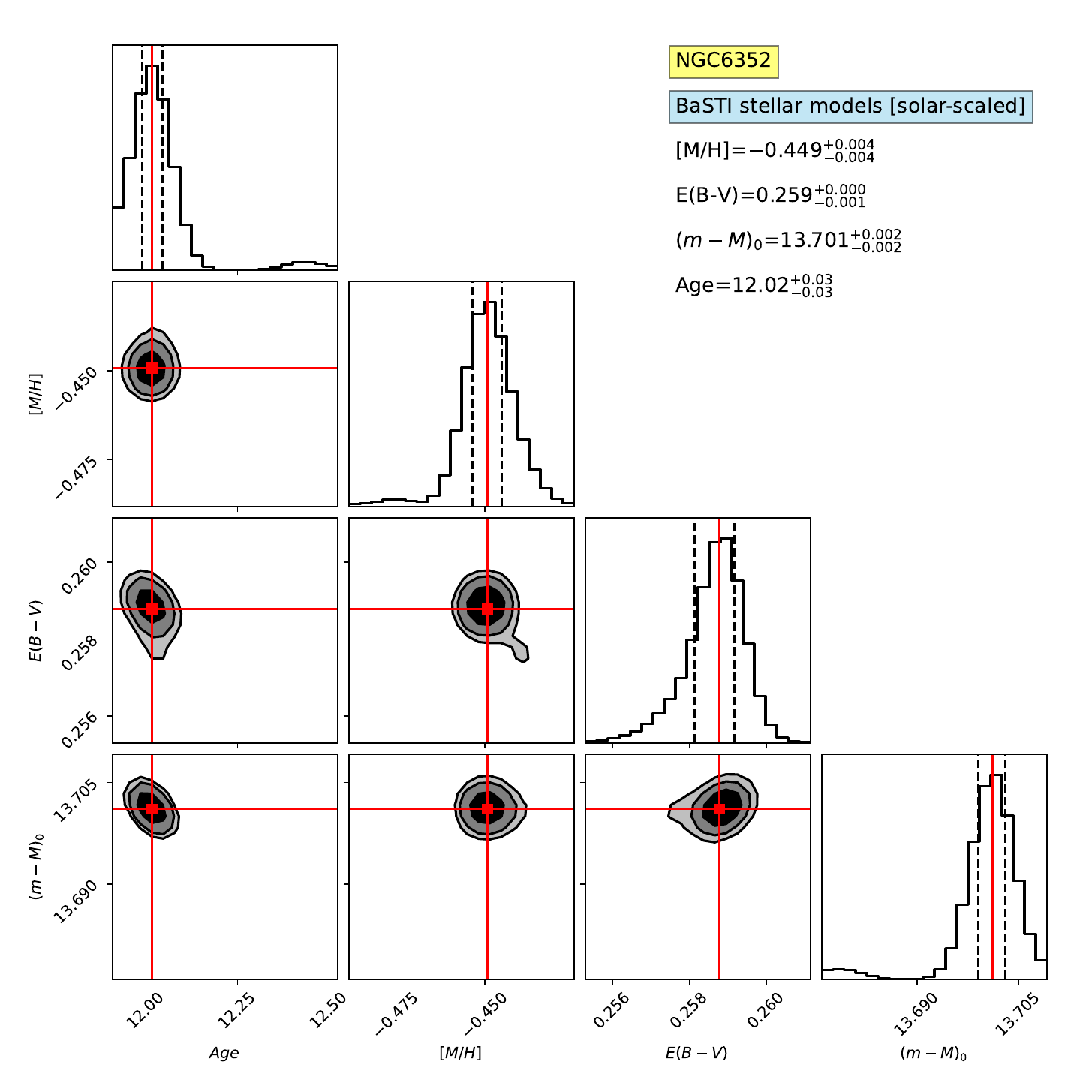}
            \caption[]%
            {{\small }}    
        \end{subfigure}
        \hfill
        \begin{subfigure}[b]{0.45\textwidth}   
            \centering 
            \includegraphics[width=\textwidth]{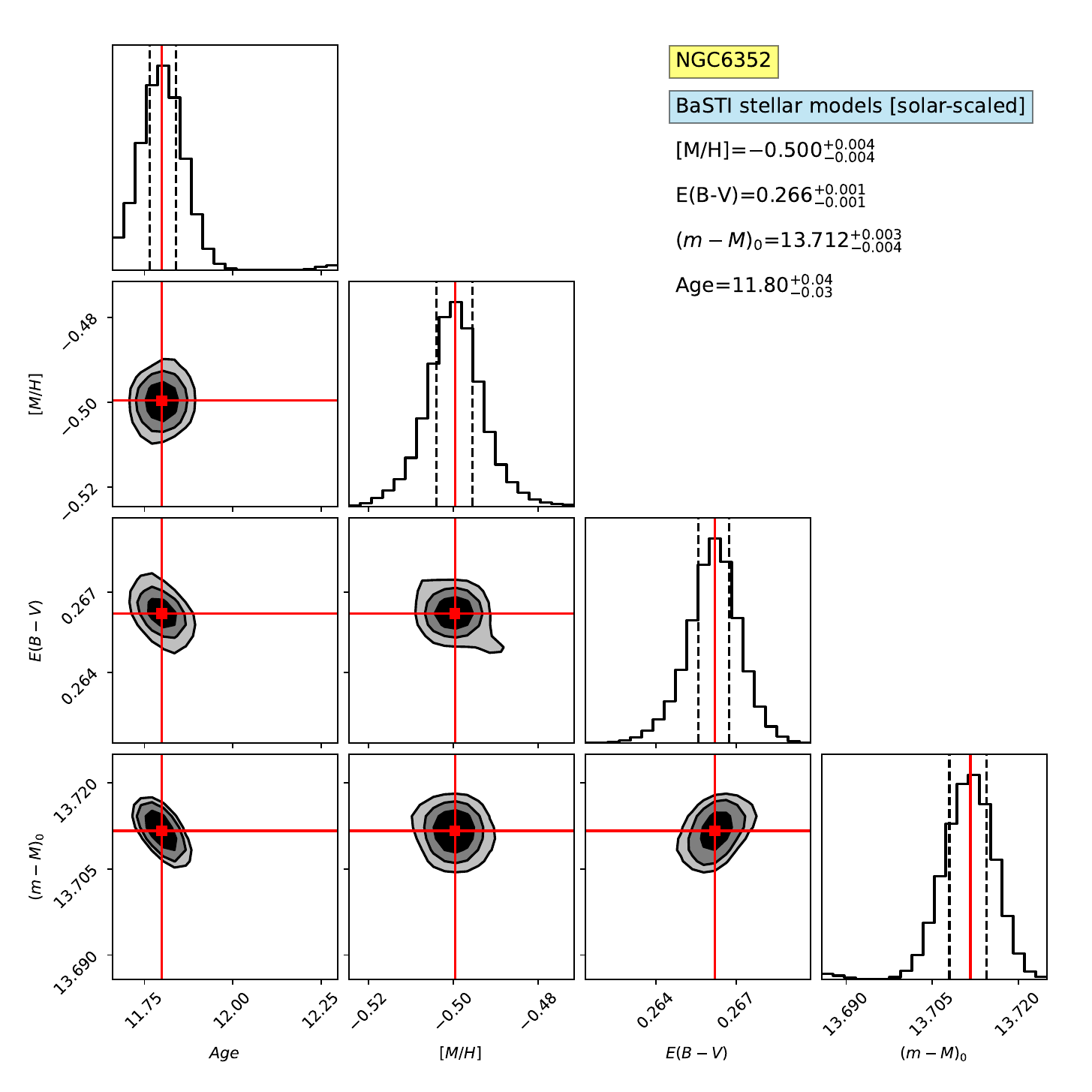}
            \caption[]%
            {{\small }}    
        \end{subfigure}
        \caption[]
        {\small Results for NGC~6352. The meaning of the panels is the same as in Fig.~A.1} 
    \end{figure*}

\begin{figure*}
        \centering
        \begin{subfigure}[b]{0.45\textwidth}
            \centering
            \includegraphics[width=\textwidth]{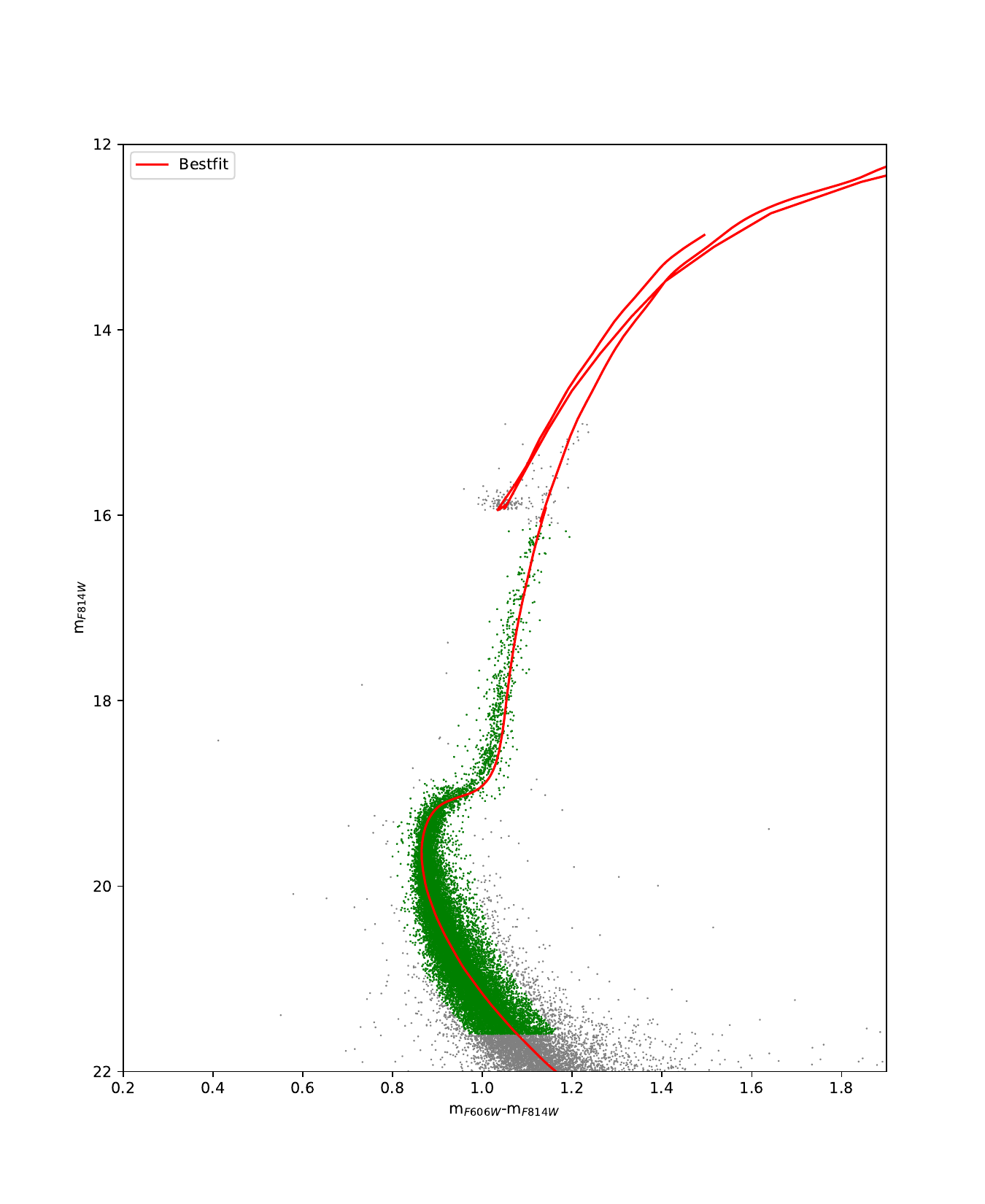}
            \caption[]%
            {{\small }}    
        \end{subfigure}
        \hfill
        \begin{subfigure}[b]{0.45\textwidth}  
            \centering 
            \includegraphics[width=\textwidth]{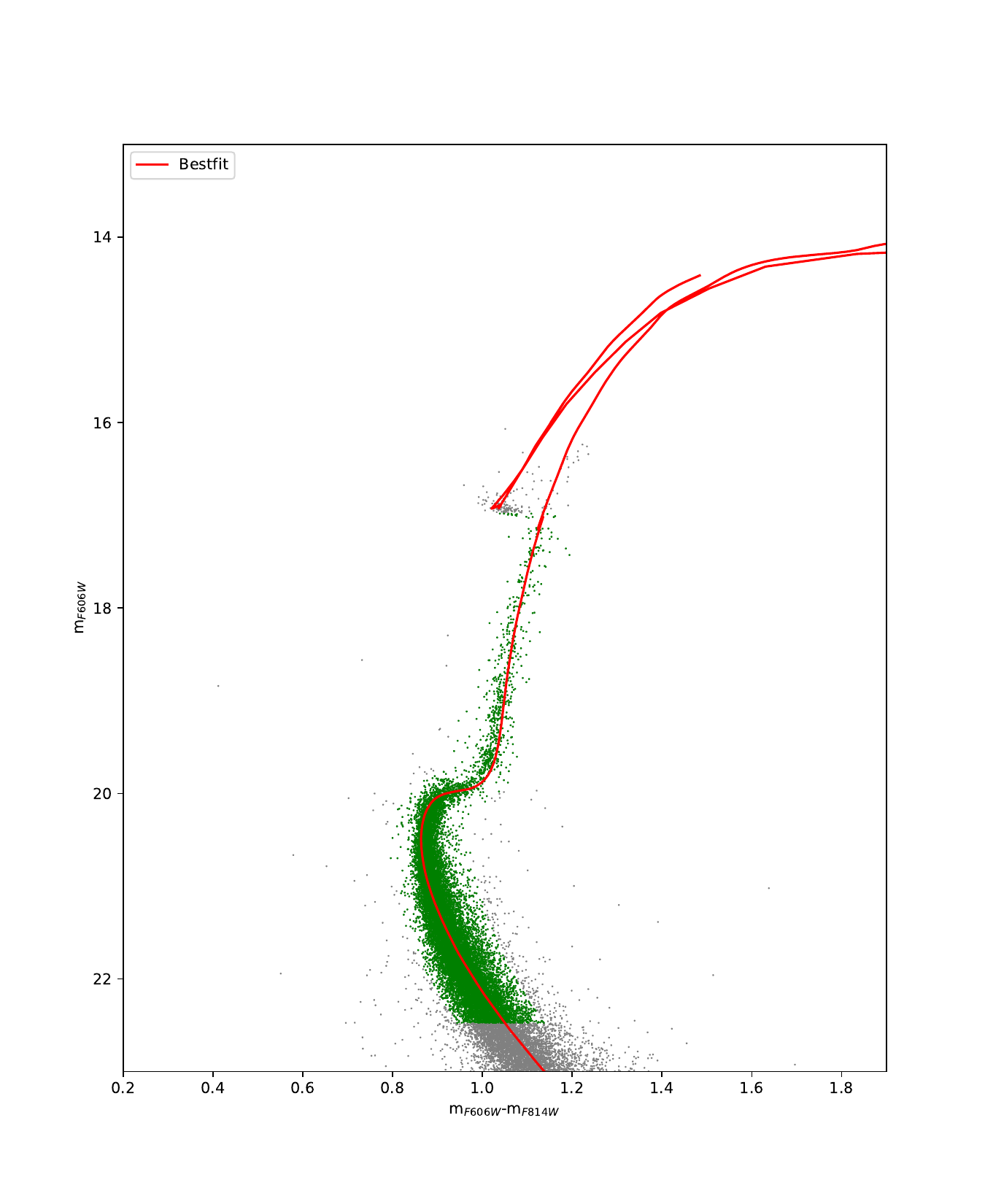}
            \caption[]%
            {{\small }}    
        \end{subfigure}
        \vskip\baselineskip
        \begin{subfigure}[b]{0.45\textwidth}   
            \centering 
            \includegraphics[width=\textwidth]{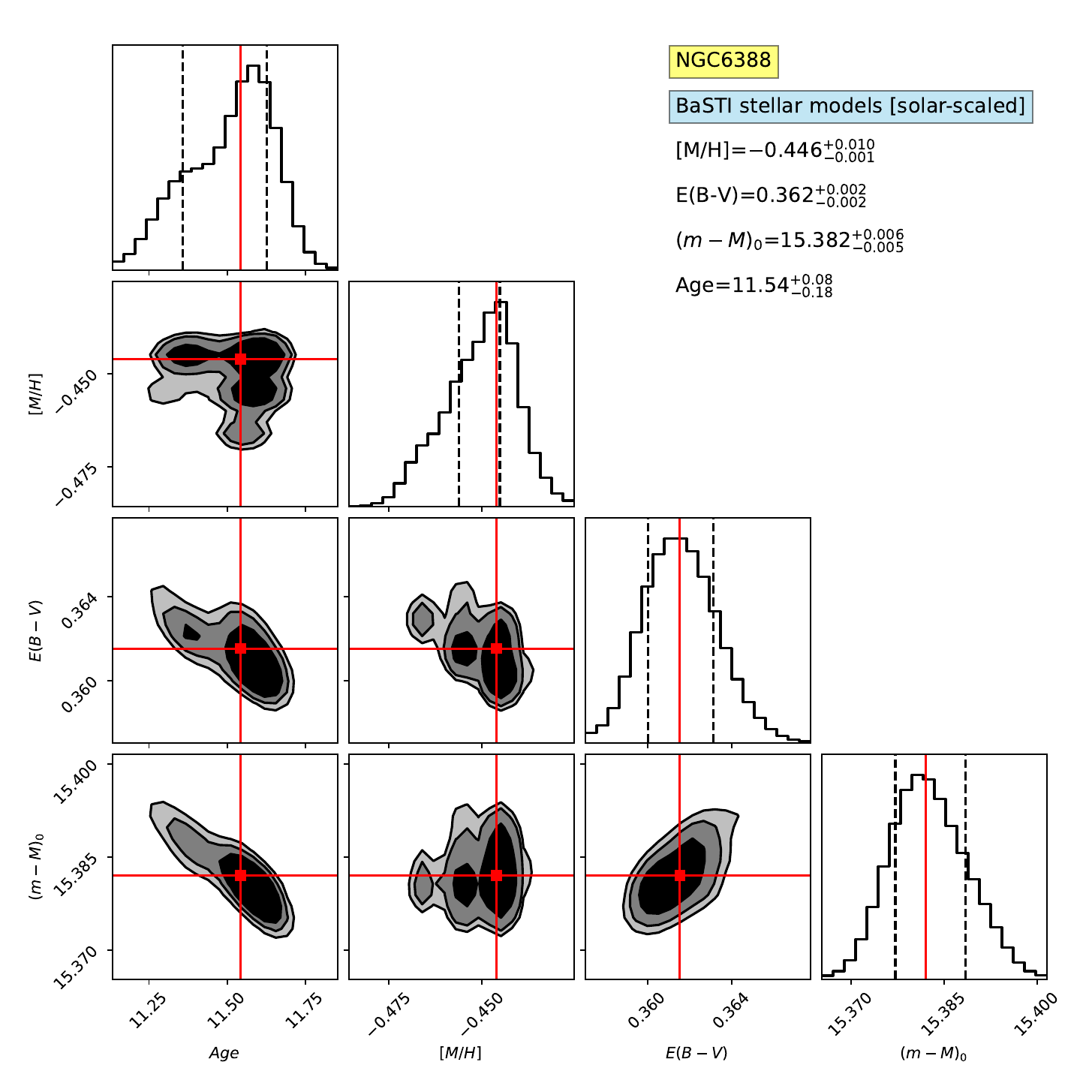}
            \caption[]%
            {{\small }}    
        \end{subfigure}
        \hfill
        \begin{subfigure}[b]{0.45\textwidth}   
            \centering 
            \includegraphics[width=\textwidth]{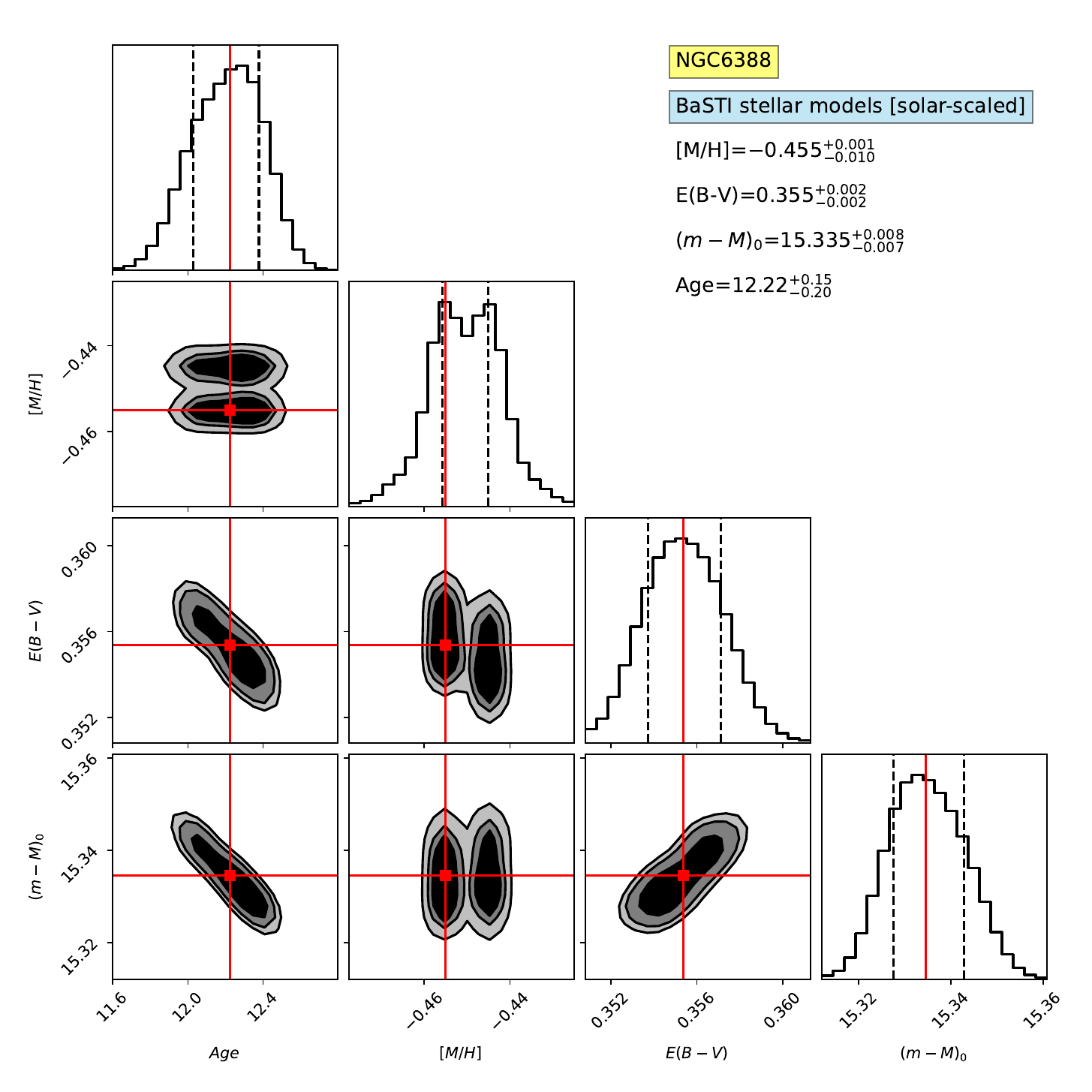}
            \caption[]%
            {{\small }}    
        \end{subfigure}
        \caption[]
        {\small Results for  NGC~6388. The meaning of the panels is the same as in Fig.~A.1} 
    \end{figure*}

\begin{figure*}
        \centering
        \begin{subfigure}{0.45\textwidth}
            \includegraphics[width=\textwidth]{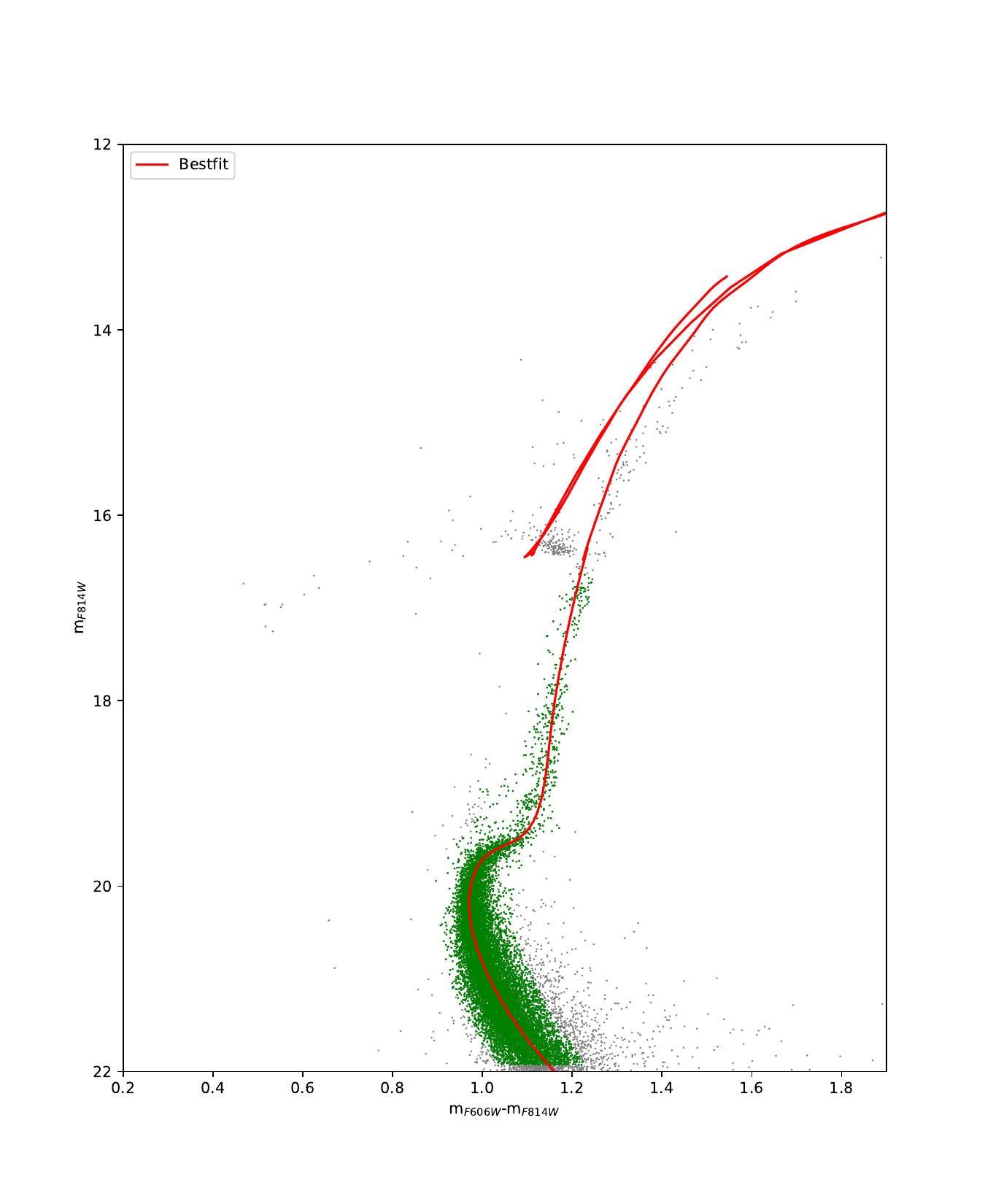}
            \caption[]%
            {{\small }}    
        \end{subfigure}
        \begin{subfigure}{0.45\textwidth}  
            \includegraphics[width=\textwidth]{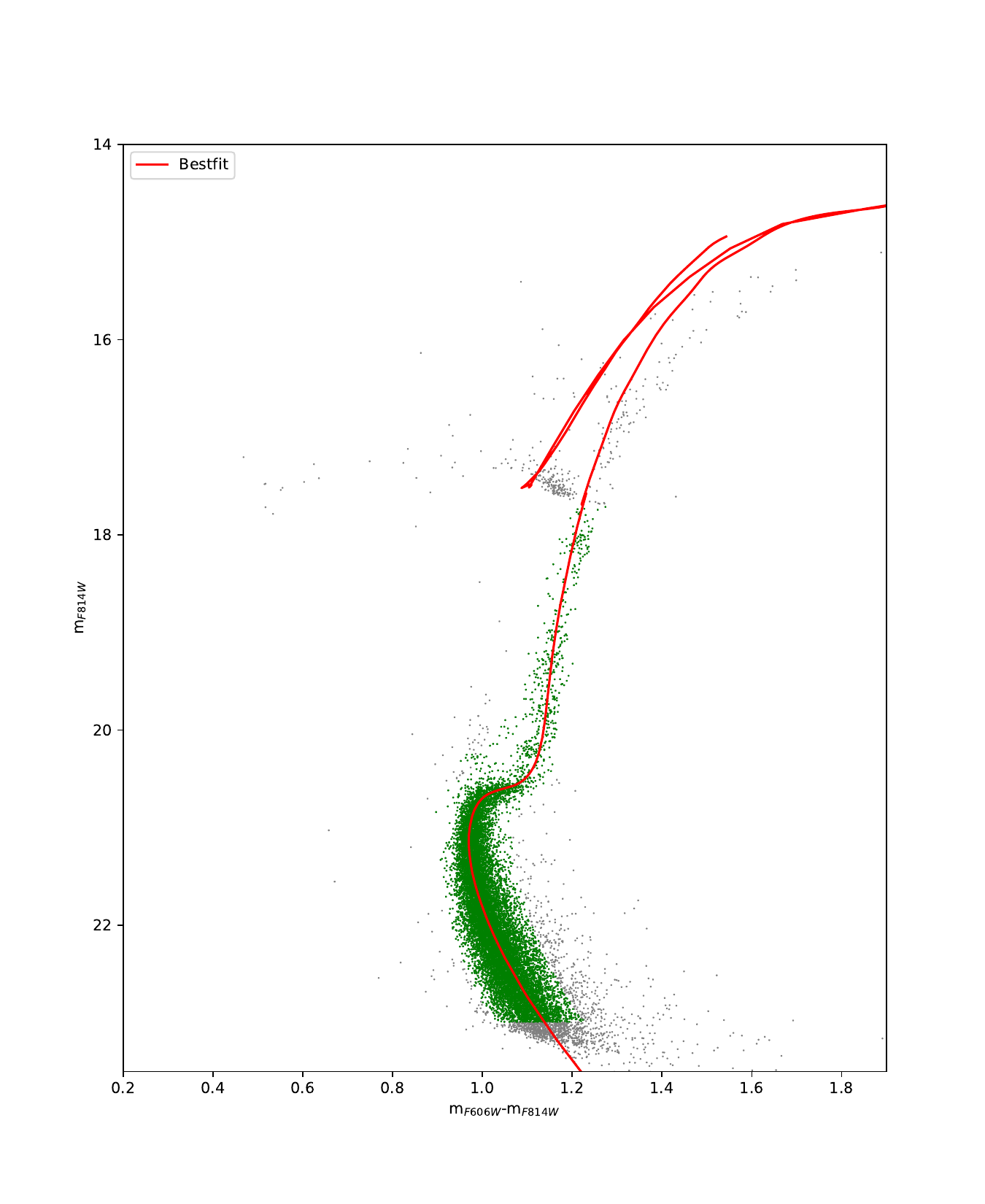}
            \caption[]%
            {{\small }}    
        \end{subfigure}
        \vskip\baselineskip
        \begin{subfigure}{0.45\textwidth}   
            \includegraphics[width=\textwidth]{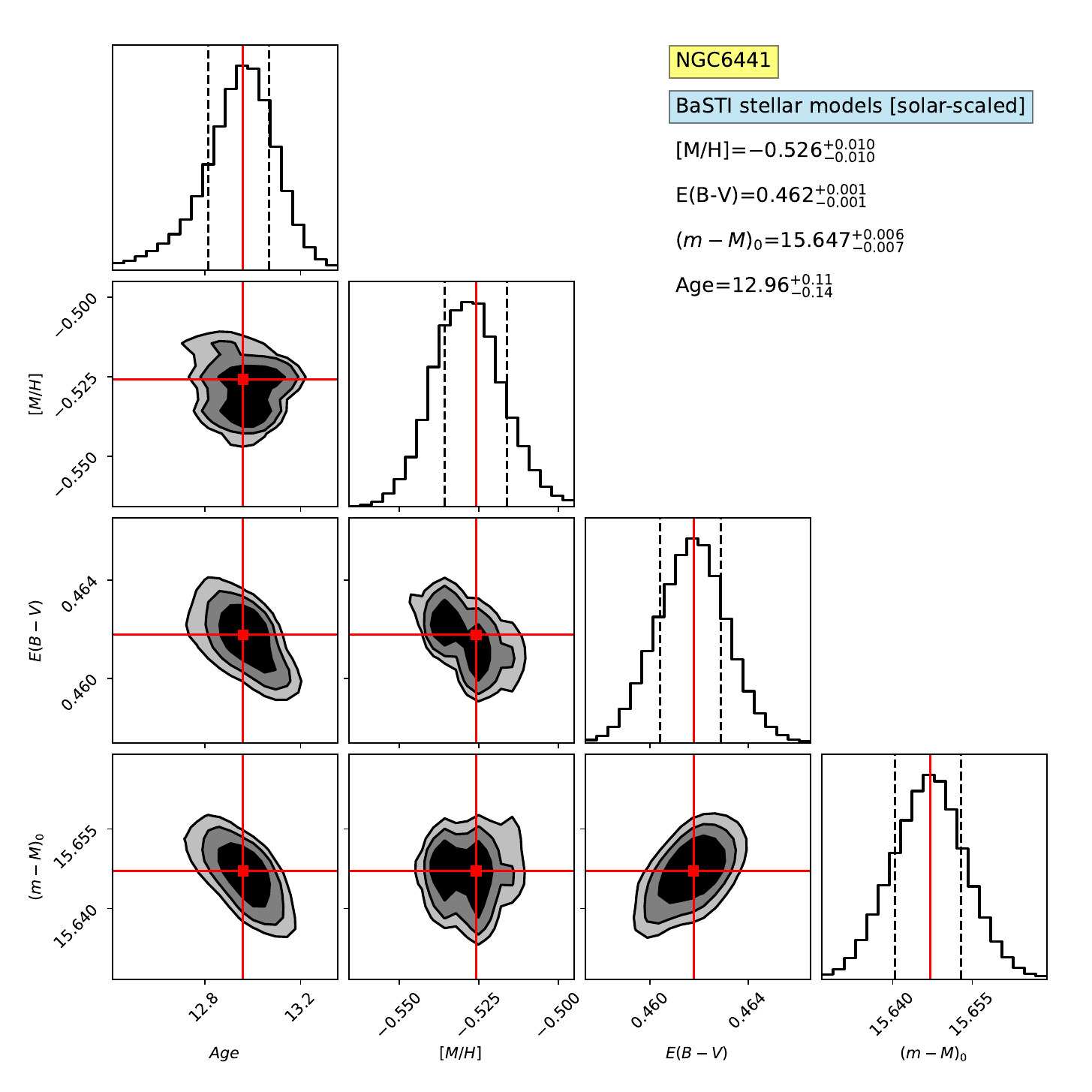}
            \caption[]%
            {{\small }}    
        \end{subfigure}
        \begin{subfigure}{0.45\textwidth}   
            \includegraphics[width=\textwidth]{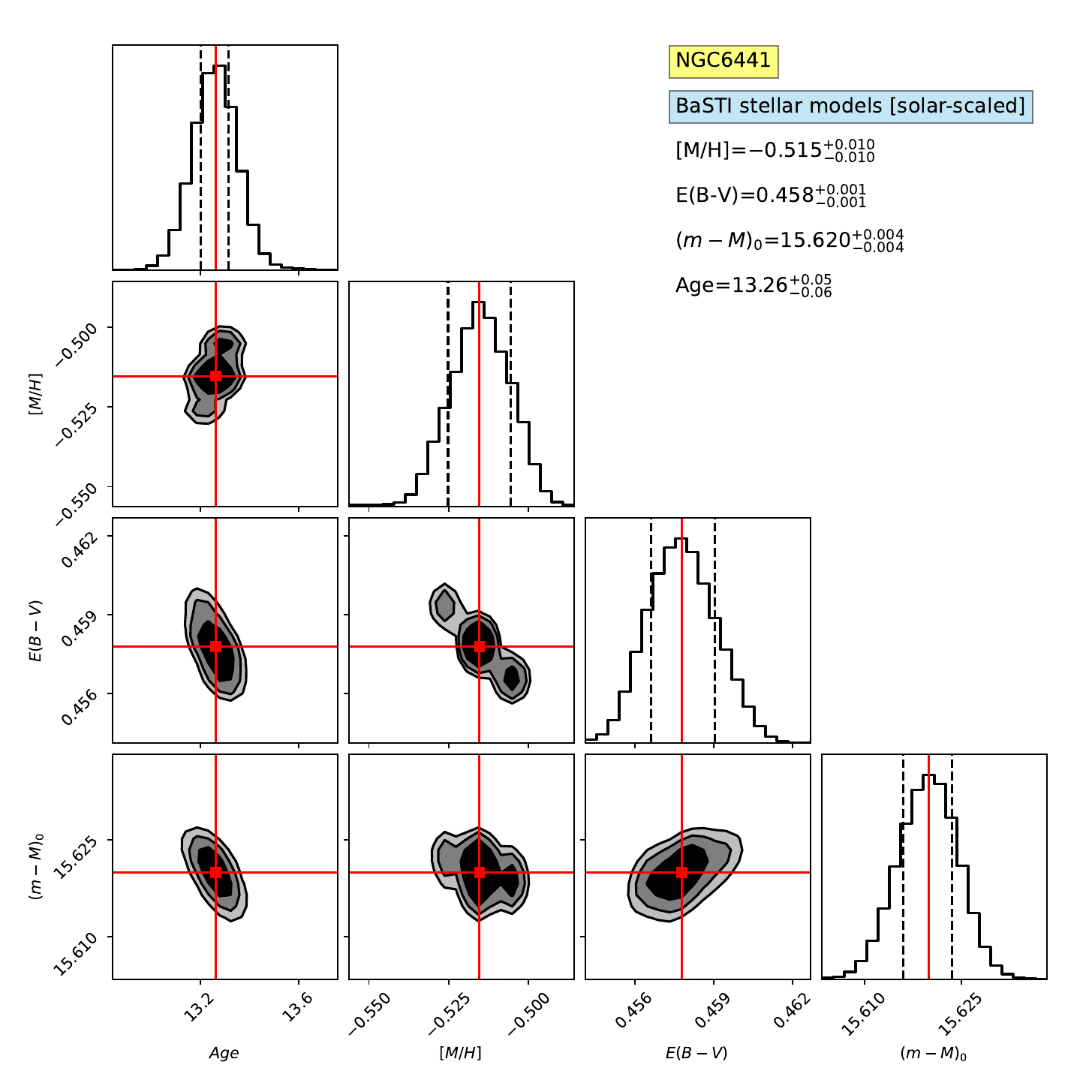}
            \caption[]%
            {{\small }}    
        \end{subfigure}
        \caption[]
        {\small Results for NGC~6441. The meaning of the panels is the same as in Fig.~A.1} 
    \end{figure*}

\begin{figure*}
        \centering
        \begin{subfigure}[b]{0.45\textwidth}
            \centering
            \includegraphics[width=\textwidth]{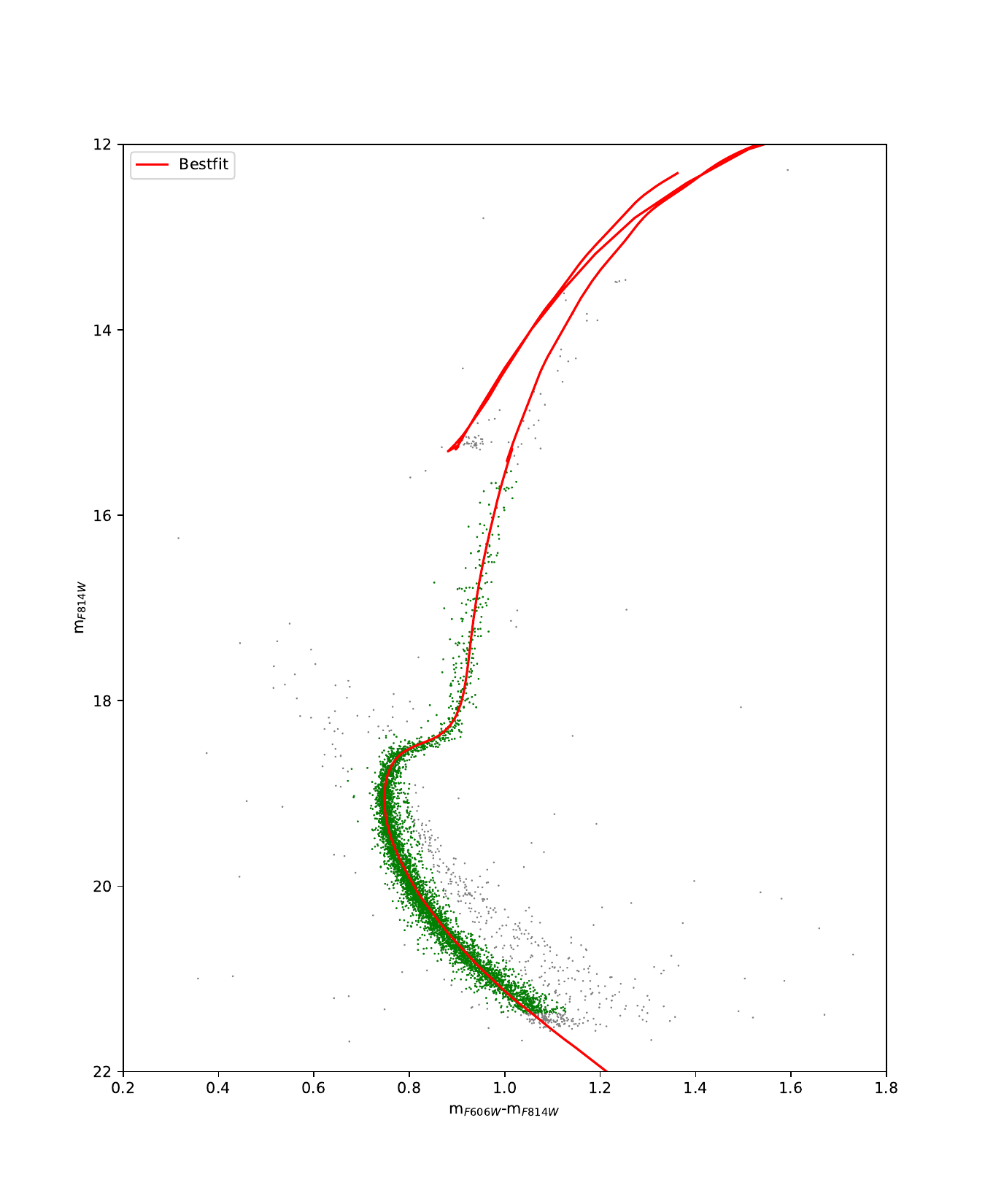}
            \caption[]%
            {{\small }}    
        \end{subfigure}
        \hfill
        \begin{subfigure}[b]{0.45\textwidth}  
            \centering 
            \includegraphics[width=\textwidth]{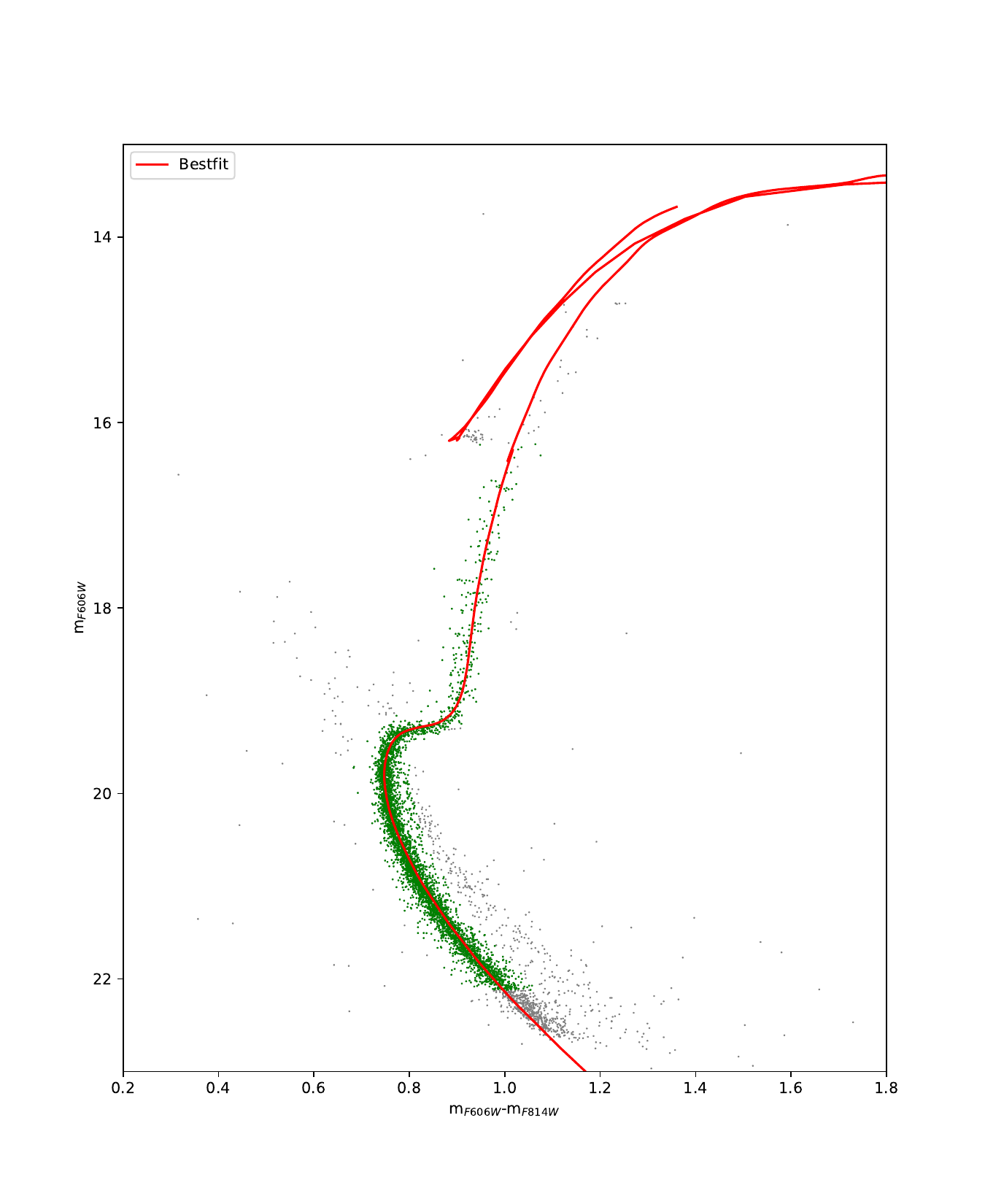}
            \caption[]%
            {{\small }}    
        \end{subfigure}
        \vskip\baselineskip
        \begin{subfigure}[b]{0.45\textwidth}   
            \centering 
            \includegraphics[width=\textwidth]{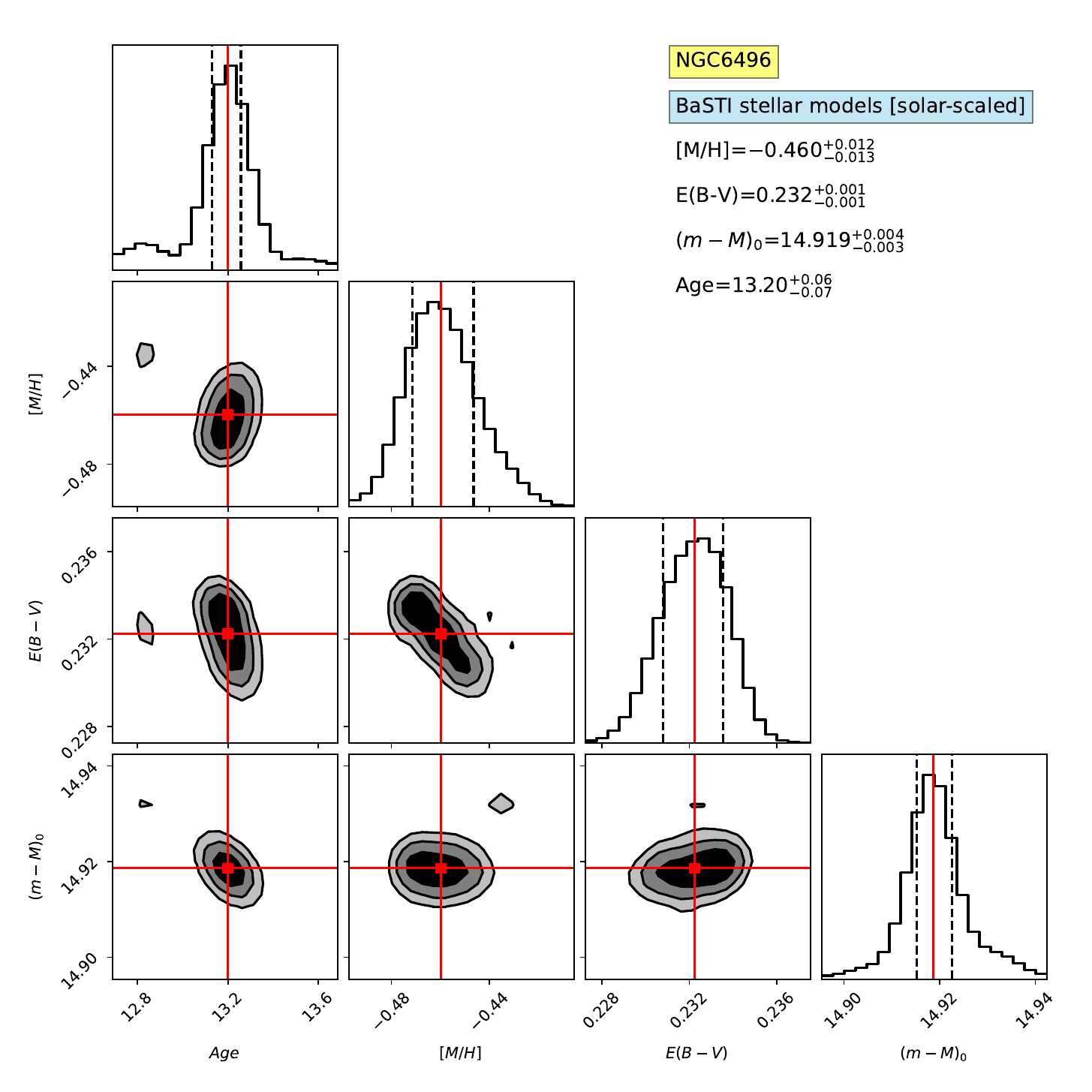}
            \caption[]%
            {{\small }}    
        \end{subfigure}
        \hfill
        \begin{subfigure}[b]{0.45\textwidth}   
            \centering 
            \includegraphics[width=\textwidth]{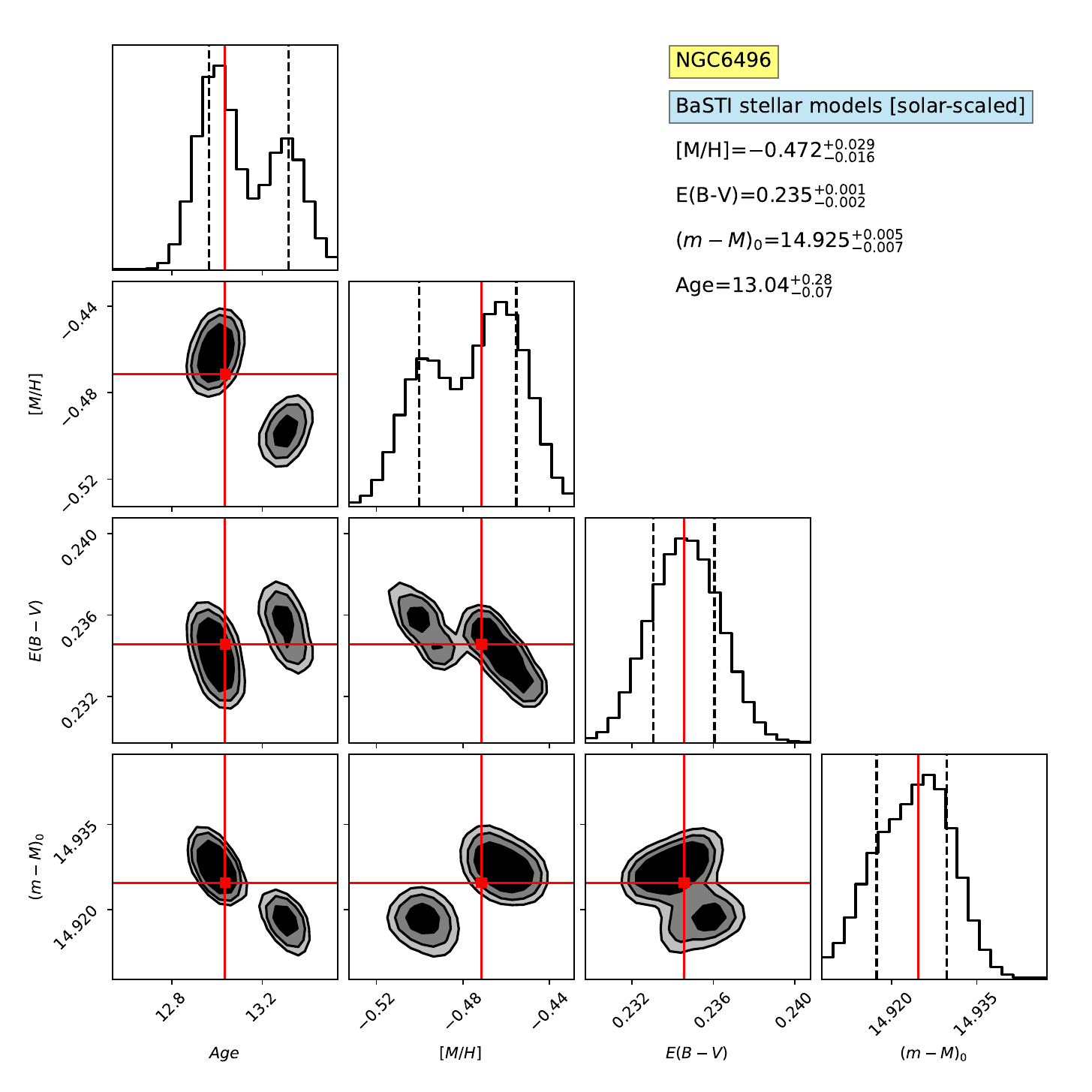}
            \caption[]%
            {{\small }}    
            \label{fig:mean and std of net44}
        \end{subfigure}
        \caption[]
        {\small Results for NGC~6496. The meaning of the panels is the same as in Fig.~A.1} 
    \end{figure*}
  
\end{appendix}
\end{document}